\documentclass[fleqn,usenatbib]{mnras}

\usepackage{newtxtext,newtxmath}
\usepackage[T1]{fontenc}

\usepackage{graphicx}	% Including figure files
\usepackage{amsmath}	% Advanced maths commands
\usepackage{natbib}
\usepackage{array}
\usepackage{url}
\usepackage{color}
\usepackage[dvipsnames]{xcolor}

\title[Stellar-based $M_{\rm BH}$ measurement of NGC 6958]{Cross-checking SMBH mass estimates in NGC~6958 - I: Stellar dynamics from adaptive optics-assisted MUSE observations}

\author[S. Thater et al.]{Sabine Thater,$^{1,2}$\thanks{E-mail: sabine.thater@univie.ac.at}
Davor Krajnovi\'{c},$^{2}$
Peter M. Weilbacher,$^{2}$
Dieu D. Nguyen,$^{3,4}$
\newauthor
Martin Bureau,$^{5,6}$
Michele Cappellari,$^{5}$
Timothy A. Davis,$^{7}$
Satoru Iguchi,$^{8}$
\newauthor
Richard McDermid,$^{9,10}$
Kyoko Onishi,$^{11}$
Marc Sarzi,$^{12}$
Glenn van de Ven$^{1}$
\\
$^{1}$ Department of Astrophysics, University of Vienna, T\"urkenschanzstraße 17, 1180 Vienna, Austria\\
$^{2}$ Leibniz-Institute for Astrophysics Potsdam (AIP), An der Sternwarte 16, 14482 Potsdam, Germany\\
$^{3}$ Department of Physics, International University, Quarter 6, Linh Trung Ward, Thu Duc City, Ho Chi Minh City, Vietnam\\
$^{4}$ Vietnam National University, Ho Chi Minh City, Vietnam\\
$^{5}$ Sub-department  of  Astrophysics,  Department  of  Physics,  University  of  Oxford,  Denys  Wilkinson  Building,  Keble  Road,  Oxford  OX13RH, UK\\
$^{6}$ Yonsei Frontier Lab and Department of Astronomy, Yonsei University, 50 Yonsei-ro, Seodaemun-gu, Seoul 03722, Republic of Korea\\
$^{7}$ School of Physics \& Astronomy, Cardiff University, Queens Buildings, The Parade, Cardiff CF24 3AA, UK\\
$^{8}$ National Astronomocial Observatory of Japan (NAOJ), National Institute of Natural Sciences (NINS), 2-21-1 Osawa, Mitaka, Tokyo 181-8588, Japan\\
$^{9}$ Department of Physics and Astronomy, Macquarie University, NSW 2109, Australia\\
$^{10}$ Astronomy, Astrophysics and Astrophotonics Research Centre, Macquarie University, Sydney, NSW 2109, Australia\\
$^{11}$ Department  of  Space,  Earth  and  Environment,  Chalmers  University  of  Technology,  Onsala  Observatory,  SE-439  92  Onsala,  Sweden\\
$^{12}$ Armagh Observatory and Planetarium, College Hill, Armagh, BT61 9DG, UK
}

\date{Accepted 2021 November 2. Received 2021 November 2; in original form 2021 May 13}

\pubyear{2021}

\begin{document}
\label{firstpage}
\pagerange{\pageref{firstpage}--\pageref{lastpage}}
\maketitle
\begin{abstract}
Supermassive black hole masses ($M_{\rm BH}$) can dynamically be estimated with various methods and using different kinematic tracers. Different methods have only been cross-checked for a small number of galaxies and often show discrepancies. To understand these discrepancies, detailed cross-comparisons of additional galaxies are needed. We present the first part of our cross-comparison between stellar- and gas-based $M_{\rm BH}$ estimates in the nearby fast-rotating early-type galaxy NGC 6958. 
The measurements presented here are based on ground-layer adaptive optics-assisted Multi-Unit Spectroscopic Explorer (MUSE) science verification data at around 0\farcs6 spatial resolution. The spatial resolution is a key ingredient for the measurement and we provide a Gaussian parametrisation of the adaptive optics-assisted point spread function (PSF) for various wavelengths. From the MUSE data, we extracted the stellar kinematics and constructed dynamical models. Using an axisymmetric Schwarzschild technique, we measured an $M_{\rm BH}$ of $(3.6^{+2.7}_{-2.4}) \times 10^8\,$M$_{\odot}$ at $3\sigma$ significance taking kinematical and dynamical systematics (e.g., radially-varying mass-to-light ratio) into account. We also added a dark halo, but our data does not allow to constrain the dark matter fraction. Adding dark matter with an abundance matching prior results in a 25\% more massive black hole.
Jeans anisotropic models return $M_{\rm BH}$ of $(4.6^{+2.5}_{-2.7}) \times 10^8M_{\odot}$ and $(8.6^{+0.8}_{-0.8}) \times 10^8M_{\odot}$ at $3\sigma$ confidence for spherical and cylindrical alignment of the velocity ellipsoid, respectively. 
In a follow-up study, we will compare the stellar-based $M_{\rm BH}$ with those from cold and warm gas tracers, which will provide additional constraints for the $M_{\rm BH}$ for NGC 6958, and insights into assumptions that lead to potential systematic uncertainty.

\end{abstract}
%%%%%%%%%%%%%%%%%%%%%%%%%%%%%%%%%%%%%%%%%\textbf{}%%%%%%%%%

%%%%%%%%%%%%%%%%% BODY OF PAPER %%%%%%%%%%%%%%%%%%

\section{Introduction}\label{s:intro}
The developments in astronomical instrumentation over the last two decades have substantially improved the capability of astronomical observations: remarkably, it is now possible to robustly measure the mass of supermassive black holes (SMBH) in nearby galaxies with a precision of less than a factor of two using a variety of different methods \citep[e.g., review by][]{Kormendy2013}. Determining robust black hole masses is a challenging task that requires the best possible spatial resolution for both photometric and spectroscopic observations and sophisticated modelling methods. As black holes are per se not visible, we need to trace the motion of the material that is sensitive to the gravitational potential of the SMBH. Popular tracers are individual stars (in the Milky Way; \citealt{Ghez2008, Gillessen2009, Gillessen2017}), masers \citep[e.g.,][]{Miyoshi1995,Kuo2011,Greene2016,Gao2017}, ionised \citep[e.g.,][]{Beifiori2012,Walsh2013}, molecular \citep[e.g.,][]{Davis2013,Onishi2015,Onishi2017,Davis2018, Boizelle2019, Boizelle2021, Davis2020, Nguyen2020,Nguyen2021b} or atomic gas \citep{Nguyen2021} and unresolved stellar systems \citep[e.g.,][]{Rusli2013,Saglia2016,Nguyen2017a,Nguyen2018,Krajnovic2018,Nguyen2019a,Thater2019}. While a variety of different tracers and methods are available, it is not possible to use a single modelling method to measure $M_{\rm BH}$ for all different types of galaxies. Stellar-based methods can be best used for early-type galaxies which usually do not have strongly varying
stellar populations nor sub-components like bars or spiral arms. On the other hand, gas is typically found in late-type galaxies and can be used as a tracer if the gas follows circular orbits and is not strongly disturbed. Other methods require the presence of nuclear maser emission or an active galactic nucleus. Checks for inconsistencies between the different mass determination methods are important for identifying systematic uncertainties associated with the techniques and deriving robust massive black hole masses. 

Hitherto, cross-checks between different dynamical modelling methods have only been performed for a handful of objects. While the checks give consistent results in a number of cases \citep{Shapiro2006, vandenBosch2010, Davies2007a, Pastorini2007, Neumayer2007, Cappellari2009, Feldmeier2014,FeldmeierKrause2017,Krajnovic2018}, many cross-checks reveal clear discrepancies \citep[e.g.,][]{VerdoesKleijn2002, deFrancesco2006, Gebhardt2011, Walsh2012, Walsh2013, Onken2014,Barth2016, Ferrarese1996, Boizelle2021}. Figure 2 in \cite{Thater2020} shows that $M_{\rm BH}$ determined from ionised and molecular gas-based measurements seem to be systematically lower than those derived from stellar dynamical models for $M_{\rm BH}$ greater than a few $10^8M_{\odot}$. 
The origin of these discrepancies cannot be pinned down easily, because different methods, assumptions, and wavelength ranges are used for different tracers, which probe the gravitational potential in different ways. The inhomogeneity of the mass measurements and the low-number statistics make it challenging to quantify the effect of the different methods on the scatter in $M_{\rm BH}$ scaling relations. Hence, providing a measure of the systematics from the different methods is mandatory for an in-depth understanding of the origin of the black hole relations and thus of the growth of supermassive black holes. Particularly, we need to answer the questions: How do systematics change the slope of the scaling relations? How much of the scatter in the black hole scaling relations can be attributed to inconsistencies between the various dynamical measurement methods?

As we slowly start to reach a statistically robust galaxy sample with measured $M_{\rm BH}$, now is the best time to revisit known black hole mass measurements with different methods for two reasons. Firstly, the high spatial resolution of the {\it Hubble Space Telescope (HST)} allowed for a systematic determination of $M_{\rm BH}$ in a large sample of galaxies. Still, the usage of long-slit rather than modern integral-field spectroscopy limited the precision of early measurements. The spatial resolution achieved by adding adaptive optics (AO) to integral-field spectroscopy was another substantial step forward in measuring black hole masses \cite[e.g.,][]{Krajnovic2005}. Secondly, even most recent dynamical mass measurements are affected by systematic biases associated with the modelling assumptions. Discussed are for example radially varying versus constant mass-to-light ratio \citep{Thater2017, Thater2019}, the inclusion of dark matter \citep{Gebhardt2009,Rusli2013}, radially varying versus constant anisotropy \citep{Drehmer2015} and axisymmetric versus triaxial shapes of galaxies \citep{vandenBosch2010, Ahn2018}. 
It is thus essential to understand and quantify the systematics as detailed as possible, to evaluate the robustness of the mass measurements and mitigate the associated systematic uncertainties. We decided to perform this test by comparing the $M_{\rm BH}$ derived with the widely applied techniques of using stars, ionised gas and molecular gas as tracers of the gravitational potential.

\begin{table}
\caption{Basic properties of NGC~\,6958.}
\centering
\begin{tabular}{lcc}
\hline\hline
Property &   &  Notes \\
\hline
Morphological type     &  S0   & 1   \\
Distance [Mpc]   &  $35 \pm 8$ & $2$    \\
Physical scale [pc arcsec$^{-1}$] & $170 \pm 10$ \\
Inclination [ $^{\circ}$] & $45 \pm 4$ & 3 \\
Position angle [ $^{\circ}$] & $109 \pm 5$ & 4\\
Sersic index &  3.3 & 5 \\
Effective radius [kpc]    &  2.59 & 5    \\
$\sigma_{\rm e,star}$ [km s$^{-1}$]    &  $168\pm 5$ & 6  \\
$\sigma_{\rm 0,star}$ [km s$^{-1}$]    &  $220\pm 5$ & 6  \\ 
Bulge mass [M$_{\odot}$]   &  (3.6$\pm 1.4) \times 10^{10} $ & 7  \\
\hline
\end{tabular}
\\
{Notes. - 1: The galaxy was misclassified in \cite{deVaucouleurs1991rc3} and we adopt the classification by \cite{Sandage1994} and \cite{Laurikainen2010}. 2: Mean distance based on dynamical scaling relations from the
NASA/IPAC Extragalactic Database (NED). 3: Inclination of the molecular gas disk of NGC 6958 derived in the follow-up publication. 4: Derived from the MUSE velocity field within a field-of-view of 5\arcsec. 5: Derived from the light model in Section \ref{ss:mge_nicmos}. 6: Derived by co-adding the spectra of the MUSE data cube in elliptical apertures with an ellipticity of 0.15 and a semi-major axis  of the effective radius $R_{\rm eff}$ ($15\farcs3$) and $R_{\rm eff}/8 = 1\farcs9$, respectively. 7: Using the total mass derived from the Jeans Anisotropic models (Section~\ref{ss:jam}) of this work and the bulge-to-total ratio (=0.45) from \cite{Laurikainen2010}.}
\label{properties}
\end{table}

We first needed to identify a galaxy which offers the possibility to apply the different modelling methods. Early-type galaxies with bright nuclear molecular gas discs are prime candidates.  
We found such an object in the mm-Wave Interferometric Survey of Dark Object Masses (WISDOM) sample \citep[e.g.,][]{Onishi2017,Davis2017a,Davis2018} that provides high-resolution Atacama Large Millimeter/submillimeter Array (ALMA) observations for a large variety of galaxies.  In this work, we targeted the massive fast-rotating early-type galaxy NGC 6958, which shows clear signs of a regularly rotating nuclear molecular gas disc (see \citealt{Thater2020}). The main properties of NGC 6958 are given in Table~\ref{properties}.  

NGC 6958 is an isolated galaxy \citep{Madore2004}. There is evidence of a recent minor merger \citep{Malin1983,Saraiva1999,Tal2009}, but the merger does not affect our $M_{\rm BH}$ measurement as the central stellar kinematics show very regular features (see Section 3). NGC 6958 was also classified as a low-ionisation nuclear emission-line region (LINER) galaxy showing large equivalent width of H$\alpha$ and [NII]$\lambda$6584 emission lines \citep{Saraiva2001, Annibali2010}, which we will use to estimate $M_{\rm BH}$ accounting for the non-circular motions of the ionised gas via assymmetric drift correction. Based on the galaxy's effective velocity dispersion of 168 km s$^{-1}$, the $M_{\rm BH}-\sigma_{\rm e,star}$ relation \citep{Saglia2016} predicts an SMBH of mass $M_{\rm BH}= 1.1\times 10^8 $M$_{\odot}$ which at a rather uncertain distance of 35 Mpc (see Table~\ref{properties}) is at the limit to be detectable ($R_{\rm{SoI},1e8\,\rm{M}_{\odot}}= 0\farcs1$)\footnote{The sphere of influence (SoI) is defined as $R_{\rm SoI}=GM_{\rm BH} / \sigma^2_{\rm e,star}$ where G is the gravitational constant. Within $R_{\rm SoI}$ the gravitational potential is dominated by the SMBH.} with AO-assisted and interferometric facilities. As shown in Fig. 2 of \cite{Thater2020}, this is the mass region where gas- and stellar-based $M_{\rm BH}$ seem to be discrepant. 

This publication is the first part of our study of using independent kinematic tracers to derive the black hole mass in NGC 6958 and check whether the different methods give consistent results. This paper will focus on the use of stars as dynamical tracers, and is composed of five sections. We begin by presenting the adaptive-optics assisted MUSE integral-field spectroscopic and {\it HST} photometric observations in Section~\ref{s:observations}. We then explain the stellar and ionised gas kinematics extraction in Section~\ref{s:kinematics}, where we also include a detailed evaluation of the MUSE+AO PSF. In Section~\ref{s:dynamics}, we derive the galaxy's stellar mass distribution and perform dynamical Jeans Anisotropic and Schwarzschild modelling of the stellar kinematics to obtain the massive black hole mass. We conclude this paper by putting our results in context with the $M_{\rm BH}$ scaling relations and providing a short outlook to the second paper in this series (Thater et al. in prep).

\begin{figure}
  \centering
    \includegraphics[width=0.48\textwidth]{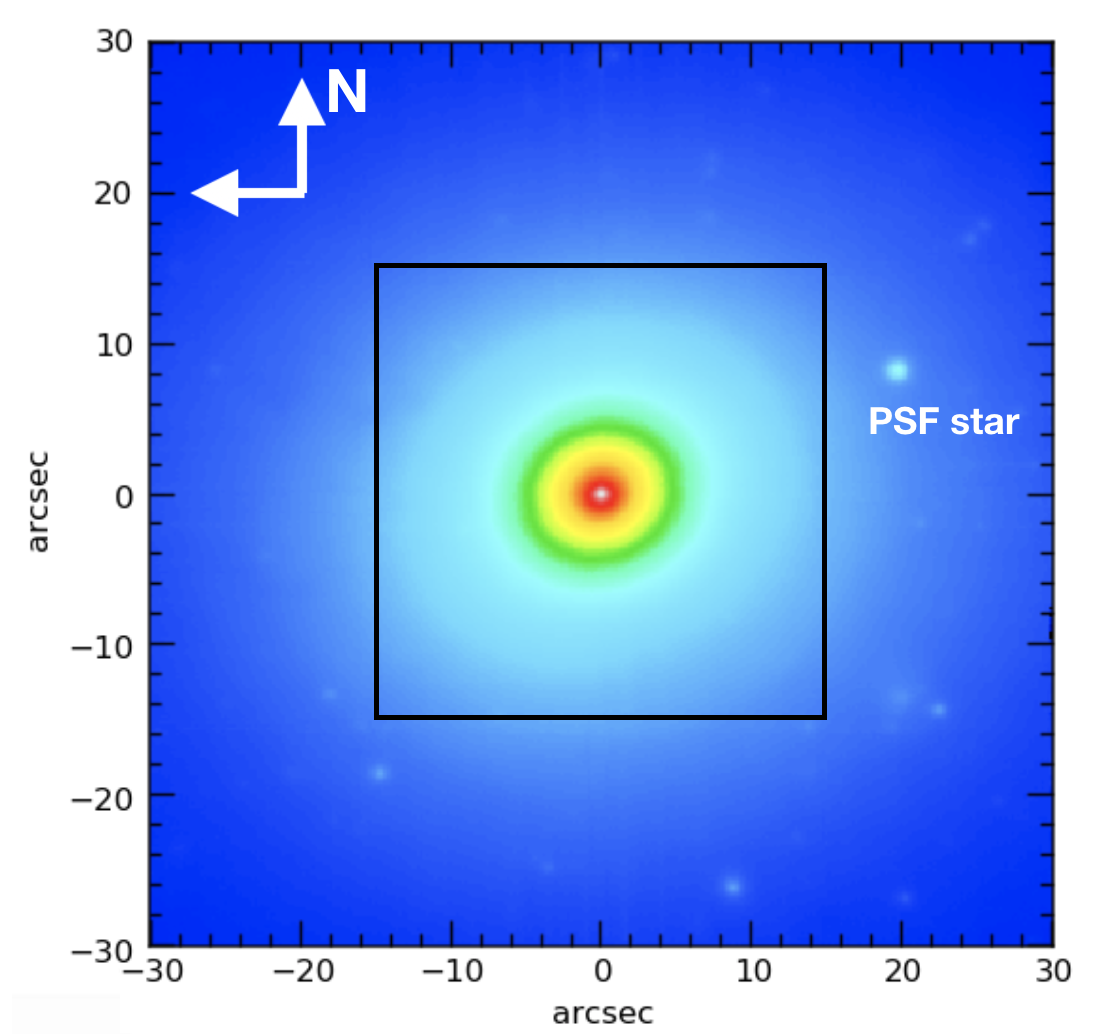}
      \caption{Full FoV white-light image of our MUSE observations covering $1\arcmin \times 1\arcmin$. The rectangle indicates the cut-out $30\arcsec \times 30\arcsec$ FoV used in this study. The bright star in the north-west was used for the PSF estimations in Section~\ref{ss:muse_psf}. 10 arcsec correspond to 1.7 kpc in physical scales. }
      \label{ff:muse_fov}
\end{figure}

%-----------------------------Observations
\section{Observations}\label{s:observations}
\subsection{MUSE integral field spectroscopic data}
We obtained AO-assisted Multi-Unit Spectroscopic Explorer \citep[MUSE;][]{Bacon2010} wide-field mode science verification data under the science program 60.A-9193(A) (PI: Krajnovi\'c) at the Very Large Telescope (VLT) in the night of the 18th of September 2017. The Ground Atmospheric Layer Adaptive Optics for Spectroscopic Imaging (GALACSI) AO-system \citep{Stroebele2012} was developed to optimize the performance of MUSE and consists of four sodium laser guide stars, a deformable secondary mirror on the VLT UT4 and an infrared low-order sensor to provide near-diffraction-limited observations at visible wavelength. In addition to the four laser guide stars, we used a slow-guiding star 39 arcsec and a tip-tilt star 65 arcsec from the nucleus. Due to bad weather conditions during the observations\footnote{https://www.eso.org/sci/activities/vltsv/musesv.html}, we could not make use of the full AO capabilities and achieved a spatial resolution of $0.6''$ (Section~\ref{ss:muse_psf}). Our MUSE observations have a total exposure time of 2040 seconds divided into four 510 second on-source integrations observed in the sequence O-S-O-O-S-O where O are the observations of the target and S of the sky. 

We performed the data reduction using the MUSE data reduction pipeline \citep{Weilbacher2020}, version 2.6. The pipeline includes bias and sky subtraction, flatfield correction, wavelength, and flux calibration and telluric correction of each on-source observation. Furthermore, new with version 2.6, wiggles that are visible in the spectral direction of high signal-to-noise (S/N) data in MUSE AO observations are appropriately corrected for. After the data reduction, we merged the individual exposures with the MUSE pipeline, taking the respective offsets into account. In the final data cube, each spaxel has a size of $0.2''\times 0.2''$ and spectral sampling of 1.25\,\AA . The total wavelength range covered by our data goes from 4700 to 9300\,\AA . However, during the observation the spectral region between 5800 and 5970 Å was blocked by a NaD notch filter to avoid light contamination  by  the  sodium  lasers  of  the  AO  system and we do not have any spectral data in this region. The spectral resolution of the MUSE data varies between 2.5 and 2.9\,\AA\, \citep{Guerou2017}.

We show the white-light image of the MUSE observation of NGC 6958 covering the full field-of-view (FoV) of  $1\arcmin \times 1\arcmin$  in Fig.~\ref{ff:muse_fov}. In the following analysis, we used the central $30''\times 30''$ of the MUSE FoV as we noted a kinematic twist for radii larger than $15''$ (whereas the kinematics are very regular within this radius). The cut-out MUSE data cube was then Voronoi-binned \citep{Cappellari2003} to a target S/N of 70 \AA$^{-1}$ for each bin, resulting in mostly unbinned spaxels in the galaxy centre and bin sizes of $1''-2''$ at a distance more than $7$ arcsec from the centre. Figure~\ref{ff:muse_fov} also shows a bright star at a projected distance of about 20 arcsec away from the galaxy centre. We used this bright star for our PSF estimations in Section~\ref{ss:muse_psf}.

\subsection{Imaging data}
\label{ss:imaging}
NGC 6958 was observed with {\it HST} several times. In the archive, we found a 400s exposure in F814W band (PI: P. Goudfrooij, PID:  8686) obtained with the Wide Field Camera of the Wide Field and Planetary Camera 2 \citep[WFPC2; ][]{Holtzman1995} and a 1152s exposure in the {\it H}-band (F160W, PI: A. Capetti, PID: 11219) of the Near Infrared Camera and Multi-Object Spectrometer (NICMOS). 
NGC 6958 contains a nuclear dust disc, which is less impacted by dust attenuation when using a near-infrared image (F160W). In addition, the galaxy was only observed with one of the Wide Field Camera chips of WFPC2, which has a lower sampling ($0\farcs1$/pixel) than the NICMOS image ($0\farcs076$/pixel).

The best possible spatial resolution and proper treatment of the nuclear dust are crucial for measuring the black hole mass in all applied dynamical methods discussed in this work. We, therefore, decided to use the F160W NICMOS imaging data for the main dynamical models and the image in the F814W band to test how alternative mass models affect our dynamical modelling results (See Section~\ref{ss:systematics_mass}). Deeper large-scale images are additionally needed to trace the galactic gravitational potential up to large scales. This is important for the construction of the orbit library of the Schwarzschild models (Section~\ref{ss:schwarzschild}). 
Here, we used an F160W Wide Field Camera 3 (WFC3) image of NGC 6958 (PI: B. Boizelle, PID: 15909) with a spatial sampling of ($0\farcs13$/pixel), and an {\it i}-band image from the Carnegie Irvine Galaxy Survey Project \citep[CGS; ][]{Ho2011,Li2011,Huang2013}.

%-----------------------------Stellar Kinematics
\section{Stellar  \& ionised gas kinematics}\label{s:kinematics}
\begin{figure}
  \centering
    \includegraphics[width=0.48\textwidth]{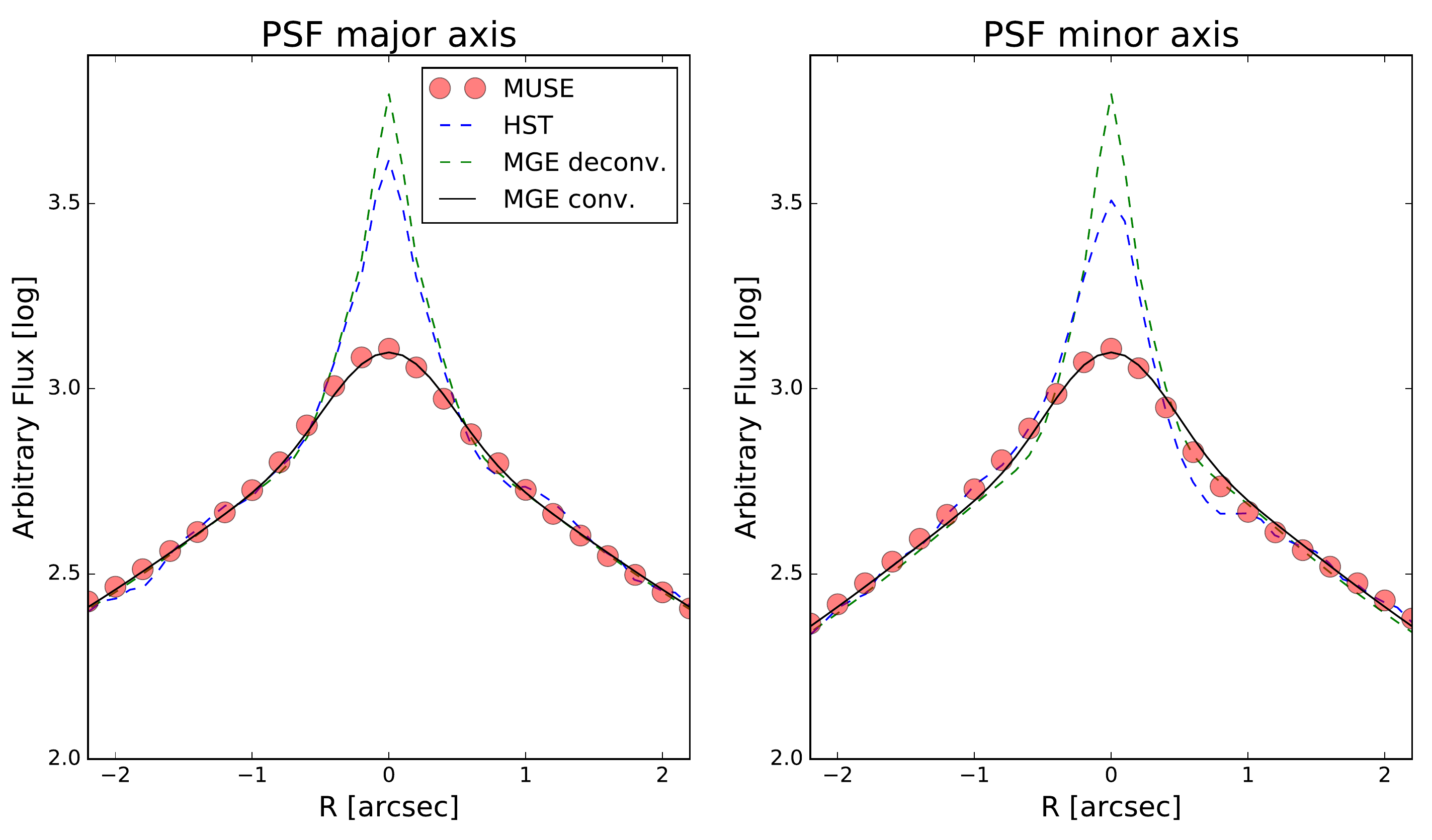}
      \caption{Spatial resolution of our MUSE observations derived by comparison of the MUSE white-light image with the {\it HST}/WFPC2/F814W image. We note that the dust contamination (well visible in the plot of the minor axis) can broaden the broad component of the Gaussian to unrealistic scales.}
      \label{ff:muse_psf}
\end{figure}

\begin{figure}
  \centering
    \includegraphics[width=0.48\textwidth]{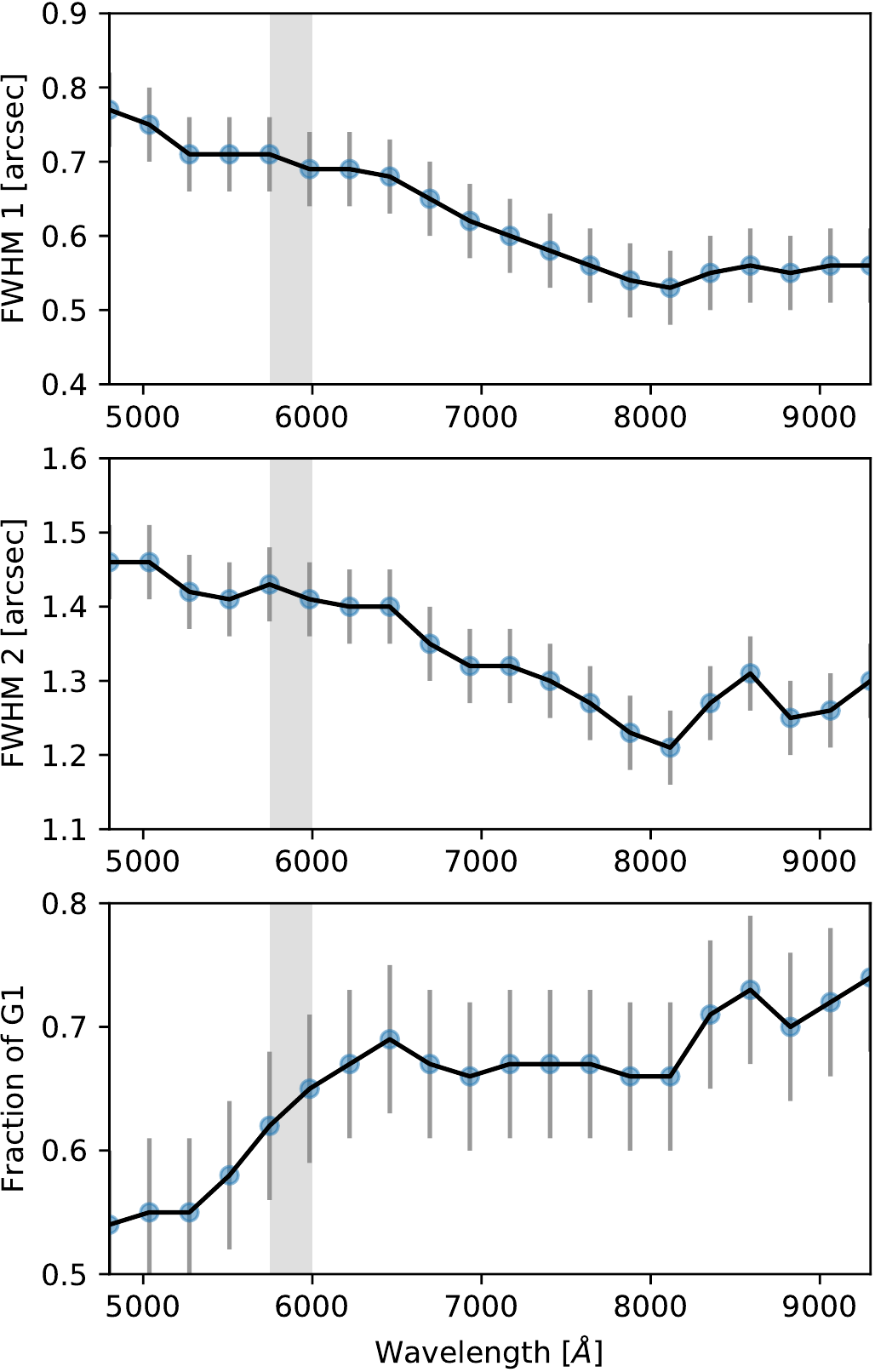}
      \caption{Parametrisation of the AO-assisted MUSE PSF as a function of wavelength. Full width at half maximum (FWHM) of the two Gaussian components and the fraction of the narrow Gaussian (G1) where derived from the PSF star fit. The two Gaussians were fitted for each wavelength bin (width of 500 \AA) between 4800 and 9300 \AA\,. The gray shaded area indicates the wavelength range in which the laser guide star light is blocked by the NaD notch filter.}
      \label{ff:muse_psf_wave}
\end{figure}

In this section, we will show the kinematic extraction of our ground-layer AO-assisted MUSE observations. So far, MUSE has only been used for one other black hole mass measurement \citep{Mehrgan2019} and this is the first paper to present a stellar-based massive black hole mass measurement using the AO mode of MUSE. 

\begin{figure*}
  \centering
    \includegraphics[width=\textwidth]{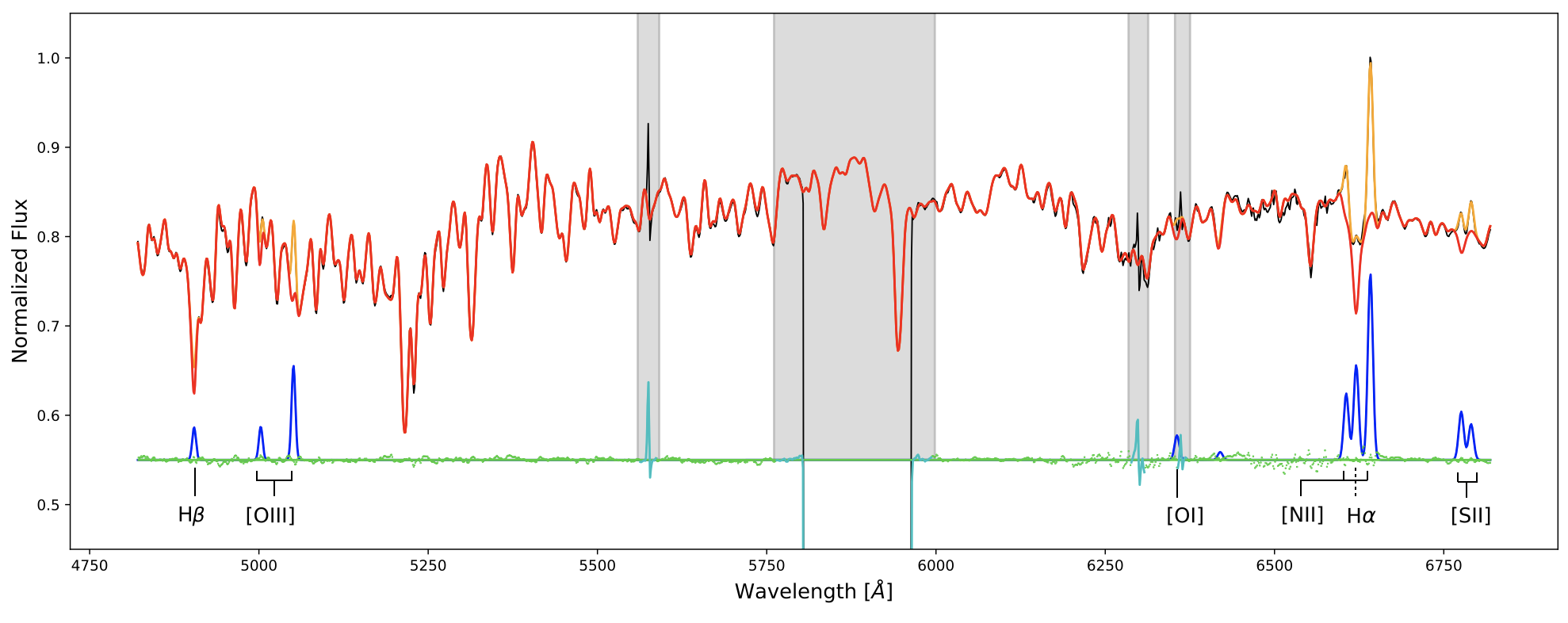}
      \caption{Integrated MUSE spectrum at observed wavelength and best-fitting pPXF fit of NGC 6958 displayed in the MILES spectral range. The integrated spectrum (black solid line) was obtained by summing up all spectra of the IFU data cube within a radius of 15\arcsec. This integrated spectrum was fitted using the pPXF routine in order to derive an optimal stellar template (red line). We simultaneously fitted the emission lines (blue) and the fit composed of gas and stellar continuum is shown in orange. The fitting residual between spectrum and best fitting model are shown as green dots and are shifted up by 0.55. Regions which were masked in the fit, owing to either the AO NaD notch filter or insufficient sky correction, are indicated as grey shaded regions. }
      \label{ff:optimal_template}
\end{figure*}

\subsection{Spatial resolution in the GALACSI adaptive optics mode}
\label{ss:muse_psf}
The quality of the MUSE data can be assessed by estimating the spatial resolution, which is composed of the instrumental and atmospheric point spread function (PSF). As the atmospheric PSF changes depending on the observational conditions, it needs to be carefully evaluated for each observation. This is a crucial step as the spatial resolution limits how far we can probe the dynamics in the centre of the galaxy. For determining the PSF, a typical method in dynamical $M_{\rm BH}$ estimation is to use a high-resolution image of the galaxy at similar wavelength and degrade it via PSF convolution until it matches the integrated light image of the integral-field unit (IFU) data (e.g. \citealt{McDermid2006, Krajnovic2009, Thater2017}). 

We first followed this approach by convolving the light model of the F814W WFPC2 image (derived in Appendix~\ref{ss:mge_f814}) with a PSF parametrised as the sum of two concentric Gaussians (Fig.~\ref{ff:muse_psf}). We used the F814W image because it is closest to the wavelength of our MUSE observations. From the fit to the white-light image, we recovered a narrow component of the PSF of 0\farcs61. As the quality of the AO correction, and therefore the PSF, is wavelength-dependent \citep{Bacon2017}, we derived the PSF for the different wavelength cuts that we used to extract the stellar kinematics (see Sections~\ref{ss:kinematics} and \ref{ss:systematics_kin}). The PSF fit is very sensitive to the dust content of the galaxy, and we noted large and unrealistic full width at half maximum (FWHM) values ($\approx 10\arcsec$) for the broad Gaussian component in the optical and blue spectral region. We therefore carefully masked the dust-affected regions and chose (based on the results of the PSF star in the next subsection) an upper boundary of $1\farcs7$ for the broad Gaussian. The PSF parametrisations of the different wavelength ranges used for the dynamical modelling are shown in Table~\ref{tt:psf}. 

An alternative approach is to fit the sum of two concentric Gaussians to one of the stars within the full FoV of our MUSE data cube. The brightest star in our FoV with a projected distance of 20\arcsec provides a good handle on the central PSF of our observations. We subtracted the galaxy light background and then fitted two concentric Gaussians to the PSF star profile along the x- and the y-axis. While the MUSE PSF is usually fitted with a MOFFAT profile \citep{Moffat1969}, the two concentric Gaussians also well reproduce our PSF star profile. A few examples of the fit are shown in the Appendix, the PSF parametrizations are also shown in Table~\ref{tt:psf}. From the PSF star measurement in the white-light image, we derived a narrow Gaussian PSF of $0\farcs60$ which is in agreement with the PSF that we measured using the nucleus of the galaxy. 

Using the PSF star, we also measured the change of the PSF over the full wavelength range using 20 regularly and equally spaced wavelength channels (of 500 \AA\, width). In Fig.~\ref{ff:muse_psf_wave}, we show the average of the PSF parameters along the major and minor axes and used the differences as uncertainties. The spatial resolution clearly improves when going from the blue to the red end of the MUSE data by about 35 per cent. We noticed a deterioration in the quality of the PSF at around 8000\,\AA\, which likely resulted from incomplete skyline removal in that region.
All of our measurements are in agreement with the study of the MUSE PSF by \cite{Fusco2020} if we assume a poor natural seeing of about 1.2, as recorded in the raw data, and translate our Gaussian measurements into MOFFAT parametrisation. Being taken in mediocre weather conditions, our data set does not reach the expected resolution that could be obtained with full adaptive optics-assisted MUSE observations \cite[e.g.,][]{Knapen2019}, but is still a significant improvement over what would have been achieved in these conditions without AO. Owing to the strong priors on the broad Gaussian when estimating the PSF in the galaxy nucleus we used the PSF values from the PSF star in the dynamical modelling of NGC 6958 (Section~\ref{s:dynamics}).

\begin{table}
\caption{Gaussian parametrisation of the MUSE PSF}
\centering
\begin{tabular}{c|ccc|ccc}
\hline\hline
Method 
& Nucleus  &   &  & Star  &   & \\
\hline
Sp.  range 
& fwhm$_1$  &  fwhm$_2$ & f1 &  fwhm$_1$  &  fwhm$_2$ & f1 \\
(\AA)     & (\arcsec)  & (\arcsec)    &  & (\arcsec) & (\arcsec) \\
(1)     & (2)  & (3)    & (4)  & (5) & (6) & (7) \\
\hline
4699-9298     &  0.61   & 1.69 &   0.48 & 0.60 & 1.32 & 0.64 \\ 
4820-6820     &  0.99 & 0.99 & 0.93 &  0.69 & 1.41 & 0.62  \\
4820-5750     &  0.64 & 1.69 & 0.29 & 0.72 & 1.43 & 0.57\\
8500-8800     &  0.57 & 1.41 & 0.49 &  0.56 & 1.30 & 0.72 \\
\hline
\end{tabular}
\\
{Note - Parametrisation of the MUSE PSF as double Gaussian centered on the galaxy center (columns 2 - 4) and double Gaussian centered on the PSF star (columns 5 - 7). In detail: Column 1: Spectral range collapsed for the integrated light image. The spectral regions are ordered as "white-light", "optical", "blue" and Calcium triplet (CaT) range. Columns 2, 3, 5 and 6: Full width at half maximum of the two Gaussians. Columns 4 and 7: relative flux of the narrow Gaussian (G1).
} 
\label{tt:psf}
\end{table}
\begin{figure}
  \centering
    \includegraphics[width=0.48\textwidth]{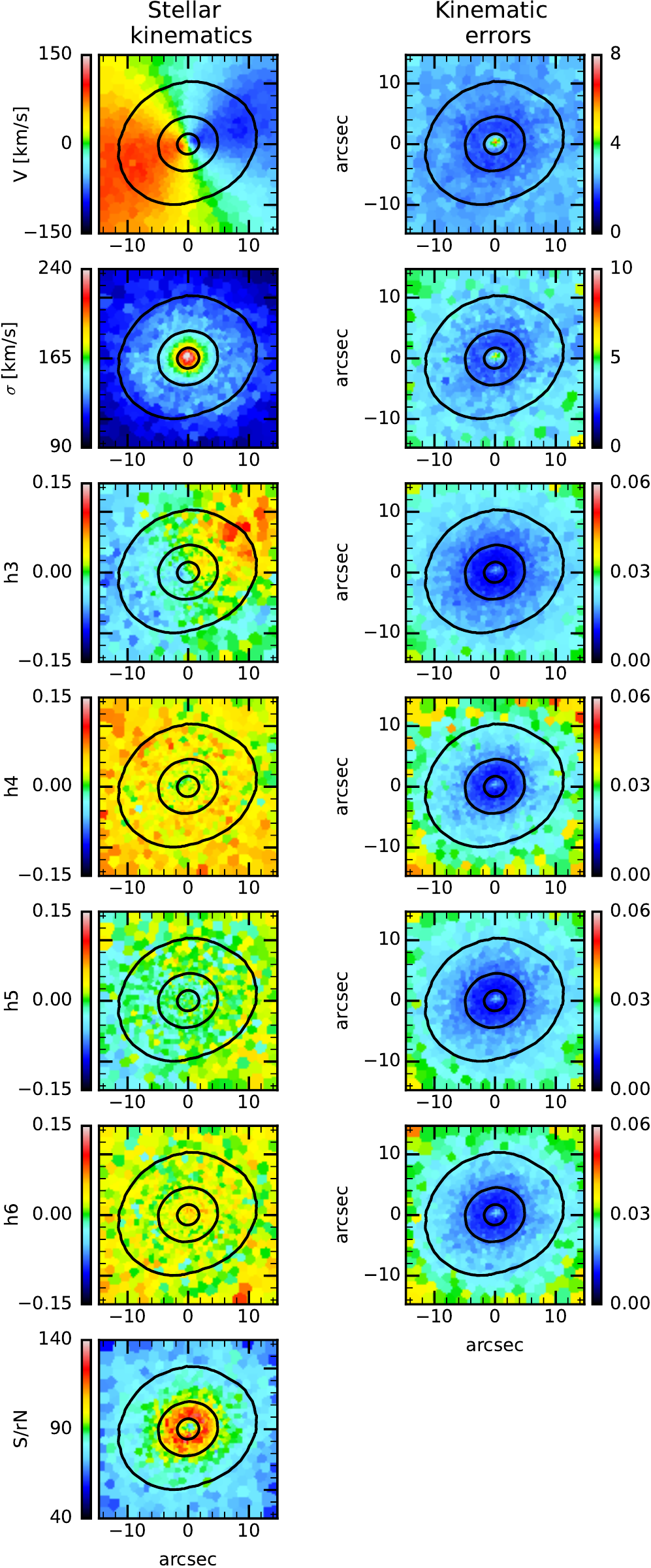}
      \caption{MUSE stellar kinematic maps (left) and kinematic errors (right) extracted from our full-spectrum fit. From top to bottom the panels show maps of signal-to-residual noise (S/rN), mean velocity (V), velocity dispersion ($\sigma$) and the Gauss-Hermite moments $h_{3}$, $h_4$, $h_{5}$ and $h_6$. The black contours indicate the isophotes from the collapsed data cube. North is up and east to the left.}
      \label{ff:ppxf_overview}
\end{figure}
\begin{figure}
  \centering
    \bigskip
    \includegraphics[width=0.48\textwidth]{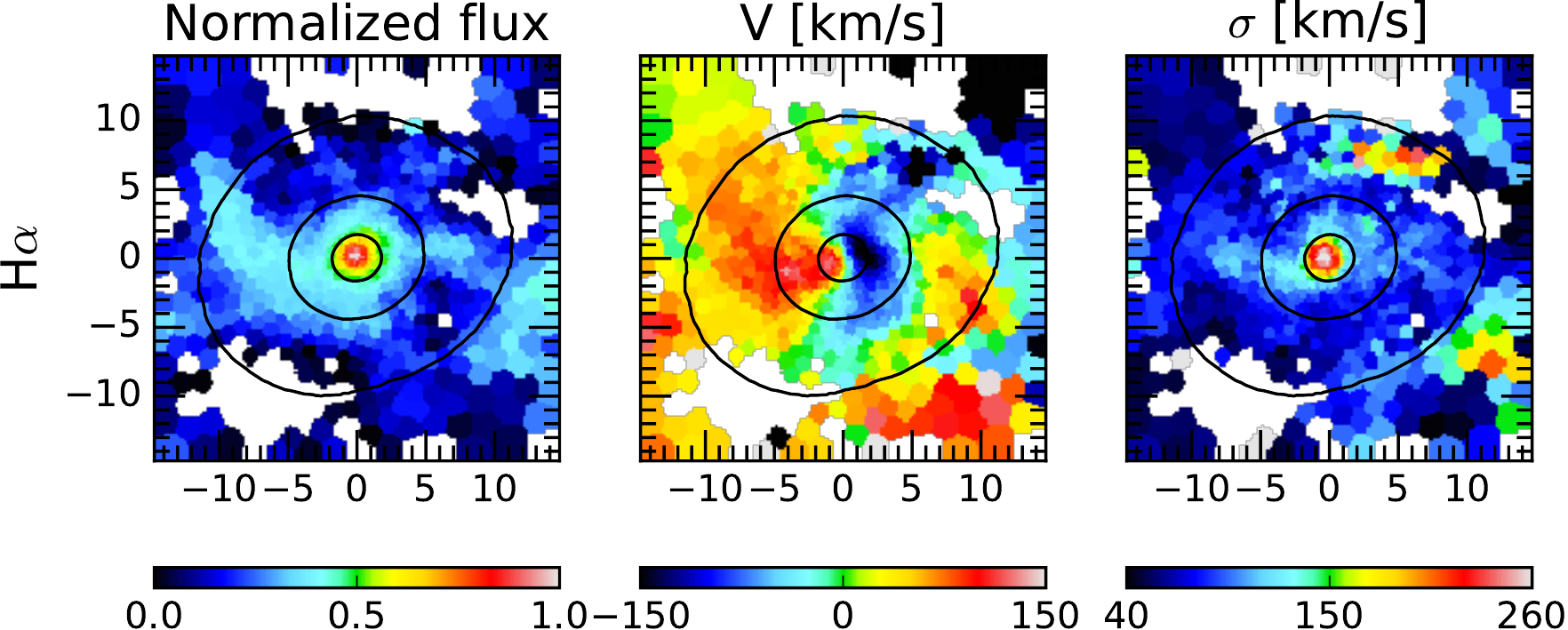}
      \bigskip
   \includegraphics[width=0.48\textwidth]{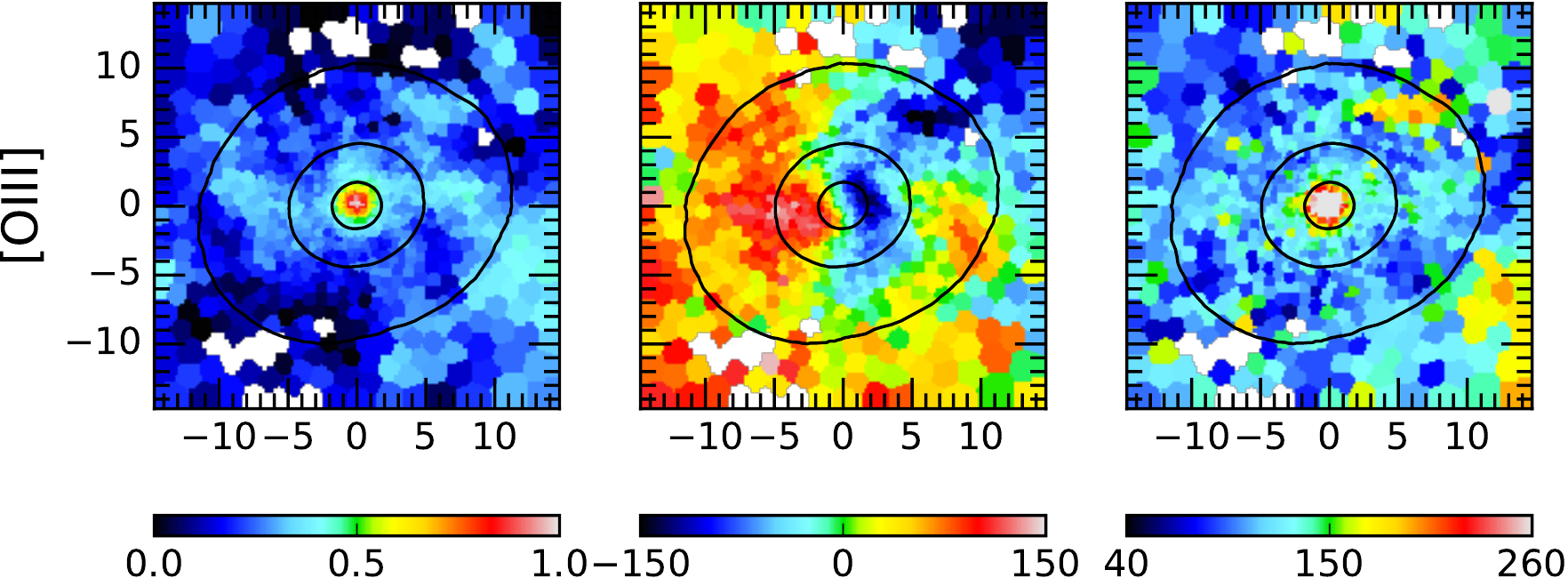}
    \includegraphics[width=0.48\textwidth]{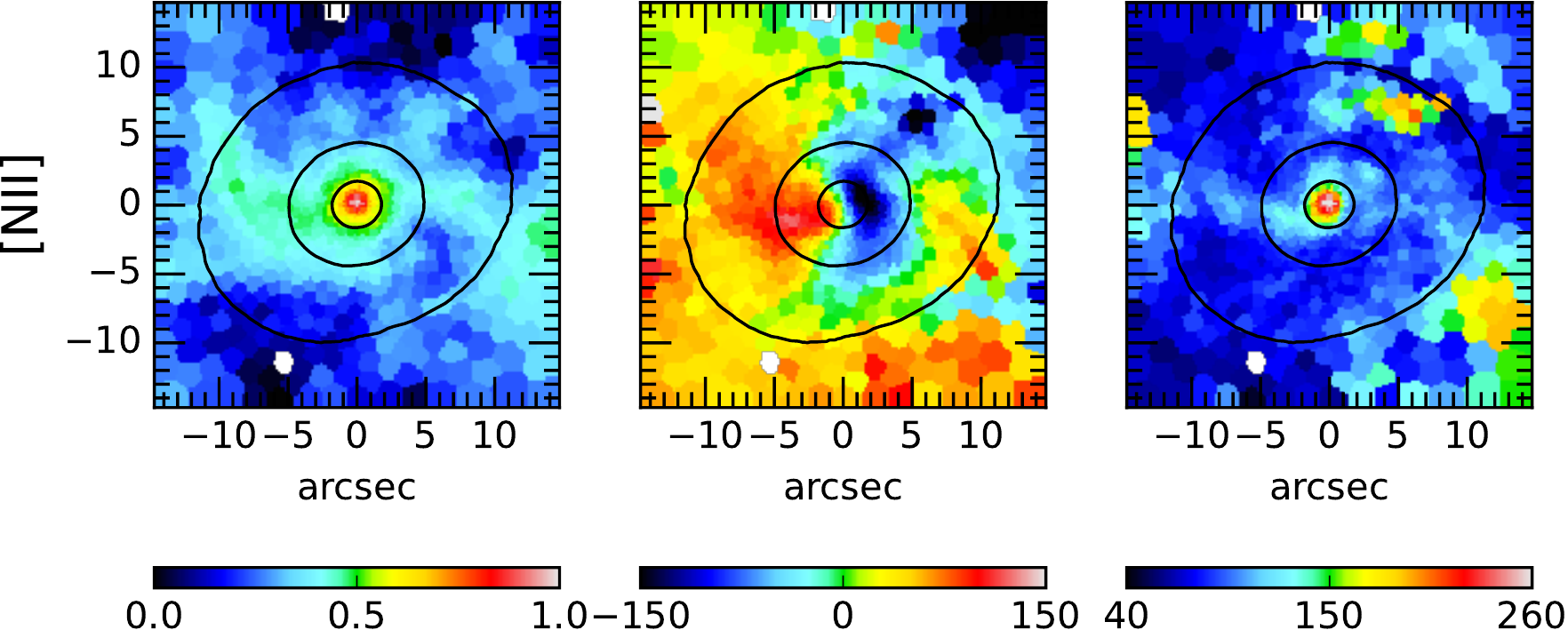}
      \caption{MUSE ionised-gas kinematics derived from the simultaneous fit of stellar continuum and emission lines. The panels show the ionised-gas distribution in log scale, mean velocity [km/s] and velocity dispersion [km/s] maps traced by H$\alpha$, [OIII] and [NII]. The maps were masked at a line significance A/rN below $3$. North is up and east to the left.}
      \label{ff:ppxf_overview_gas}
\end{figure}
\subsection{Kinematics extraction }\label{ss:kinematics}
We used the Python implementation of the penalised Pixel Fitting method\footnote{\url{https://pypi.org/project/ppxf/}} \citep[pPXF, ][]{Cappellari2004, Cappellari2017} to measure the line-of-sight velocity distribution (LOSVD) of each Voronoi bin of the MUSE data. pPXF fits the observed galaxy spectrum by convolving a linear combination of stellar templates with the best-fitting LOSVD. The stellar templates were taken from the Medium-resolution Isaac Newton Telescope Library of Empirical Spectra \citep[MILES, ][]{Sanchez-Blazquez2006, Falcon-Barroso2011} stellar library (version 9.1). We used the full MILES library consisting of 985 stars and fitted the integrated MUSE spectrum to derive an optimal template. As the MILES templates have a similar spectral resolution to the MUSE spectra\footnote{Note that the MUSE LSF is not uniform over the complete wavelength range. We have analysed the effect of the non-uniformity on the extracted kinematics in \cite{Thater2019}. By not convolving the MILES spectra adaptively to the spectral resolution of MUSE, we impose a systematic uncertainty of about 3 km s$^{-1}$ in the velocity dispersion which we took into account in the dynamical modelling.}, we did not need to degrade either of the two datasets. MILES stellar templates span the wavelength range 4760 to 7400\,\AA\, and were fitted to the wavelength range from 4820 to 6820\,\AA\, in the galaxy spectrum. We chose this wavelength range in the galaxy spectrum as we noticed an increased template mismatch at wavelengths redder than 7000\,\AA. The region between 5750 and 6000\,\AA\, was blocked in the observation to avoid contamination and saturation of the detector by the strong sodium laser light. During the pPXF fit, we masked the blocked region and insufficiently removed skylines. We performed two runs: In one run, we parametrised the LOSVD as a simple Gaussian (V, $\sigma$) and used the kinematics for the JAM modelling (Section~\ref{ss:jam}). In the second run, we parametrised the LOSVD as Gauss-Hermite polynomial of the order of 6 regulated by a bias of 0.8. We used the second set of extracted kinematics in the Schwarzschild modelling (Section~\ref{ss:schwarzschild}). The stellar continuum was modelled with a seventh-order additive Legendre polynomial.

The residuals from the stellar fit showed a richness of emission line features (see Fig.~\ref{ff:optimal_template}), such that we decided to fit for the emission lines and the stellar continuum simultaneously. We were thus able to detect and measure the H$\alpha\lambda$6563 and H$\beta\lambda$4861 Balmer lines and the [OIII]$\lambda\lambda$4959, 5007, [OI]$\lambda\lambda$6364, 6300, [NII]$\lambda\lambda$6548, 6583 and [SII]$\lambda\lambda$6716, 6731 forbidden line doublets over the whole MUSE FoV. We first derived a best-fitting optimal template to the integrated spectrum of the full MUSE FoV (see Fig.~\ref{ff:optimal_template}). After having found the optimal template (using both stellar continuum and gas emission information), we re-ran pPXF on the individual spectra of each bin of the MUSE observations and used the global-spectrum optimal template to extract the stellar and ionised-gas kinematics simultaneously. During the fit, we treated the stars and different gas elements as separate components and assigned individual LOSVD to each of them. We made sure that doublets were treated as one gas component with a single value for V and $\sigma$. We then estimated the uncertainties using Monte Carlo simulations (with 500 realisations) on each Voronoi bin as described in \cite{Thater2019}. We therefore used the standard deviation of the residuals between the galaxy spectrum and the best-fitting pPXF model to define a residual noise (rN) for each Voronoi bin. The signal-to-residual noise (S/rN) measures then not only the quality of the data, but also the quality of the spectral fit. Due to the high S/rN of the observations (see Fig.~\ref{ff:ppxf_overview}), we obtained very small stellar kinematic errors of typically 2.5 km s$^{-1}$ for the mean velocity, 4 km s$^{-1}$ for the velocity dispersion and 0.02, 0.03, 0.02 and 0.02 for the higher-order Gauss-Hermite moments. The errors of the ionised gas kinematics from the Monte Carlo simulations reached typical values of 4-8 km s$^{-1}$ for the mean velocity and  5-9 km s$^{-1}$ for the velocity dispersion. We also noticed a decrease of the S/rN for $R<2$ arcsec that led to increased errors in this region. The lower S/rN is caused by the nuclear dust disk and we discuss its effect on our $M_{\rm BH}$ measurement in Section~\ref{ss:systematics_kin}. 

We show our extracted stellar kinematics maps in Fig.~\ref{ff:ppxf_overview} and the kinematic maps of the ionised gas in Fig.~\ref{ff:ppxf_overview_gas}. As expected from the selection criteria for this galaxy, the extracted stellar kinematics features of NGC 6958 are very smooth and do not show any substantial irregularities in the central $15$ arcsec. After subtracting the systemic velocity of 2630 km s$^{-1}$, the rotational velocities reach up to 130 km s$^{-1}$, and a clear velocity dispersion peak is visible reaching up to 250 km s$^{-1}$. The h$_3$ moment also shows the clear anti-correlation to the mean velocity, and the h$_4$ moment increases slightly asymmetrically with increasing radius. Similar features are also seen within the higher moments, albeit they do not strongly differ from 0.  All in all, NGC 6958 has very regular stellar kinematics at radii <15 arcsec and is therefore ideally suited for the tests that we want to perform in this study. The extracted emission-line maps show a different picture to the stellar kinematic maps. For each of the three rows in Fig.~\ref{ff:ppxf_overview_gas}, we present the ionised gas distribution, the mean velocity and the velocity dispersion of the LOSVD. Based on the arguments in \cite{Sarzi2006}, we decided our emission line fits to be unreliable for amplitude-to-residual-noise ratios (A/rN) < 3 and in Fig.~\ref{ff:ppxf_overview_gas} for each emission line map masked the bins at lower A/rN (where rN was measured from the residual noise of the pPXF fit). The morphology of the ionised gas distribution and kinematics are very similar between the different emission lines but differ strongly from the stellar kinematics maps. The gas rotates at faster velocities, but has a similar velocity dispersion in the galaxy centre. It is notable that, while the ionised gas shows regular features in the central $5$ arcsec, outside of this region, we see very irregular structures largely dominated by receding motion. The irregular motion of the gas could be due to its recent acquisition \citep{Malin1983,Saraiva1999}, but we will postpone this analysis to the follow-up publication.  We will also investigate in the follow-up publication, if it will be possible to derive the black hole mass from the ionised gas to have a comparison of the effects of cold versus warm gas tracers. For now, we will focus on the stellar-based dynamical model, described in the next section.
 
\begin{table}
\caption{{\it HST}/NICMOS F160W + {\it HST}/WFC3 F160W MGE model.}
\centering
\begin{tabular}{lcccrr}
\hline\hline
j &   $\log\,(I_j$) & $\sigma_j$ & q$_j$ & $\log\,(M_{j}^{\rm const}$)  & $\log\,(M_{j}^{\rm var}$)\\
  & (L$_{\sun , H}$ pc$^{-2}$) & (arcsec) & & (M$_{\sun}$) &  (M$_{\sun}$) \\
(1)  & (2) &  (3) & (4) & (5) & (6)  \\
\hline
1 & 5.751 & 0.053 & 0.91  & 8.373 &  8.462 \\
2 & 5.131 & 0.207 & 0.91  & 8.941 &  9.028 \\
3 & 4.562 & 0.292 & 0.91  & 8.669 &  8.754 \\
4 & 4.57 & 0.704 & 0.87  & 9.424 &  9.504 \\
5 & 4.252 & 1.52 & 0.89  & 9.785 &  9.856 \\
6 & 3.903 & 3.04 & 0.89  & 10.036 &  10.09 \\
7 & 3.145 & 6.728 & 0.85  & 9.946 &  9.968 \\
8 & 2.88 & 13.311 & 0.84  & 10.269 &  10.261 \\
9 & 2.164 & 31.543 & 0.82  & 10.294 &  10.277 \\
10 & 1.443 & 65.187 & 0.89  & 10.239 &  10.221 \\
\hline
\end{tabular}
\\
{Note - Column 1: Index of the Gaussian component. Column 2: Surface brightness. Column 3: Projected Gaussian width along the major axis. Column 4: Projected axial ratio for each Gaussian component. Columns 5 and 6: Total mass of Gaussian component. In column (5) the constant dynamical $M/L$ =0.91\,M$_{\odot}$/L$_{\rm \odot,H}$ from the Schwarzschild modelling (Section~\ref{ss:schwarzschild}) was used to determine the mass of each Gaussian component and in column (6) the radially-varying stellar $M/L$ from Section~\ref{ss:systematics_mass} was used. The model has a uniform position angle of 110.8$^{\circ}$ for all Gaussian components.  }
\label{tt:mge}
\end{table}
\section{Dynamical Modelling}
\label{s:dynamics}
We modelled the stellar kinematics of NGC 6958 using the two independent methods: axisymmetric Schwarzschild modelling and Jeans anisotropic modelling. Both methods are commonly used for $M_{\rm BH}$ determinations and a cross-comparison of the respective results can serve as check on the robustness of the measurement \citep[e.g.,][]{Ahn2018,Krajnovic2018,Thater2019,Brok2021}. We refer to \cite{Thater2017} and \cite{Thater2019} for a detailed description of the methods and repeat here just the main assumptions and parameters used in our models.

\begin{figure}
  \centering
    \includegraphics[width=0.42\textwidth]{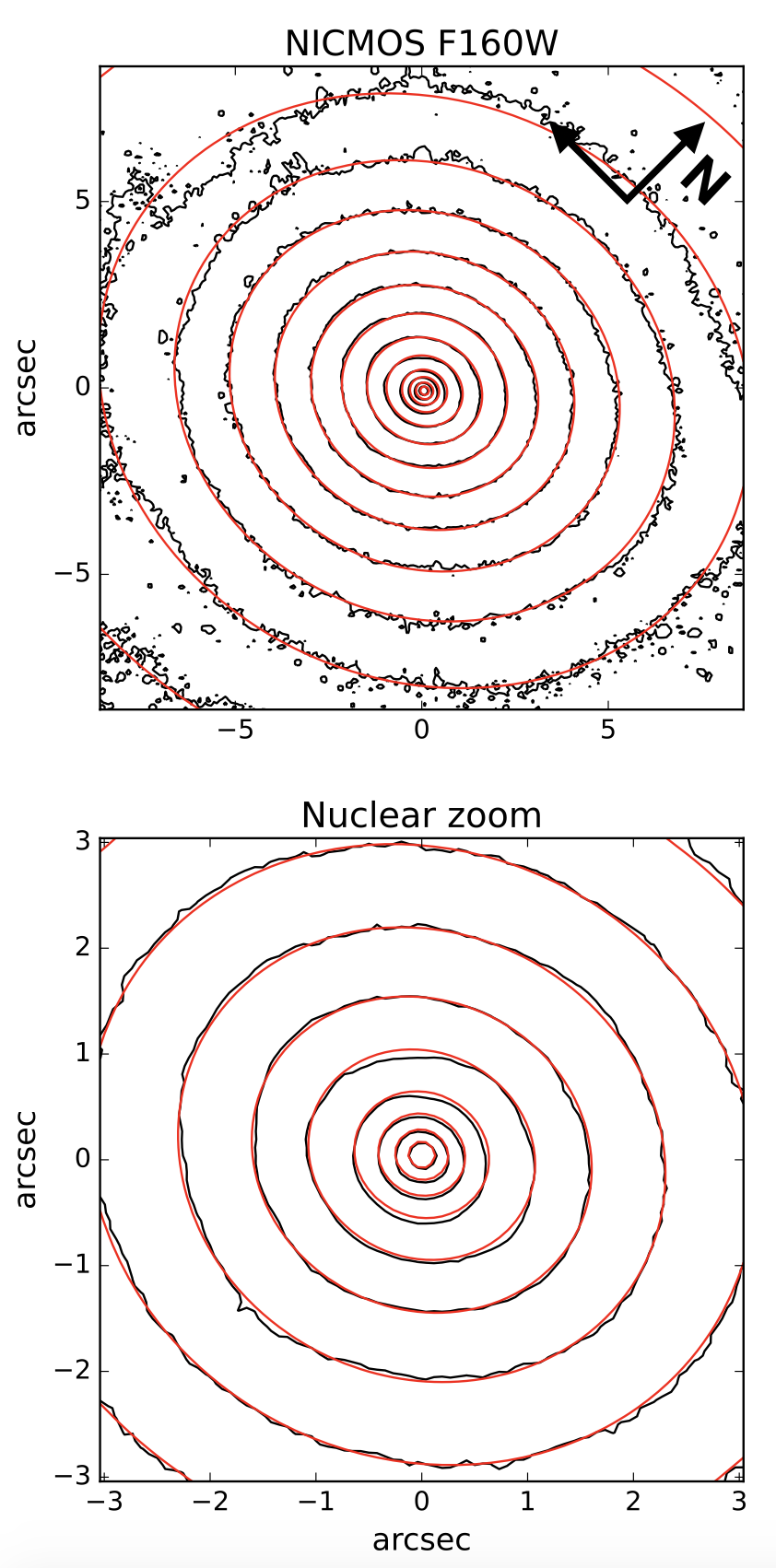}
      \caption{Isophotes of the NICMOS image of NGC 6958 within a FoV of $20\arcsec \times 20$\arcsec (top) and a cutout of the central $6\arcsec \times 6$\arcsec (bottom).  The contours of our best-fitting MGE model (red) are superimposed on the {\it HST} images (black). }
      \label{ff:mge_alma}
\end{figure}

\begin{figure*}
  \centering
    \includegraphics[width=0.47\textwidth]{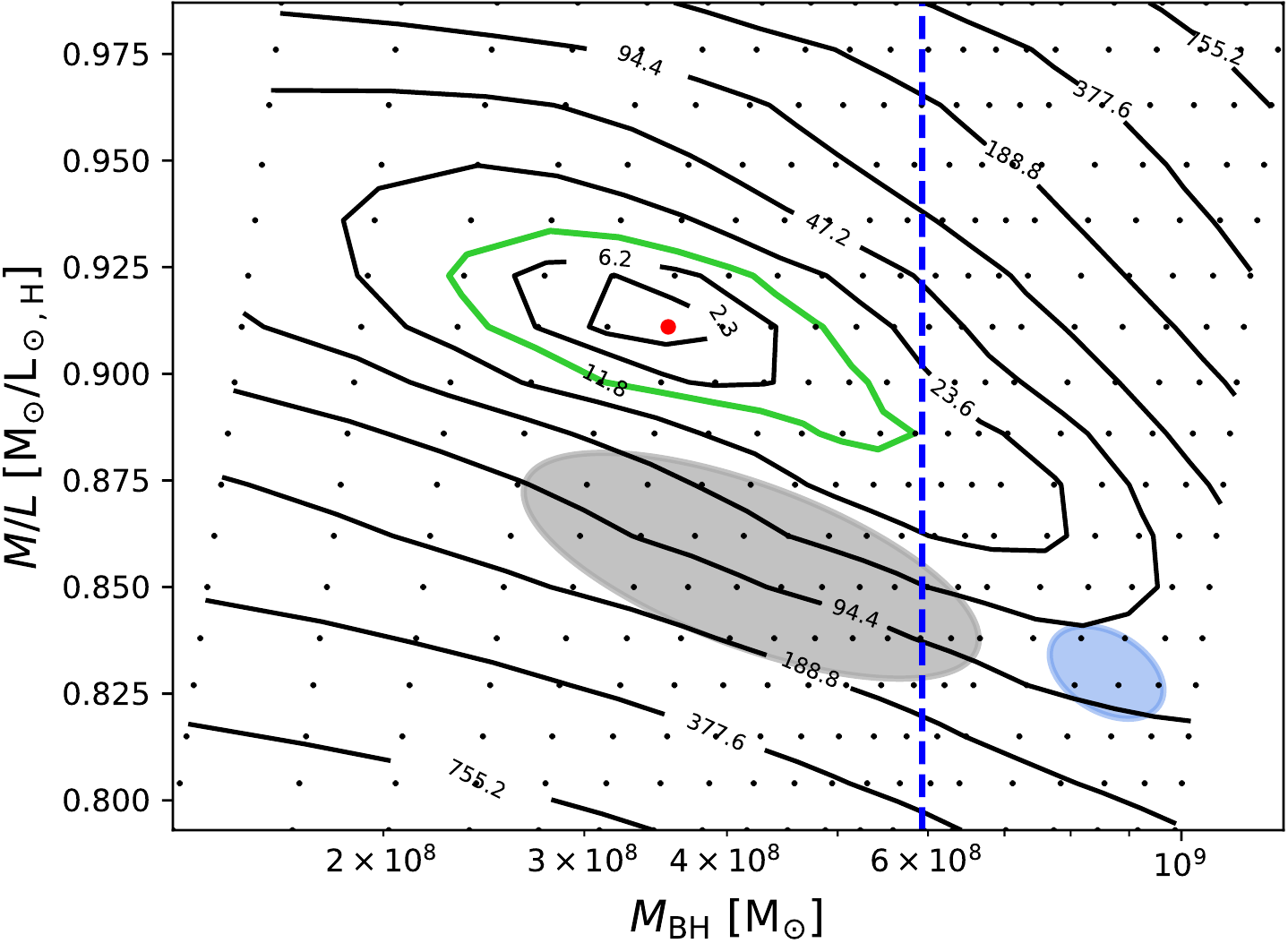}
    \hspace{2em}%
    \includegraphics[width=0.49\textwidth]{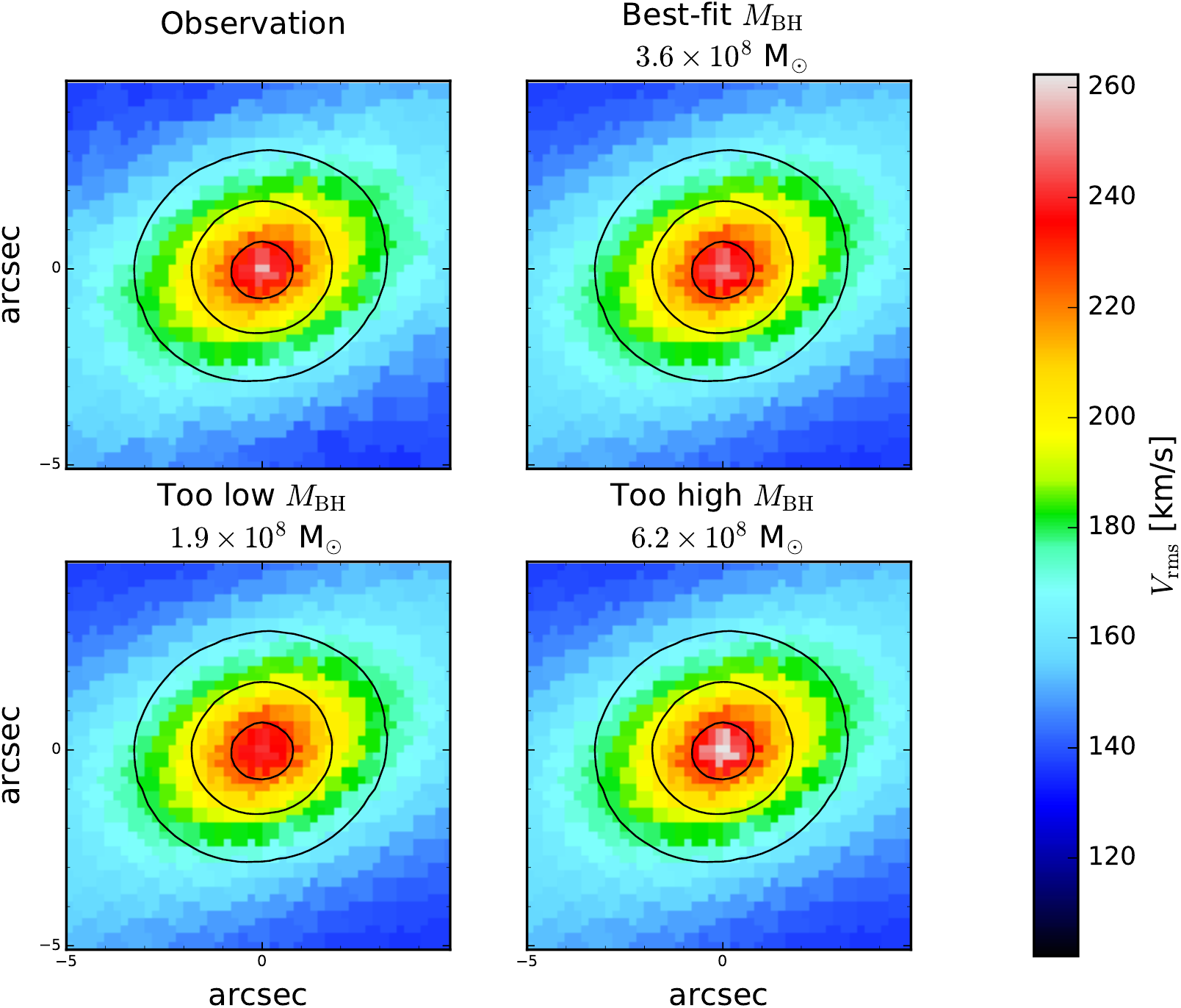}

      \caption{Results of the axisymmetric Schwarzschild modelling. Left panel: Grid of our Schwarzschild models (indicated as black dots) over various mass-to-light ratios and black hole masses. The overplotted contours indicate the $\Delta\chi^2=\chi^2-\chi^2_{\rm min}$ levels, the green contour indicating the $3\sigma$ confidence interval. The best-fitting model was derived as the minimum of the $\chi^2$-distribution and is shown as red dot. The grey and blue shaded areas indicate the $3\sigma$ confidence intervals for $M_{\rm BH}$ and $M/L$ that we have estimated using JAM$_{\rm sph}$ and JAM$_{\rm cyl}$, respectively (Section~\ref{ss:jam}). Note that the M/L from Schwarzschild is larger than from JAM. This is due to a radial dynamical M/L gradient as explained in Section~\ref{ss:jam}.
      The dashed blue line indicates the smallest black hole mass that we expect to robustly detect based on the SoI argument and the achieved resolution. Right panel: Visual comparison between the symmetrised observed $V_{\rm rms}$ map, the $V_{\rm rms}$ map of the model with best best-fitting model parameters, a too high and a too low black hole mass at fixed best-fitting $M/L$. The black contours indicate the observed light distribution of the galaxy. North is up and east to the left.} 
      \label{ff:schwarzschild_grid}
\end{figure*}

%-----------------------------MASS MODEL
\subsection{Mass model}\label{ss:mge_nicmos}
Constructing dynamical models and deriving black hole masses requires an estimate of the gravitational potential of the galaxy. We inferred the stellar potential directly from the luminosity of the galaxy multiplied with its (radially-varying)  mass-to-light ratio $M/L$. 
For a precise model of the stellar luminosity, a combination of high-resolution {\it HST} and deep large-scale imaging data is essential.

We used the Multi-Gaussian Expansion \citep[MGE;][]{Cappellari2002} fitting routine\footnote{\url{https://pypi.org/project/mgefit/}} to parametrise the surface brightness of NGC 6958 with a sum of two-dimensional Gaussians. We simultaneously fitted the sky-subtracted NICMOS and the WFC3 images; within a radius of 7 arcsec the light distribution was constrained with the high-resolution NICMOS image and for larger radii with the WFC3 image. We matched the  surface  brightness profiles of the two data sets by re-scaling the WFC3 imaging data to the central NICMOS light profiles and used the NICMOS imaging for the photometric calibration.
During the fit, we took the NICMOS PSF into account to obtain the intrinsic light distribution of the galaxy. This is a crucial step as the accuracy of our black hole mass measurement depends on how well we can describe the stellar mass in the centre of the galaxy. We generated the NICMOS PSF using the TinyTim PSF modelling tool \citep{Krist2001} and parametrised the PSF as a sum of Gaussians. Figure~\ref{ff:mge_alma} shows our best-fitting MGE model overplotted on the observed surface brightness distribution of NGC 6958. While the central parts of the {\it HST} image are well fitted with the MGE model, we noticed a clear isophote twist by almost 30 degrees at $R>15''$. 
Our dynamical models do not account for isophotal twists as they assume axisymmetry. However, the centre probed by our stellar kinematics shows no significant isophotal twist and relaxed stellar kinematics (see Fig.~\ref{ff:ppxf_overview}). We, therefore, kept the position angle constant while fitting the surface brightness. The profile of our final MGE is shown in Figure 1 of the supplementary material. Our final MGE consists of ten concentric Gaussian components. We converted the flux units  
into physical units of L$_{\odot}$ pc$^{-2}$ following the guideline and zero point given by \cite{Thatte2009}. For the conversion, we adopted a value of 4.64\,mag \citep{Willmer2018} for the absolute AB-magnitude of the sun in the $H$-band. We also took the Galactic extinction of A$_{\rm F160}=0.026$ mag \citep{Schlafly2011} into account. The converted MGE parameters are shown in Table~\ref{tt:mge} and describe the luminosity of NGC 6958.

The two-dimensional light parametrisation is then (assuming an axisymmetric potential and the inclination of the galaxy) deprojected into three-dimensional space. Multiplied with the (radially-varying) mass-to-light ratio in the given band, we thus obtained a model of the mass density from which the gravitational potential can be calculated via the Poisson equation. 

We also used our MGE model to derive the effective radius of the galaxy. We followed the approach described in \cite{Cappellari2013} and used the routine \verb@mge_half_light_isophote@ of the Python Jeans Anisotropic Modelling package described in Section~\ref{ss:jam}. The derived effective radius is $15\farcs3$ that translates into 2.59 kpc at a distance of 35 Mpc.

\subsection{Axisymmetric Schwarzschild models}
\label{ss:schwarzschild}
In this first approach, we modelled the collective motion of stars within a FoV of $30'' \times 30''$ using the axisymmetric \cite{Schwarzschild1979} orbit superposition modelling method, with the software implemenation described in \cite{Cappellari2006}. In this method, we calculated the predictions for the Gauss-Hermite polynomials up to h$_6$ and compared them with the observed stellar kinematics. For velocity, velocity dispersion, h$_3$ and h$_4$, we used the measured kinematic errors, while for h$_5$ and h$_6$ we set the errors to a constant 0.15 to account for systematics in the kinematic extraction and reduce their influence on the $\chi^2$ distribution. We also bi-symmetrised the MUSE kinematics along PA$_{\rm kin}$=109$^{\circ}$ as the models are bi-symmetric by construction. PA$_{\rm kin}$ was derived from the MUSE velocity field within a radius of 5 arcsec using the routine \verb@fit_kinematic_pa@\footnote{\url{https://pypi.org/project/pafit/}} \citep{Krajnovic2006}. The Schwarzschild models were computed as described in \cite{Thater2019} by running a grid of models of the two free parameters ($M_{\rm BH}$, $M/L$). 

Axisymmetric Schwarzschild models become highly degenerate at low inclinations, so we adopt a fixed $i=45^{\circ}$, which is the inclination of the nuclear gas disk that likely lies in the galaxy midplane (Thater et al. in prep.). We need to use the inclination of the gas disc as it is not possible to constrain the inclination of the galaxy with axisymmetric Schwarzschild models \citep{Krajnovic2005,Lipka2021}. However, in Section~\ref{ss:systematics_dyn}, we discuss how the results will change for a more edge-on ($i=90^{\circ}$) model. We also took the spatial PSF and binning of the kinematics into account, before comparing the Schwarzschid models to the observations. For our orbit library, we sampled the orbits via 41 logarithmically-spaced orbit energies, 11 linearly-spaced orbit angular momenta L$_{\rm z}$ and 11 linearly-spaced non-classical third integral values I3. In order to improve the smoothness of the model, each orbit was split into $6^3$ sub-orbits with similar initial conditions. Additional smoothing was applied by setting a moderate regularisation of 4  \citep{vanderMarel1998}. 

We ran a first coarse grid along ($M_{\rm BH}/$M$_{\odot}) \in [10^6, 5\times 10^9]$ and ($M L^{-1}/$M$_{\odot} $L$^{-1}_{\rm \odot ,H})  \in [0.1, 3.0]$ to get an indication of the global minimum of the $\chi^2$ distribution. Then, we centered a refined grid on that global $\chi^2$ minimum with 20 $M_{\rm BH}$ and 16 $M/L$ sampling locations.  Figure~\ref{ff:schwarzschild_grid} shows our final grid of Schwarzschild models for NGC 6958 with overplotted $\chi^2$ contours. From the $\chi^2$ distribution we derived the best-fitting parameters to be $M_{\rm BH} = (3.6^{+2.5}_{-1.3} )\times 10^8\,$M$_{\odot}$ and $M/L = 0.91 \pm 0.04\,$M$_{\odot}/L_{\rm \odot ,H}$ within $3\sigma$ significance ($\Delta \chi^2=11.8$). Our data revealed a single high-velocity dispersion pixel at 257 km s$^{-1}$ that can be fitted well by a model with $M_{\rm BH} = 4.7\times 10^8\,$M$_{\odot}$ and $M/L = 0.91\,$M$_{\odot}/$L$_{\rm \odot ,H}$ . However, this Schwarzschild model has too high velocity dispersion in the surrounding pixels and therefore a considerably higher $\chi^2$. As this plausible higher mass is included in the $3\sigma$ uncertainties of our measurement, we decided for $3.6\times 10^8\,$M$_{\odot}$ as final result of our Schwarzschild models. Such a black hole has a SoI of 57 parsec which corresponds to $0\farcs34$ at a distance of 35 Mpc.

The grid in Fig.~\ref{ff:schwarzschild_grid} also indicates the formally lowest black hole mass detectable with the spatial resolution of our MUSE observations (blue dashed line). While properly spatially resolving the SoI was perceived to be a strict condition for the robustness of black hole mass estimates for a long time \citep[see, e.g. discussion by][]{Kormendy2013}, \cite{Krajnovic2009} and \cite{Thater2019} have shown that when using high-quality IFU data and the sophisticated Schwarzschild modelling method, it is still possible to constrain the black hole mass, albeit with more care required around possible systematics and larger final error budgets \citep[e.g.,][]{Rusli2013}. Figure~\ref{ff:schwarzschild_grid} implies that we can measure a black hole mass which is half of the nominally minimal detectable black hole mass of $\sim 6 \times 10^8$M$_{\odot}$. The robustness of our measurement is illustrated by the $V_{\rm rms}=\sqrt{V^2+\sigma^2}$ maps in Fig.~\ref{ff:schwarzschild_grid}, given that the model $V_{\rm rms}$ of the too low and too high black hole masses significantly deviate from the observed $V_{\rm rms}$. A comparison of the remaining kinematic moments for the best-fitting Schwarzschild model and the observation is also shown in Fig. 2 and Fig. 3 of the supplementary material.

\begin{figure}

    \includegraphics[width=0.48\textwidth]{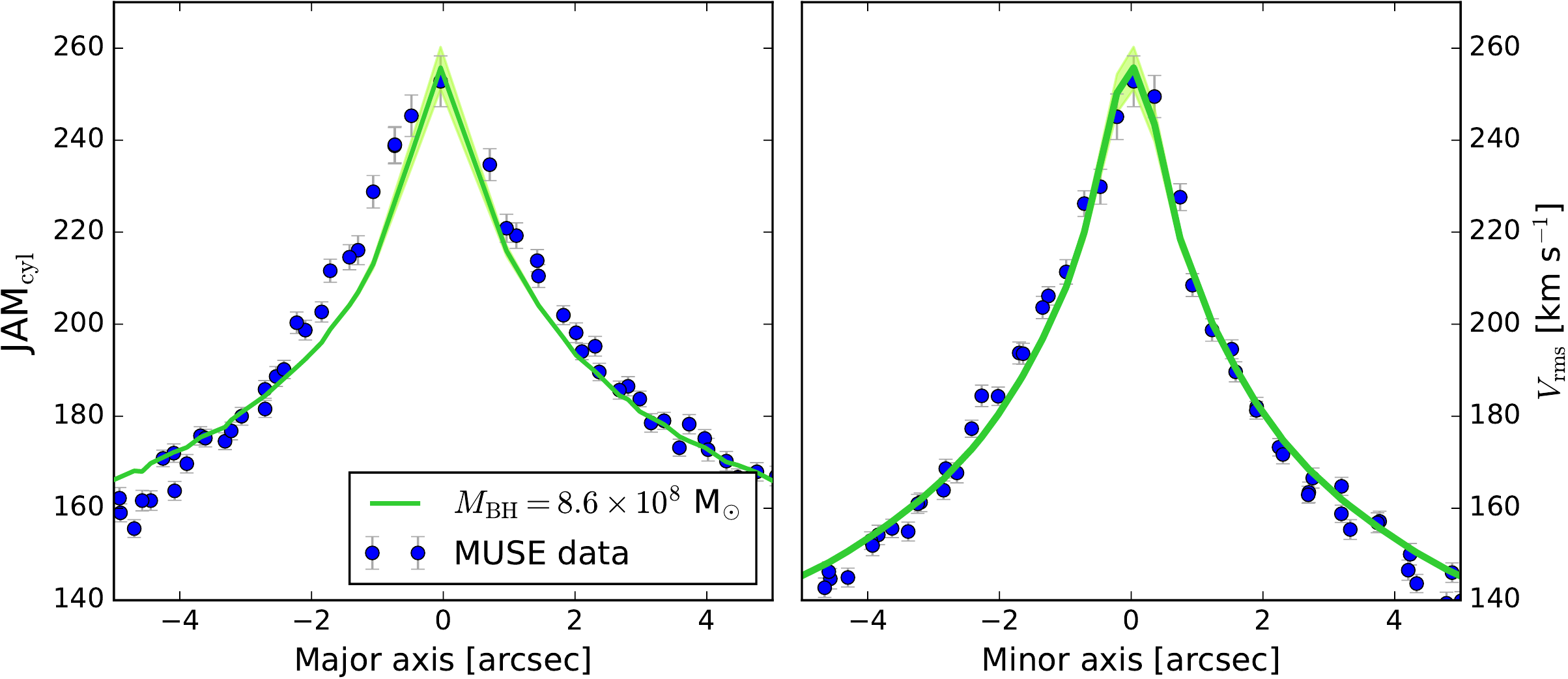}
      \includegraphics[width=0.48\textwidth]{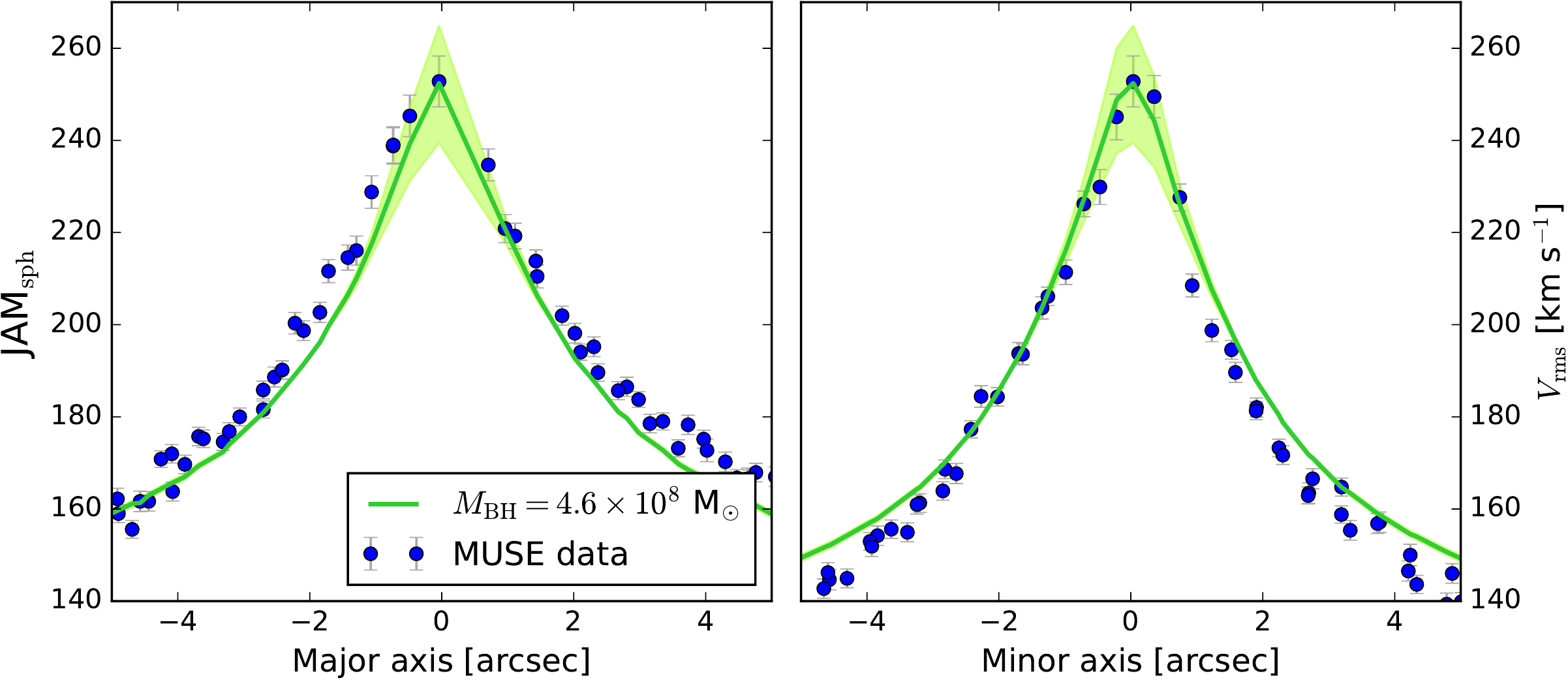}
      \caption{Results of the JAM modelling with cylindrical (JAM$_{\rm cyl}$) and spherical (JAM$_{\rm sph}$) velocity ellipsoid alignment. The panels show cuts of the observed $V_{\rm rms} =\sqrt{V^2+\sigma^2}$ (blue dots) along the  and minor axis of NGC 6958. Overplotted are the $V_{\rm rms}$ of the best-fit JAM models (solid green) and models which are within the $M_{\rm BH}$ uncertainty range from our MCMCs (green shaded area).}
      \label{ff:jeans}
\end{figure}

\subsection{Jeans anisotropic models}
\label{ss:jam}
We used the axisymmetric Jeans anisotropic modelling method\footnote{We used the \url{jam_axi_proj.py} routine, version 6.3.1 of the %{\verb@JAMPY} %JAMPY supposed to be in verbatim
JAMPY Python software package which is downloadable at \url{https://pypi.org/project/jampy/}} \citep[JAM;][]{Cappellari2008, Cappellari2020} to obtain a second independent $M_{\rm BH}$. 
Assuming an axisymmetric galaxy shape, JAM predicts the average second velocity moment along the line-of-sight $\langle V^2_{\rm los} \rangle^{1/2}$. The model $\langle V^2_{\rm los} \rangle^{1/2}$ is then compared to the observed $V_{\rm rms}$ of NGC 6958 (taking the spatial PSF into account). The modelling depends on the black hole mass $M_{\rm BH}$, the M/L and the anisotropy parameter that describes the flattening of the velocity ellipsoid along the minor axis. In JAM, the alignment of the velocity ellipsoid can be chosen to be cylindrical or spherical. We used both extreme configurations to test the robustness of our models and call them "cyl" and "sph". The definition of the anisotropy parameter depends on alignment of the velocity ellipsoid: $\beta^{\rm cyl}_{z} (R,z) = 1-(\langle v_z \rangle /\langle v_R \rangle)^2$ versus $\beta^{\rm sph} (r,\theta) = 1-(\langle v_{\theta} \rangle/\langle v_{r} \rangle)^2$.

Because of the low inclination of NGC 6958, we did not keep the inclination as a free parameter. Instead, we set it to $45^{\circ}$ derived from the molecular gas analysis (same inclination as for the Schwarzschild models). Compared to the Schwarzschild method discussed above, our JAM models are fitted to the central $10\arcsec \times 10\arcsec$ of the stellar kinematics parametrised by a pure Gaussian (without Hermite moments). The 10$\arcsec$ FoV was chosen to balance the weight of the very central kinematics that are affected by the dynamical potential of the black hole and the use of more extended kinematics to reduce the $M_{\rm BH}$ - M/L degeneracy.

We derived the best-fitting JAM model using a Bayesian framework as implemented in emcee \citep{Foreman-Mackey2013} as described in \cite{Thater2019}. 200 walkers explored the parameter space within the burn-in phase (500 steps) and were tracked during the post-burn-in phase (500 steps) to generate posterior distributions.  The parameter space was defined by uniform priors in the ranges: $\log (M_{\rm BH}/$M$_{\odot}) \in [4.8,9.8]$, $\beta \in [-1,+1]$, and ($M L^{-1}$/M$_{\odot}$L$^{-1}_{\rm \odot ,H}) \in [0.1,20]$. To derive robust results, we carefully ensured that our Markov chain Monte Carlo (MCMC) chain converged by visually checking the burn-in plots and running several Markov chains.  
We obtained the best-fitting parameters within 99.7 per cent confidence level (which is in accordance with $3\sigma$ in a normal distribution) from the posterior distributions. We fitted JAM models using the formal kinematic uncertainties derived in Section~\ref{ss:kinematics}, but the $\chi^2$ of these models was dominated by the fit to the many bins at large radii rather than by the kinematics inside the black hole SoI. For this reason the fit was driven by the inaccuracies of the modelling assumptions (e.g., constant anisotropy) rather than the black hole mass and failed to reproduce the kinematics in the black hole SoI within the formal errors. These fits gave too large $M_{\rm BH}$ and correspondingly too large central $V_{\rm rms}$ values. 

To be able to interpret the $M_{\rm BH}$ returned by JAM it is essential for the models to fit the kinematics within the uncertainties inside the SoI, where the effect of the SMBH dominates. To achieve this, we used a heuristic approach to deal with systematic uncertainties proposed by \cite{vandenBosch2009}. It consists of expanding the confidence level on $\chi^2$ by an amount equal to its variance. When using a Bayesian approach, this same result is achieved by multiplying the kinematic errors by $(2N)^{1/4}$, where $N$ is the number of constraints, as shown in \cite{Mitzkus2017}. A similar approach was also applied by \cite{Drehmer2015} to improve the JAM fit of their central $V_{\rm rms}$. However, here we differ from previous works by keeping unchanged the kinematic uncertainties inside the approximate black hole SoI ($R<0\farcs5$), while only increasing the uncertainties at larger radii. This ensures that we have a proper scaling of the $M_{\rm BH}$ errors and additionally makes sure that the fit is maximally sensitive to the black hole's kinematic influence while still matching the kinematics at larger radii. 

Our model with modified kinematic errors resulted in $M_{\rm BH}=(8.6^{+0.8}_{-0.8}) \times 10^8$~M$_{\odot}$, $\beta^{\rm cyl}_{\rm z}$ of $-0.02 \pm 0.06$ and a dynamical $M/L$ of $0.83\pm 0.02\,$M$_{\odot}$/L$_{\rm \odot,H}$ for JAM$_{\rm cyl}$. On the other hand, JAM$_{\rm sph}$ gave a significantly lower $M_{\rm BH}=(4.6^{+2.5}_{-2.7}) \times 10^8$~M$_{\odot}$, $\beta^{\rm sph}$ of $0.38 \pm 0.17$ and a dynamical $M/L$ of $0.86\pm 0.02\,$M$_{\odot}$/L$_{\rm \odot,H}$. The final posterior distributions of our JAM models are shown in Fig. \ref{ff:jam_corner_plots} of the Appendix. The cuts of the best-fitting JAM models (Fig.~\ref{ff:jeans}) show that our models with modified uncertainties for larger radii reproduce the central kinematics very well. The $M_{\rm BH}$ measurement from JAM are larger than the mass measurement from the Schwarzschild models and only the black hole mass from JAM$_{\rm sph}$ is (within $3\sigma$) consistent with the one derived with Schwarzschild. Also the $M/L$ measured with both JAM implementations is lower than with the Schwarzschild models. It is important to note that we fit the Schwarzschild models to the full MUSE kinematics, while we fit the JAM models only to the nuclear kinematics. For this reason, given that we assume mass-follow-light models in both cases, this M/L difference suggest that the M/L increases with radius. A possible reason for this increase could be the dark matter, which we ignored here, but which may starts affecting the total density slope at the largest radii probed by the MUSE data. We demonstrate the correctness of this interpretation in Section~\ref{ss: darkmatter}.

\subsection{Systematic uncertainties of the measurement}\label{ss:systematics}
We summarise the main results of the Schwarzschild and JAM models in Table~\ref{tt:results} and from now on refer to them as fiducial models. In the following section, we show several tests for systematics which should be considered when deriving robust black hole mass measurements. The following sections are ordered such that we first look for systematics in the stellar kinematics data, then systematics from the dynamical model assumptions and finally systematics arising from the mass model of the galaxy. The tests have a similar effect on both the Schwarzschild and JAM models. In order to keep the section on systematics clear, we explain the results of the Schwarzschild models in the main text and provide the results for the two JAM implementations in Table~\ref{tt:mge_results} of the Appendix.

\subsubsection{Systematics in the stellar kinematics extraction}\label{ss:systematics_kin}
  \begin{table}
  \label{tt:results}
\caption{Results of the different $M_{\rm BH}$ measurement methods - fiducial models with constant (top) and radially-varying (bottom) M/L}
\centering
\begin{tabular}{lcccc}
\hline\hline
Method & FoV   & $M_{\rm BH}$  &  $M/L$ & $\chi^2$/d.o.f.\\
     & (arcsec)& ($\times 10^8\,$M$_{\odot}$) & (M$_{\odot}/$L$_{\rm \odot ,H}$)  & \\
 (1) & (2) & (3) & (4) & (5)      \\
\hline
Schwarzschild & 30 & 3.6$^{+2.5}_{-1.3}$ & 0.91$\pm 0.04$ & 0.86 \\
JAM$_{\rm cyl}$     & 10 & 8.6$^{+ 0.8}_{-0.8}$   & 0.83$\pm 0.02$  &   0.24 \\
JAM$_{\rm sph}$     & 10 & 4.6$^{+2.5}_{-2.7}$   & 0.86$\pm 0.02$  &  0.21 \\
\hline
Schwarzschild & 30 & 2.9$^{+2.2}_{-1.8}$ & 0.87$\pm 0.04$ & 0.87 \\
JAM$_{\rm cyl}$     & 10 & 7.3$^{+ 1.2}_{-1.1}$  & 0.80$\pm 0.02$  &   0.20 \\
JAM$_{\rm sph}$     & 10 & 4.1$^{+ 2.2}_{-1.8}$   & 0.82$\pm 0.02$  &   0.18 \\

\hline
\end{tabular}
\\
{\textbf{Notes.} Column 1: $M_{\rm BH}$ measurement method using an inclination of $45^{\circ}$. Column 2: Field-of-view (FoV) of the data used in the methods. Columns 3, 4 and 5: Parameters of the best-fitting models (black hole mass $M_{\rm BH}$, mass-to-light ratio (M/L) and the $\chi^2$ over the degree of freedoms (using the modified kinematic} uncertainties derived in Section~\ref{ss:jam}).
\label{tt:results}
\end{table}
\textit{Applying radially dependent kinematic errors:} We noticed that within a radius of 2 arcsec from the centre of NGC 6958, our MUSE kinematic maps showed relatively low S/rN ($\approx 70$) compared to further out ($\approx 100$) leading to elevated kinematic errors in this region. For detailed maps, see Fig. 5 of the supplementary material. We carefully inspected the fitting of this region by masking and de-masking the spectrum but could not improve the S/rN significantly. As the low S/rN region can also be seen in the blue range of the fit but disappears in the redder CaT region (next subsections), we believe that the higher S/rN originates from the about 4 arcsec in diameter extended nuclear dust disk. The accuracy of the black hole mass measurement is strongly driven by these central pixels. We therefore carried out a Schwarzschild modelling test in which we applied radially increasing kinematic errors: ($V_{\rm err}$/km s$^{-1})=3.5 + 0.5\times (R/$arcsec) and ($\sigma_{\rm err}$/km s$^{-1})=4.0+0.5\times (R/$arcsec). This change reduced the velocity errors from 4.9 km s$^{-1}$ to 3.5 km s$^{-1}$ (at $R=0\arcsec$) and increased from 2.5 km s$^{-1}$ to 6.0 km s$^{-1}$ (at $R=5\arcsec$) and for the velocity dispersion errors from 5.0 km s$^{-1}$ to 4.0 km s$^{-1}$ (at $R=0\arcsec$) and 2.2 km s$^{-1}$ to 6.5 km s$^{-1}$ (at $R=5\arcsec$). This procedure is similar to applying higher weights to the central region of the kinematics during the dynamical modelling, but is a strong modification of the data set and therefore should only serve as a test for the accuracy of the measurement. Schwarzschild models using modified central kinematic errors have the following best-fitting parameters: $M_{\rm BH}=(4.4^{+1.8}_{-1.9}) \times 10^8$~M$_{\odot}$ and $M/L =0.88\pm 0.04\,$M$_{\odot}$/L$_{\rm \odot, H}$. The derived $M_{\rm BH}$ is slightly higher than the fiducial, but consistent with the results presented in Section~\ref{ss:schwarzschild} .
\begin{figure}
  \centering
      \includegraphics[width=0.45\textwidth]{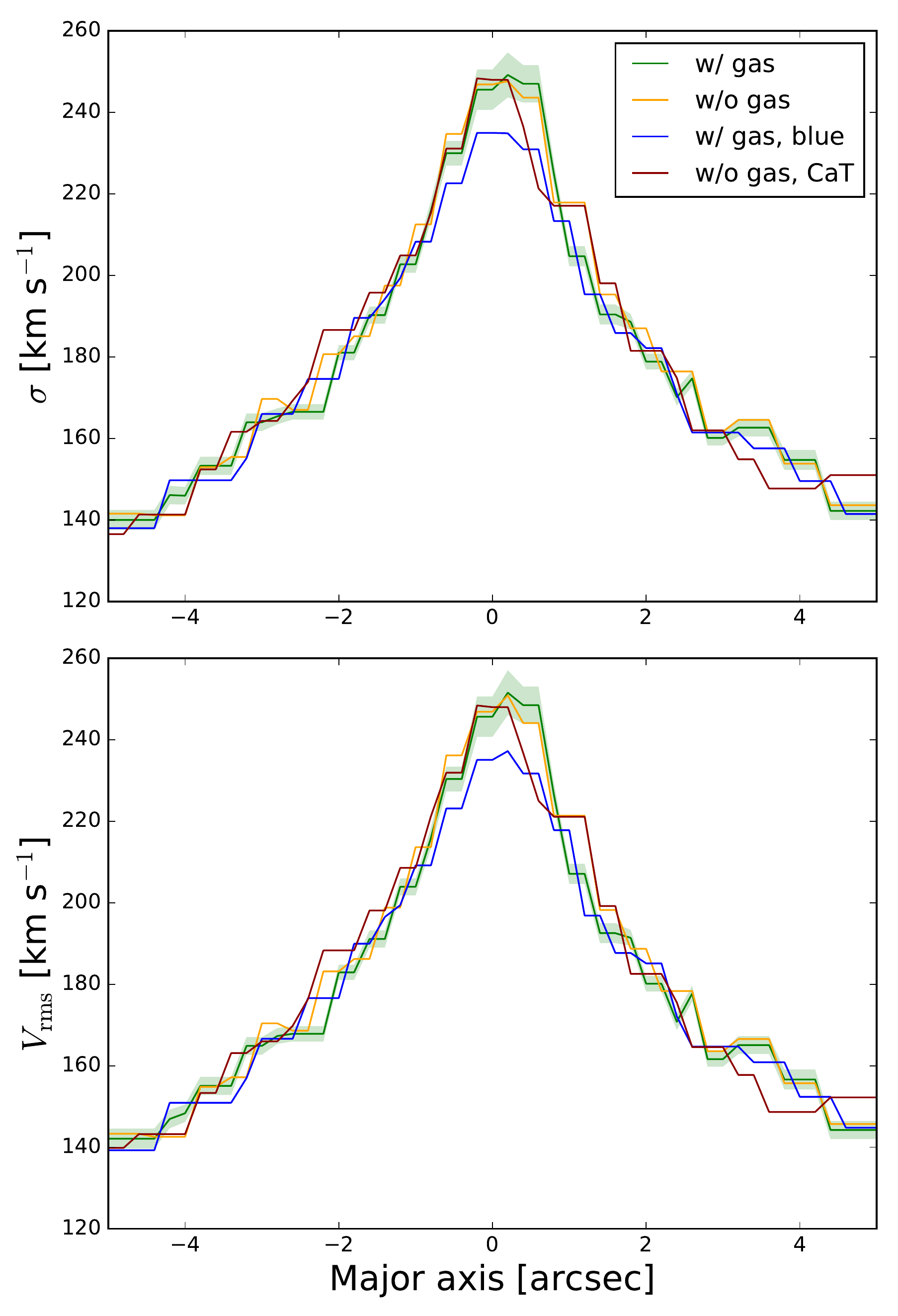}
      \caption{Comparison between the different stellar kinematics extractions with pPXF. Shown are the derived velocity dispersion (top panel) and the $V_{\rm rms}$ (bottom panel) of the extraction including gas emission lines (green), without including gas emission lines (orange), fitting the blue part of the spectrum with emission lines (blue) and fitting the Calcium triplet (red). We also show the kinematic errors for the extraction including gas emission lines as green shaded area. Except for that of the blue part of the spectrum, the extractions are consistent within their uncertainties within the central region.}
      \label{ff:vrms_comparison}
\end{figure}
\\\\
\textit{Exclusion of emission lines in the kinematics fit:} Looking at Fig.~\ref{ff:optimal_template}, it is evident that NGC 6958 shows a richness of ionised gas emission lines in its central region. While we have simultaneously fitted the emission lines and the stellar continuum in Section~\ref{ss:kinematics}, it is also common practice to mask the emission lines during the spectral fit. In this test, we extracted stellar kinematics maps by keeping the same spectral coverage but masking the emission lines. The resulting kinematics are very similar to the kinematics extracted with fitting the gas emission lines and deviations are within the very small kinematics errors (see Fig.~\ref{ff:vrms_comparison}). Running Schwarzschild models on this different set of kinematics, the best-fitting model parameters are: $M_{\rm BH}=(3.6^{+1.7}_{-1.3})\times 10^8$~M$_{\odot}$ and $M/L=0.91\pm 0.04$ M$_{\odot}$/L$_{\rm \odot, H}$, which is very close to our main result from the Schwarzschild modelling. While the exclusion of the emission lines during the pPXF kinematics extraction provided a slightly better S/rN, there is no change to the results of this paper when including them.
\\\\
\textit{Fitting only the blue spectral range of MUSE:} While extracting the stellar kinematics, we noticed a larger discrepancy between observed spectrum and pPXF fit in the red wavelength range (around 6500\,\AA) than in the blue (see Figure~\ref{ff:optimal_template}). We, therefore, tested the robustness of our result by only fitting the spectral region between 4820 and 5750\,\AA. Doing so, the median S/rN increased from 104 to 114. We created Schwarzschild models using the same inputs as with the fiducial model but replacing the stellar kinematic maps and the corresponding PSF (see Table \ref{tt:psf}). The best-fitting parameters are: $M_{\rm BH}= (3.1^{+2.2}_{-1.9}) \times 10^8$~M$_{\odot}$ and a dynamical $M/L$ of $0.84\pm 0.04\,$M$_{\odot}$/L$_{\rm \odot,H}$
As expected from the lower central $V_{\rm rms}$ (Fig.~\ref{ff:vrms_comparison}), the obtained black hole mass was lower in this case than for the other three kinematic extractions but consistent within $1\sigma$ of the measurement uncertainty. While we do not exactly know the reason for the lower central $V_{\rm rms}$, there are two possibilities. Either our spectral library is not fully representative for the bluer stars in this spectral range, or the dust is hiding the strongly accelerated stars close to the black hole (as this difference is mostly seen in the central 2\arcsec) leading to an underestimated black hole mass. Hence, we strongly advise against only using the blue spectral region of MUSE if there is any indication of nuclear dust in the galaxy.
\\\\
\textit{Fitting only CaT:} MUSE offers not only a wide FoV, but also has the great advantage of providing high-quality spectra over a wide wavelength range, including also characteristic features like the Ca II triplet around 8500\,\AA.  The Ca II triplet can be used to obtain a black hole mass measurement at a moderately higher spatial resolution of $0\farcs56$ instead of $0\farcs69$ (see Table \ref{tt:psf}) and in a wavelength regime less affected by dust. We therefore also derived kinematic maps from this spectral feature by only fitting the wavelength range between 8500 and 8800\,\AA. For this pPXF fit, we used the 61 stars from the Phoenix synthetic stellar library \citep{Husser2013}, which covers the spectral range 8350 to 9020\,\AA\, at a resolution of 1.0\,\AA. We made sure to degrade those template spectra to match the instrumental resolution of our MUSE data before starting the fitting procedure. Schwarzschild models using the Ca II triplet kinematics, uncertainties and PSF as input (while keeping the remaining inputs the same) yield the best-fitting parameters: $M_{\rm BH}= (3.5^{+2.7}_{-2.0}) \times 10^8$~M$_{\odot}$ and a dynamical $M/L$ of $0.95\pm 0.05\,$M$_{\odot}$/L$_{\rm \odot ,H}$. The black hole mass measurement is very consistent with the measurements using the optical wavelength range shown in Figure~\ref{ff:optimal_template}. Due to the consistency, this test serves as confirmation that the circumnuclear dust disk is not significantly affecting the stellar kinematics extraction.

\subsubsection{Systematics in the dynamical modeling}\label{ss:systematics_dyn}

\textit{Symmetrisation in Schwarzschild models:}
All of our dynamical models are by construction, axisymmetric. However, how to deal with this assumption is handled differently in Schwarzschild and JAM models. While the unmodified extracted kinematics are fitted with JAM, in Schwarzschild models the kinematics are usually symmetrised before the dynamical modelling to constrain the freedom of the models better and reduce the noise of the observations \citep{Krajnovic2005,vandenBosch2010}. We therefore also ran a Schwarzschild model with unmodified kinematics. This run resulted in wider $\chi^2$ contours than by bi-symmetrising the kinematics. However, the resulting black hole mass does not differ significantly: the best-fitting black hole mass using the unmodified kinematics is $M_{\rm BH}=(4.0^{+2.3}_{-1.7})\times 10^8$~M$_{\odot}$ and $M/L$ of $0.88\pm 0.05\,M_{\odot}$/L$_{\rm \odot ,H}$. Hence, the symmetrisation of the kinematics does not affect our main result. A similar result was also found in \cite{Walsh2012} when switching on the symmetrisation for their measurement.  
\\\\
\textit{Inclination:} When creating the dynamical models, we assumed a galaxy inclination of $45^{\circ}$ based on the estimated inclination of the nuclear molecular gas disc (Thater et al. in prep.). However, the inclination plays a significant role in the deprojection of the luminosity model and can, therefore, bias the final results of dynamical models. This is particularly the case for low-inclination galaxies like NGC 6958 \citep[e.g.,][]{Lablanche2012,Bellovary2014}. For those galaxies the deprojection of the stellar surface brightness to the luminosity density is strongly degenerate \citep{Rybicki1987,Gerhard1996} and the kinematics cannot properly constrain the anisotropy and mass \citep[e.g.,][]{Lablanche2012}. 
To evaluate possible systematics associated to the deprojection of the galaxy, we also constructed a Schwarzschild model assuming a nearly edge-on inclination ($89^{\circ}$). The corresponding Schwarzschild models resulted into $M_{\rm BH}=(5.7^{+2.0}_{-1.0})\times 10^8$~M$_{\odot}$ and $M/L$=$0.87\pm 0.03\,$M$_{\odot}$/L$_{\odot}$, which was the highest black hole mass that we got from all of our tests with the Schwarzschild models. $M_{\rm BH}=5.7 \times 10^8$~M$_{\odot}$ is still within the $3\sigma$ measurement uncertainties and NGC 6958 is clearly not an edge-on galaxy, such that we decided to not enhance the systematic uncertainties of the fiducial model due to this test. 
\\\\
\subsubsection{Systematics in the mass model}\label{ss:systematics_mass}
\textit{Stellar mass model:} As the mass model does not have associated errors, we tested how much using a different mass model changes the results. In Section~\ref{ss:imaging}, we described different imaging data that are available for NGC 6958. We constructed an alternative MGE model using the F814W image obtained with WFPC2 in combination with Carnegie Irvine Galaxy Survey \citep{Ho2011,Li2011} imaging data. As bluer images are much more prone to dust extinction, using the F814W images required a detailed dust masking. We followed all steps described in \cite{Thater2017}, applied a dust masking of attenuated pixels, fitted the surface brightness profile with MGE and then created dynamical Schwarzschild and JAM models. We used the same setup as in the fiducial model, inclination of $45^{\circ}$, increased kinematic errors for $R>2\arcsec$\, and stellar kinematics from the simultaneous fit of stellar absorption and gas emission. Evaluating these Schwarzschild models, we obtained a formal best fit of M$_{\rm BH}= (3.1^{+1.5}_{-1.3}) \times 10^8$~M$_{\odot}$ and $M/L =4.1\pm 0.06\,$M$_{\odot}$/L$_{\rm \odot,I}$, which is 15 per cent lower than using the mass model from near-infrared imaging, but consistent with the fiducial model results. 

The results of this test agree with the conclusion from e.g., \cite{Nguyen2018} and \cite{Nguyen2019a} that taking different mass models into account does not significantly alter the mass measurements provided that both mass models are created from images that have a spatial resolution close to the black hole SoI \citep{Yoon2017} and a careful treatment of dust attenuation was applied \citep[e.g.,][]{Cohn2021}.
\begin{figure}
  \centering
  \includegraphics[width=0.48\textwidth]{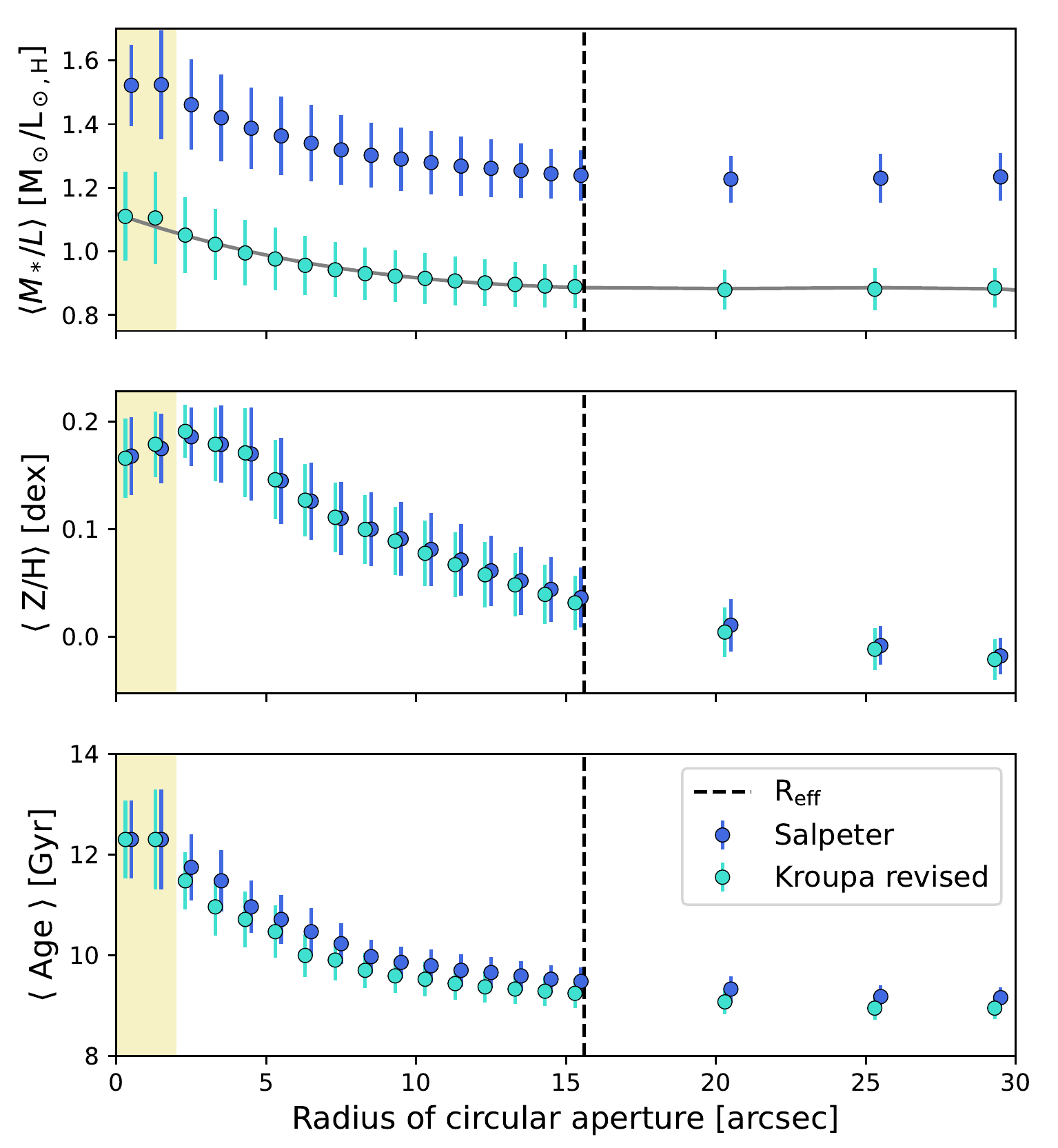}
      \caption{Stellar mass-to-light ratio in the {\it H}-band, metallicity and age profiles of NGC 6958 derived from stellar population analysis of the MUSE data. The two different colours specify whether the MILES SSP templates were calculated with a Salpeter IMF (blue) or a revised Kroupa IMF (cyan). The grey line through the Kroupa-revised IMF-based $M_*/L$ values is the 3rd-order polynomial fit derived in Section 4.4.3.The dashed line denotes the effective radius of NGC 6958, the yellow shaded area shows where dust is affecting the results ($<2\arcsec$).}
      \label{ff:ml_profile}
\end{figure}
\\\\
\textit{$M_*/L$ gradients:} In the dynamical models that we presented so far, we assumed a constant dynamical $M/L$ within the MUSE FoV. However, observations have shown that most early-type galaxies have negative stellar mass-to-light ratio ($M_{*}/L$) gradients from the centre towards larger radii \citep[e.g.,][]{Tortora2011,Li2018b}. With an effective radius of $15\farcs3$ and a galaxy mass of $8\times10^{10} M_{\odot}$, we expect a small negative gradient for NGC 6958 (Figure 7 of \citealt[][]{Li2018b}). On the other hand, the presence of nuclear molecular gas might lead to on-going nuclear star formation \citep{Crocker2011} which would indicate a lower $M_{*}/L$ in the galaxy centre \citep{Davis2017b}. A $M_{*}/L$ gradient for NGC 6958 is therefore expected. If the change in the $M_{*}/L$ profile is of the order of ten per cent, the ignorance of $M/L$ gradients in the dynamical models can lead to a significant under-estimation of the stellar mass in the centre and thus an overestimation of the black hole mass. \cite{McConnell2013b} and  \cite{Thater2019} included radially-varying $M_{*}/L$ in their dynamical models and noted a decrease in M$_{\rm BH}$ by up to 30 per cent, while \cite{Cappellari2002b} found negligible effects. 

We followed the approach of \cite{Thater2019}, derived $M_{*}/L$ profiles for NGC 6958 from the full-spectrum fitting of the MUSE observations and included the $M_{*}/L$ variation in our dynamical models. We utilized the pPXF routine and fitted a linear combination of MILES Single Stellar Population (SSP) models \citep[Version 11.0;][]{Vazdekis2016} to the co-added MUSE spectra within circular apertures with radii between $0\farcs5$ and $29\farcs5$. We used two different SSP model libraries assuming Padova isochrones \citep{Girardi2000} and 1) a \cite{Salpeter1955} initial mass function (IMF) and 2) a \cite{Kroupa2001} revised IMF. For each of the two IMF choices, we used template spectra in a regular grid of log(age) between 0.1 and 14.1 Gyrs and metallicities ([Z/H]) between -1.71 and 0.22 dex. Note, that compared to our previous study \citep{Thater2019}, we cut the grid to the safe age ranges specified at the MILES website\footnote{http://research.iac.es/proyecto/miles//pages/ssp-models/safe-ranges.php}. We then fitted those SSP templates to our MUSE spectra between 4820 and 6820\,\AA\, and masked the same regions as for the kinematics extraction (Section~\ref{ss:kinematics}). Due to a template mismatch caused by bad fits in the red wavelength range, we added a "red" mask to cover the region between 6350 and 6570\,\AA. This template mismatch remains when adding additional lines like [Fe X] to the pPXF fit. Our age determination is strongly driven by the H$\beta$ absorption line at rest-frame wavelength of 4861\,\AA\, which is contaminated by the H$\beta$ gas emission line. To obtain a robust age estimate, we did not mask the emission lines but added the Balmer lines and forbidden emission lines to the fit. The Balmer lines were assumed to follow a characteristic theoretical Balmer decrement (case B recombination with T=$10^{4}$ K, n=100 cm$^{-3}$), but were allowed to vary their relative intensities following a \cite{Calzetti2000} reddening curve. As in \cite{Thater2019}, we also made use of the regularisation to the weights in age and metallicity to derive a smooth star formation history (SFH). The uncertainties were determined from the difference between regularised and non-regularised solution. 
\begin{figure}
  \centering
      \includegraphics[width=0.45\textwidth]{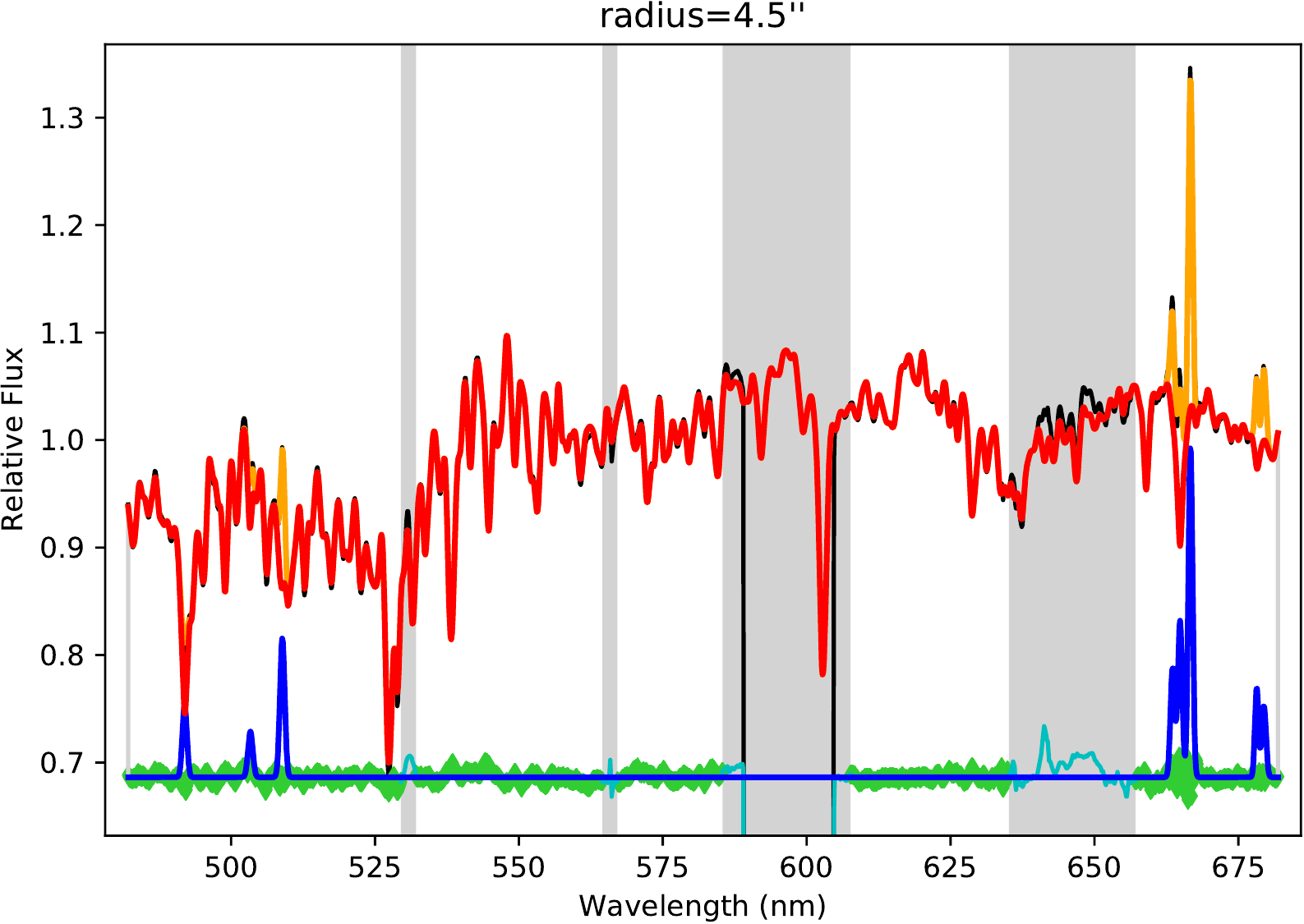}
        \includegraphics[width=0.45\textwidth]{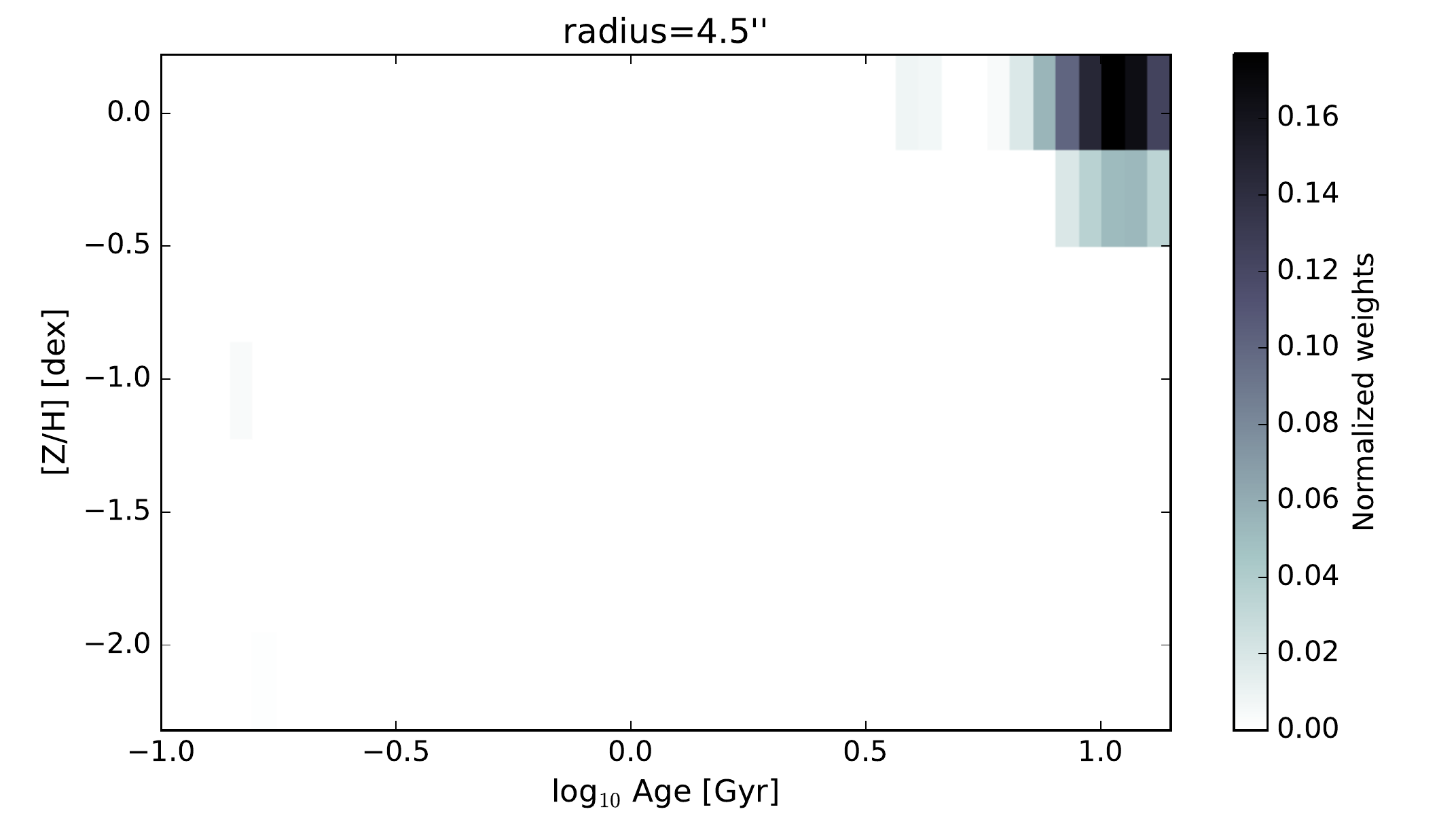}
        \includegraphics[width=0.45\textwidth]{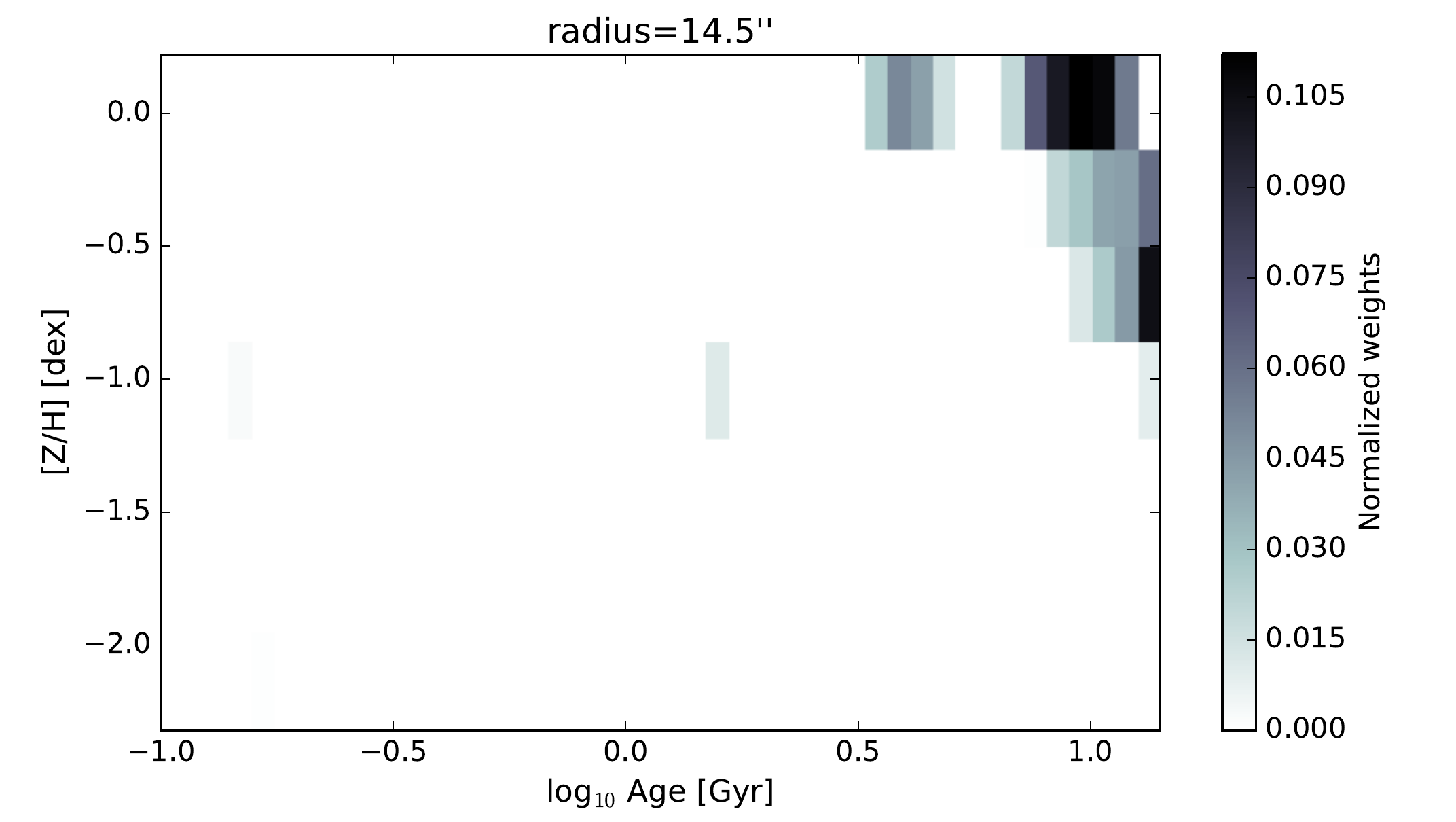}
      \caption{Results of the stellar population analysis. Top: A comparison between the integrated MUSE spectrum (black) and the spectral fit (red). The representation of the components is the same as Fig. 4. Bottom: Mass-weighted star formation histories (age and metallicity bins that contribute to the spectral fitting). The greyscale show the normalized weights of the models. Both panels were chosen to reflect the contribution at the edge of our kinematics.}
      \label{ff:stell_pop}
\end{figure}

Figure~\ref{ff:ml_profile} shows the mass-weighted $M_{*}/L$, age and metallicity profiles extracted for the two different IMFs. Like \cite{Thater2019}, we see a similar trend in the $M_{*}/L$ gradients for the two IMFs but a difference of about 0.4\,M$_{\odot}$/L$_{\odot}$, the $M_{*}/L$ of the Kroupa revised IMF being lower. The stellar $M_{*}/L$ derived from the Salpeter IMF are not consistent with the dynamical $M/L$ from our dynamical models and we will use the stellar $M_{*}/L$ from the Kroupa revised IMF for the rest of this section. A `light' Milky-Way-like IMF like a Kroupa IMF appears to better describe the low- to intermediate-dispersion ellipticals \citep{Cappellari2012,Cappellari2013,Lyubenova2016,Li2017} which is confirmed for NGC 6958 by our dynamical modelling results.  We derived stellar $M_{*}/L$ values from 1.12\,M$_{\odot}$/L$_{\odot}$ in the centre to 0.88\,M$_{\odot}$/L$_{\odot}$ at about 30\arcsec which are consistent with the dynamical $M/L$ from our dynamical models. Driven by variations in metallicity and age, we see a clear stellar $M_{*}/L$ decrease of 20 per cent within the bulge effective radius, flattening out at larger radii. The shape of the $M_{*}/L$ can be well parameterised by a third-order polynomial\footnote{$(M_{*}/L$ /M$_{\odot}$/L$_{\rm \odot,H})$= 1.12 $-$ 3.29\,$\times$\,10$^{-2} \times$  ($R$/arcsec) $+$ 1.50\,$\times$\,10$^{-3} \times$  ($R$/arcsec)$^2$ $-$ 2.24\,$\times$\,10$^{-5} \times$  ($R$/arcsec)$^3$  for $R \leq 30$ arcsec\,  and constant 0.89 M$_{\odot}$/L$_{\odot}$ for $R>30$ arcsec}. For $R<2\arcsec$, our derived values were strongly affected by the nuclear dust disk of the galaxy, and we excluded those values in the fit. We then multiplied each component j of our MGE with the interpolated stellar $M/L_{*}$ at the corresponding $\sigma_{\rm j}$ radius of the MGE component and included this mass density into the dynamical models. From these new models, we obtained a black hole mass of $(2.9^{+2.2}_{-1.8}) \times 10^8\,M_{\odot}$ in the Schwarzschild models resulting in an about 20 per cent decrease in mass. A radially varying stellar $M/L_{*}$ is usually not explicitly included in the dynamical models when deriving black hole masses. However, when the total mass profile is allowed to differ from the distribution of the tracer population (e.g. by including a dark matter profile), the change in $M/L_{*}$ can be accounted for as well. Furthermore, the difference to the models with constant $M/L$ is so small, that the models with radially-varying $M/L$ lie within the $3\sigma$ uncertainties of our reported measurements. 

When inspecting the star formation histories at different distances from the centre (Fig.~\ref{ff:stell_pop}), we noticed a second younger stellar component of about 4 Gyrs which starts to be visible at 5 arcsec. The star formation history might also reveal a third low-mass component at an age of about 1.5 Gyrs, possibly the remnant of a recent minor merger which has also been discussed in the literature \citep{Malin1983,Saraiva1999}. However, the 1.5 Gyr old stellar component is very uncertain as it appears only in a single age and metallicity bin. As it is only visible at radii larger than 15 arcsec, this young component does not affect our dynamical models.

\subsubsection{Conclusion of the systematics and alignment of the velocity ellipsoid}\label{ss:systematics_conc}
\begin{figure}
  \centering
      \includegraphics[width=0.46\textwidth]{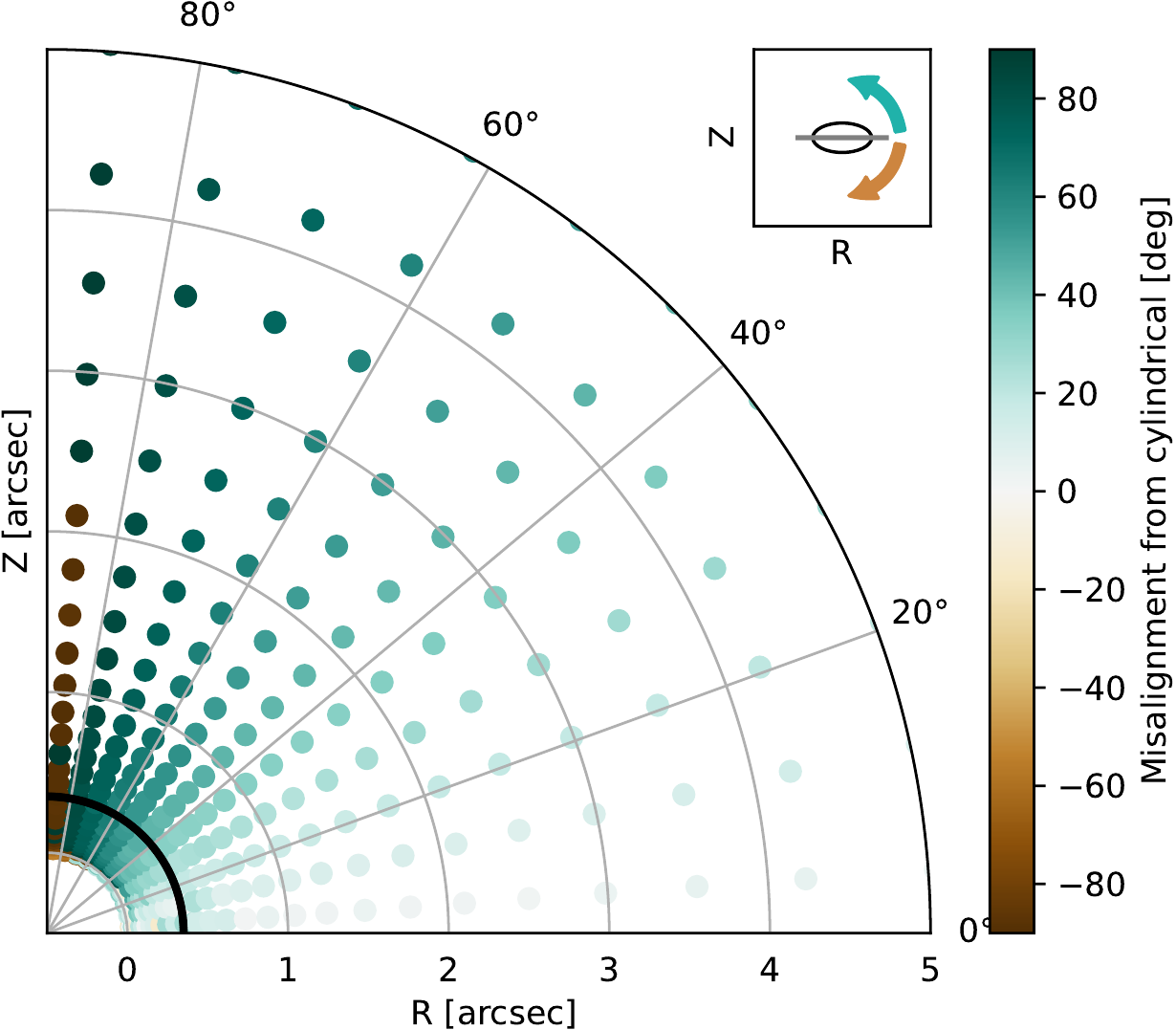}
        \includegraphics[width=0.46\textwidth]{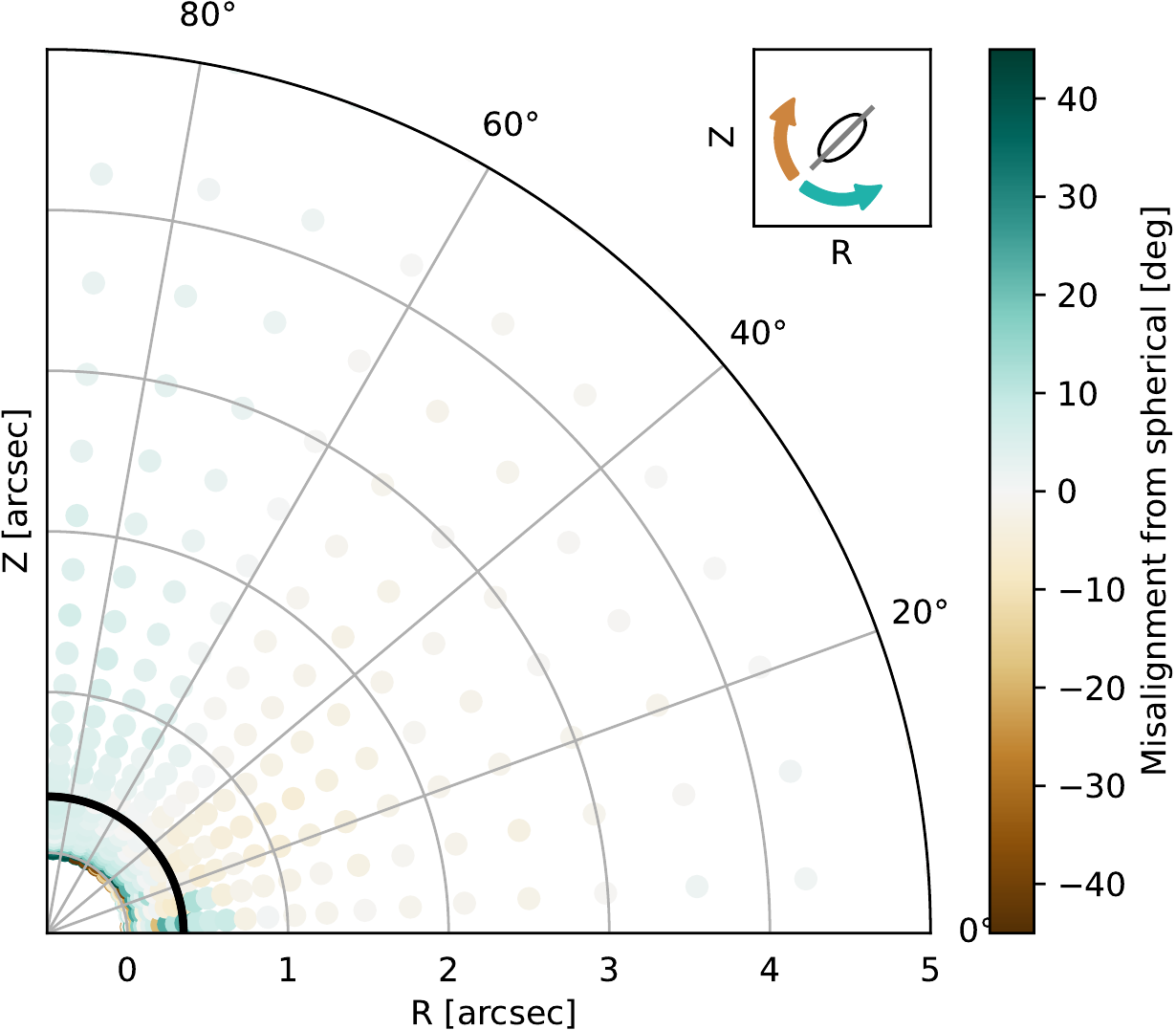}
      \caption{The alignment of the  velocity ellipsoid from the best-fitting Schwarzschild model in the meridional plane projection. The upper panel shows the misalignment from a cylindrical-aligned velocity ellipsoid. We show the non-tilted version in the inset in the top right of the panel. Both arrows in the inset indicate the direction of the misalignment. The bottom panel shows the misalignment from a spherical-aligned velocity ellipsoid (example again in the inset). The values within the black curve are affected by the PSF.  }
      \label{ff:meri_plane}
\end{figure}

In the previous sub-sections, we analysed several sources of uncertainty. While some of them gave negligible deviations from the main result $M_{\rm BH}= (3.6^{+2.5}_{-1.3}) \times 10^8 M_{\odot}$, for others we noticed significant differences which we included as additional systematic uncertainty in our final result. We found a confirmation of our main result by accounting for differences in the kinematics extraction (except for going into the blue region) and also by using different mass models. However, strong biases for both used dynamical modeling methods are evident from the mass deprojection at such a low inclination. Taking all tested systematics into account, we report $M_{\rm BH}= (3.6^{+2.7}_{-2.4}) \times 10^8 M_{\odot}$ at $3\sigma$ significance as final result of the Schwarzschild models. This measurement is consistent with our JAM$_{\rm  sph}$ models which resulted in more massive black holes of $M_{\rm BH}= (4.6^{+2.5}_{-2.7}) \times 10^8 M_{\odot}$ and inconsistent with $M_{\rm BH}= (8.6^{+0.8}_{-0.8}) \times 10^8 M_{\odot}$ from JAM$_{\rm cyl}$. One  difference between the models is the alignment of the velocity ellipsoid. Compared to JAM, the velocity ellipsoid does not have a fixed alignment in the Schwarzschild models. The tilt angle $\alpha$ of the velocity ellipsoid can be derived from the components of the velocity tensor via $\tan 2\alpha =  2\langle v_r v_{\theta}\rangle/(\langle v^2_{r}\rangle \,\text{-}\, \langle v^2_{\theta} \rangle$) where the components are an output from the Schwarzschild models \cite[see e.g.,][]{Cappellari2007}. $\alpha$ measures the deviation from spherical alignment, which corresponds to $\alpha=0^{\circ}$. An analogous tilt angle can also be defined for cylindrical coordinates (see e.g., \citealt{Cappellari2008, Smith2009}). Figure~\ref{ff:meri_plane} shows the misalignment of the Schwarzschild velocity ellipsoid from cylindrical and spherical-alignment for our best-fit model. We also explored the axis ratio of the velocity ellipsoid \citep{Cappellari2008} and found ranges between 0.65 and 0.9, which indicates that the misalignment is not neglegible. It is clearly visible that the velocity ellipsoid has a varying orientation, which was also seen in axisymmetric Schwarzschild models for other galaxies \citep{Cappellari2007}. Furthermore, the velocity ellipsoid of our Schwarzschild model of NGC 6958 is more consistent with spherical-alignment, which could explain why the spherical-aligned JAM models give a $M_{\rm BH}$ that is closer to the measurement from Schwarzschild.

\subsubsection{The infuence of dark matter}
\label{ss: darkmatter}
Our dynamical models assume self-consistence (mass follows light). Breaking this assumption by including dark  matter as additional component in the dynamical models can lead  to  systematic  changes  in  the  black  hole  mass if $R_{\rm SoI}$ is not well resolved \citep{Gebhardt2009, Gebhardt2011,Schulze2011,Rusli2013}. NGC 6958 lies in a similar mass range as the galaxies that we analysed in \cite{Thater2019}, in particular NGC~584 and NGC~2784. It is therefore expected that dark matter has only a negligible contribution within the FoV of our kinematic tracer. Analogous to our previous study, we used the radial acceleration relation \citep{McGaugh2016,Lelli2017} and calculated the total acceleration from our MGE model yielding $g_{\rm dyn}=1.3\times10^{-9}$\,m\,s$^{-2}$. As long as a galaxy stays in the linear regime of the radial acceleration relation 
($g_{\rm dyn}> g_{\rm crit}=1.2 \times 10^{-10} $m\, s$^{-2}$) it is expected that dark matter only marginally affects the dynamics. Based on the calculated $g_{\rm dyn}$, NGC 6958 lies within the linear regime of the radial acceleration relation. However, the dark matter halo will influence orbits which go beyond the probed radii, but also come close to the SMBH. This might explain the tendency for moderately larger $M_{\rm BH}$ in the JAM$_{\rm{sph}}$ models compared to Schwarzschild models that cover almost one effective radius of NGC 6958.  
\begin{figure*}
  \centering
      \includegraphics[width=0.47\textwidth]{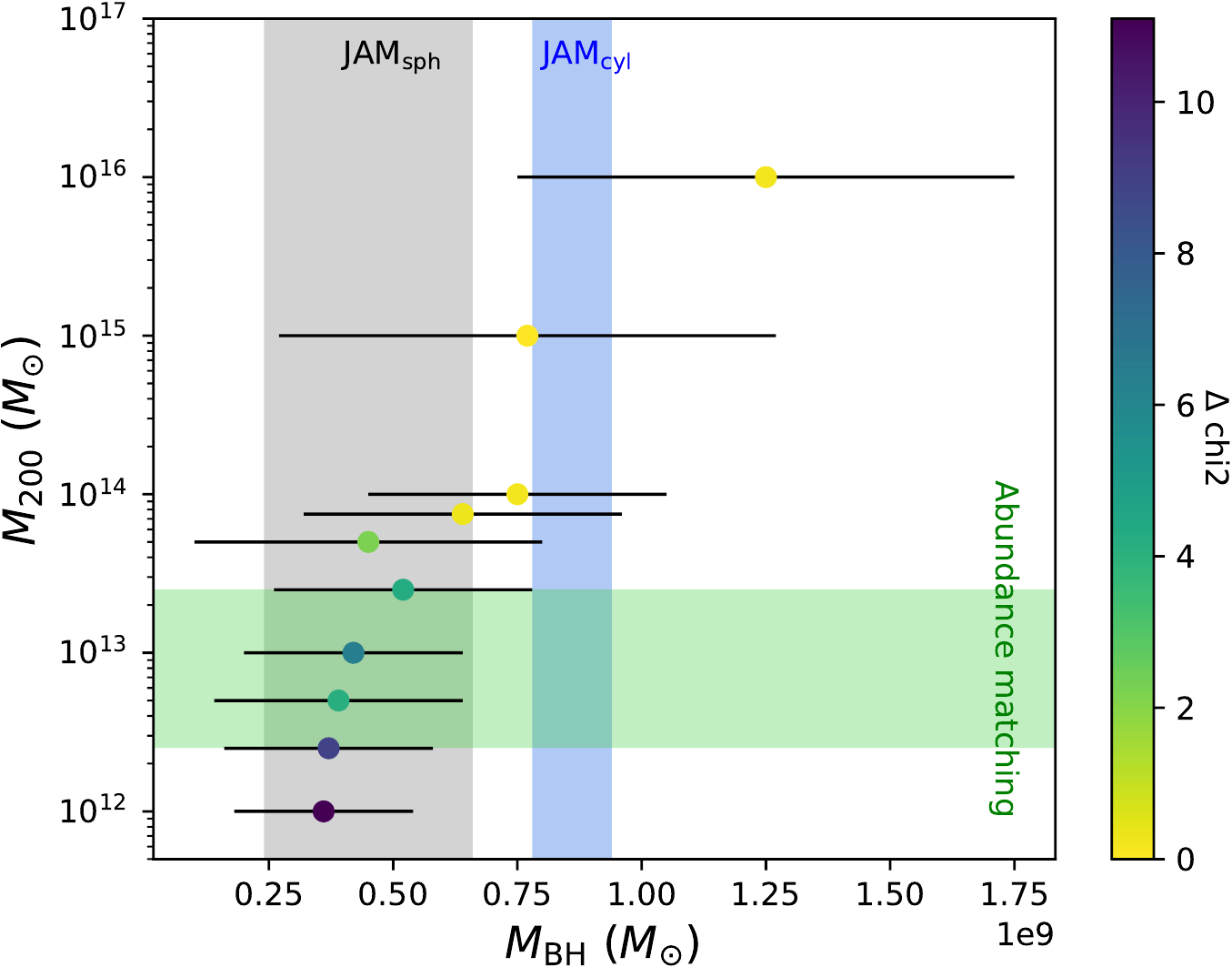}
      \hspace{0.02\textwidth}
        \includegraphics[width=0.47\textwidth]{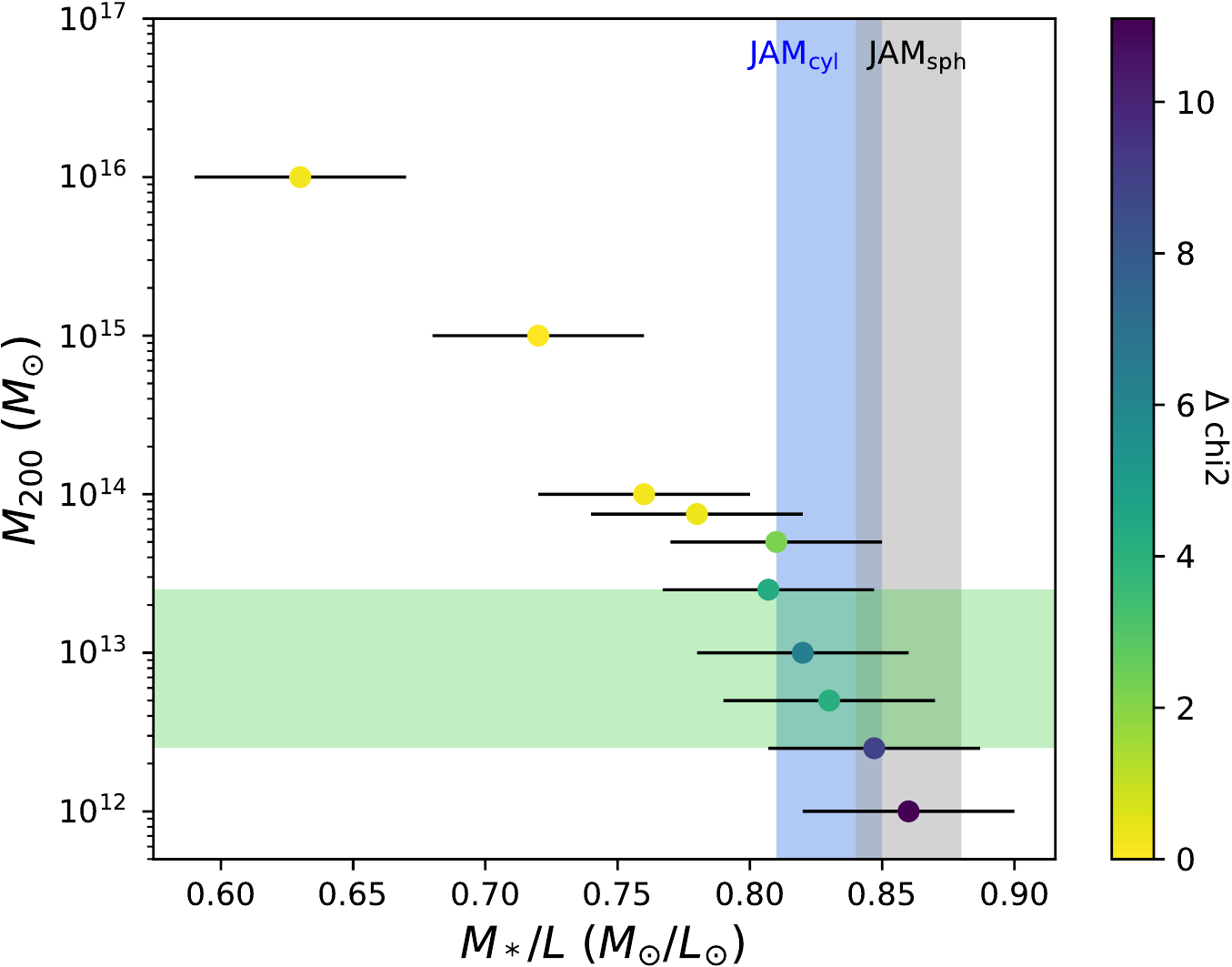}
      \caption{The effect of dark matter on the measured black hole mass and $M/L$ in the Schwarzschild models. Every point is the best-fit Schwarzschild measurement of one of the grids shown in the Appendix E. The colour specifies the $\Delta \chi^2$ between the best-fit model and the minimum $\chi^2$ of all Schwarzschild models + dark matter. All Schwarzschild results are compared with the best-fit results from JAM with spherically (grey) and cylindrially (blue) aligned velocity ellipsoid. Note that with JAM, we measured a dynamical $M/L$.}
      \label{ff:dm}
\end{figure*}

We decided to test this hypothesis by running the fiducial Schwarzschild models with a Navarro-Frank-White \citep[NFW; ][]{Navarro1996}  dark halo. We used a similar approach to the one described in \cite{Cappellari2013}, which in their study was applied to JAM models. We assumed the dark matter follows a two-parameter power law NFW profile with a spherical shape. The NFW can then be parametrised as a function of the halo mass ($M_{200}$) and the halo concentration  ($c_{200}$) which are connected via the $M_{200} - c_{200}$ relation \citep{Navarro1996}. We used equation 8 by \cite{Dutton2014} in order to make the halo profile a function of only one free parameter, $M_{200}$. We then fitted a one-dimensional MGE to this profile and added those MGE parameters to the galaxy potential in the Schwarzschild models. The Schwarzschild models were run in a three-dimensional grid ($M_{\rm BH}, M_{*}/L, M_{200}$). Note that in this run $M_{*}/L$ is the stellar mass-to-light ratio and not the dynamical mass-to-light ratio anymore. 
For $M_{\rm BH}$ and $M_{*}/L$, we kept the grid values from Section~\ref{ss:schwarzschild}, while the $M_{200}$ grid values were varied between $10^{12}$ M$_{\odot}$ and $10^{14}$ M$_{\odot}$ which is expected for a galaxy of $\approx 10^{11}$ M$_{\odot}$ stellar mass based on abundance matching \citep{Moster2013}. In order to test the effects of very massive (and physically unrealistic) halos, we extended the $M_{200}$ grid to $10^{16}$ M$_{\odot}$. Compared to the main runs in this paper, for this test we sampled the orbits in a smaller orbit library with 21 logarithmically-spaced orbit energies, 8 linearly-spaced orbit angular momenta L$_{\rm z}$ and 7 linearly-spaced non-classical third integral values I3. The smaller orbit library does not change the best-fit values but mostly has an effect on the contour shape of the $\chi^2$ distribution, and is therefore sufficient for this test.

The resulting Schwarzschild grids are shown in Fig.~\ref{ff:dm_grid}. The degeneracy between $M_{\rm BH}, M_{*}/L$ and $M_{200}$ is clearly visible. With increasing dark matter fraction, the $M_{*}/L$ decreases and $M_{\rm BH}$ increases. However, it is not possible with our MUSE data set (covering one effective radius of NGC 6958) to put constraints on the dark matter, and the best-fit $\chi^2$ values of the different grids are very similar. Small fluctuations between the $\chi^2$ values are likely caused by numerical errors. The trends of these models and a comparison with the JAM result is shown in Fig.~\ref{ff:dm}. Within a dark matter fraction that is consistent with abundance matching, the change in black hole mass is not significant and stays within the uncertainties that were given in the previous tests. Furthermore, taking into account dark matter in Schwarzschild models would remove the remaining small difference between Schwarzschild and JAM$_{\rm sph}$ models. The discrepancy with JAM$_{\rm cyl}$ cannot be explained with dark matter as it would require unrealistically high dark matter mass halos (matching those of massive galaxy clusters), and seems to predominantly follow from the assumption of the velocity ellipsoid as discussed in Section~\ref{ss:systematics_conc}.

Fig.~\ref{ff:dm} also shows the effect of the inclusion of dark matter on the measured $M_*/L$. Contrary to Fig. 8, the M/L from JAM is now larger or equal to that from Schwarzschild. This is because the $M/L$ from JAM is a total value, while that from Schwarzschild's models is the $M_*/L$ of the stars alone. Figure~\ref{ff:dm_grid} shows that when the dark halo is small and the stars dominate the total mass in the Schwarzschild models, the $M/L$ from JAM and Schwarzschild's models agree well. But when the dark matter contribution increases, the stellar $M_*/L$ must correspondingly decrease as observed. Specifically, the stellar and total $M/L$ are approximately related as (eq. 23 of \cite{Cappellari2013})
\begin{equation}
(M/L)_{\rm JAM} \approx (M_*/L)/[1 - f_{\rm DM}(r=R_{\rm JAM})],    
\end{equation}
here $f_{\rm DM}$($r=R_{\rm JAM}$) is the fraction of dark matter enclosed within the region fitted by the JAM models.
\newline

From the $\chi^2$ distribution of our Schwarzschild models, it is not possible to solve the $M_{\rm BH}$ - $M_{200}$ degeneracy. That is why we tried to quantify the effect of dark matter on the black hole mass by creating a Gaussian prior from the abundance matched dark halo values and multiplying this prior probability with the likelihood probability of the dynamical models (similar to a Bayesian analysis):
\begin{equation}
P_{\rm posterior} (\mathrm{Model}\, \vert \,  M_{\rm BH}, M_{*}/L,  M_{\rm 200}) \propto P_{\rm likelihood} \cdot P_{\rm prior} (M_{\rm 200}) 
\end{equation}

NGC 6958 has a galaxy mass of $(8.6 \pm 2.0)\times10^{10}$ M$_{\odot}$ which results in $\log (M_{200}/$M$_{\odot}) = 12.9 \pm 0.4$. Together with a scatter in the abundance matching relation of 0.1 dex at z=0 \citep{Moster2013}, we obtained a Gaussian prior with $\mu = \log (M_{200}/$M$_{\odot}) = 12.9$ and $\sigma = 0.5$. The likelihood can be directly inferred from the $\chi^2$ distribution of the dynamical models. The posterior probability is then calculated as

\begin{equation}
\ln P_{\rm posterior} \propto -\chi^2/2 + \ln P_{\rm prior}
\end{equation}
We then assumed that we can compute confidence levels on $P_{\rm posterior}$ as usually done on the likelihood alone. Adopting the minimum of $\ln P_{\rm posterior}$ as our best-fit value, we obtained $M_{\rm BH, DM} = (4.5 \pm 2)\times10^8$ M$_{\odot}$. This value is 25\% larger than $M_{\rm BH}$ from the fiducial Schwarzschild models but fully consistent with those results.
This test confirms that the inclusion of dark matter in our dynamical models does not significantly change our final results.

\section{Discussion and conclusion}
\subsection{Black hole scaling relations}
Together with the derived effective velocity dispersion our estimated black hole mass $(3.6^{+2.7}_{-2.4}) \times 10^8 M_{\odot}$ from the Schwarzschild models can be compared with dynamical black hole masses from the literature (most recent compilation by \citealt{Sahu2019b}). We first compared our $M_{\rm BH}$ measurement with predictions from the bulge effective velocity dispersion of 168 km s$^{-1}$ with different scaling relations. Using the scaling relation by \cite{Saglia2016} for power-law early-type galaxies, we estimated $(1.1 \pm 0.2) \times 10^8 $\,M$_{\odot}$ where the uncertainty was derived from the uncertainty in the velocity dispersion. A similar black hole mass was estimated for the scaling relation in \cite{Bosch2016}. However, our mass measurement turned out to be three times more massive. On the other hand, \cite{Sahu2019b} used the central velocity dispersion in their black hole scaling relation analysis to correct for possible contamination of disc rotation. We also derived the central velocity dispersion of NGC~6958 (within 1.95\arcsec) yielding $220 \pm 5 $ km s$^{-1}$. Inserting this value into their black hole mass - central velocity dispersion relation for early-type galaxies gives $(3.6 \pm 0.5) \times 10^8 $\,M$_{\odot}$. Our derived black hole mass is therefore neither strongly over- nor undermassive when compared to the bulk of literature black hole masses given NGC 6958's central velocity dispersion. We also compared our mass measurement with the black hole mass - bulge mass relation from \cite{Saglia2016} for power-law early-type galaxies. Given NGC 6958's bulge mass $(3.6 \pm 1.4) \times 10^{10}$ M$_{\odot}$ (Table 1), this relation yields a black hole mass of $(1.1 \pm 0.9) \times 10^8 $\,M$_{\odot}$. Again our measurement is over massive compared to the scaling relations but we might underestimate the bulge mass due to the limited FoV. Furthermore, our measurement is consistent with the general scatter of black hole masses at $\sigma_{\rm e,star}\approx 200$ km s$^{-1}$. We will further discuss the implications of our measurement in the context of the scaling relations and galaxy assembly in Thater et al. (in preparation), where we will use gas kinematics as an independent tracer to derive the black hole mass. \\

\subsection{Summary of our results}
We have presented our central black hole mass measurement of the lenticular galaxy NGC 6958. For that purpose, we obtained adaptive-optics assisted MUSE science verification data and extracted ionised gas and stellar kinematics maps. We used Gaussian LOSVDs for the ionised gas kinematics and LOSVDs parametrised as Gauss-Hermite polynomials up to the order of 6 for the stellar kinematics extraction. The ionised gas kinematics have a clear velocity dispersion peak of 270 km s$^{-1}$ and a regular rotational velocities within 5 arcsec. However, at greater distance from the centre, we notice strongly disturbed features in the gas rotational velocity map, dominated by receding motion. On the other hand, our stellar kinematic maps show very regular rotation within 15 arcsec with a maximal rotational velocity of 130 km s$^{-1}$  and a clear and distinct velocity dispersion peak of 250 km s$^{-1}$. We combined the extracted stellar kinematics with high-resolution NICMOS F160W images and created axisymmetric Jeans anisotropic and Schwarzschild models. Jeans anisotropic models gave best-fit black hole masses of $(4.6^{+2.5}_{-2.7}) \times 10^8 M_{\odot}$ and $(8.6^{+0.8}_{-0.8}) \times 10^8 M_{\odot}$ for spherical and cylindrical alignment of the velocity ellipsoid, respectively. From our Schwarzschild models, we estimated a black hole mass of $(3.6^{+2.7}_{-2.4}) \times 10^8 M_{\odot}$ and a constant dynamical $M/L$ of $0.91\pm 0.04$\,M$_{\odot}$/L$_{\rm \odot,H}$. Two of our three determinations are consistent within their uncertainties, while only JAM$_{\rm cyl}$ is slightly inconsistent. When using a radially-varying $M/L$ in our our dynamical models we obtained a black hole mass 20 per cent lower than the fiducal models. When adding a dark halo (based on abundance matching) to our Schwarzschild models, the black hole mass increases by 25 per cent. Our derived black hole mass is overmassive compared to most scaling relation but agrees with the $M_{\rm BH} - \sigma_{\rm e,star}$ relation within $3\sigma$.  We carefully discuss the systematics with the kinematic data, the mass model and the dynamical models in Section 5 and conclude that we fully cover the discussed systematics in our associated uncertainties. The most dominant effects were found to come from inconsistencies in the stellar kinematics extraction (when using only the "blue" spectral range), the well-known problem of the inclination - mass deprojection degeneracy in low-inclination galaxies and model dependent differences. In a companion paper, we will derive the SMBH mass in NGC 6958 using ionised and molecular gas as dynamical tracers. A cross-check of the three measurements will help to constrain the black hole mass and provide additional value in understanding whether the scatter in the black hole scaling relations is strongly affected by measurements from different measurement methods.

\section*{Acknowledgements}
We thank the anonymous referee for the detailed referee report which improved the quality of this manuscript. We thank Tadeja Ver\v{s}i\v{c} for illuminating discussions on dark matter in dynamical models. S.T. acknowledges funding from the TAIZAI Visiting Fellowship during the Spring
2018 at National Observatory of Japan and the DFG Grant KR 4548/1-1. Furthermore, part of this research was supported by the European Union's Horizon 2020 research and innovation programme under grant agreement NO 724857 (Consolidator Grand ArcheoDyn). P.M.W. was supported by BMBF Verbundforschung (MUSE-NFM Project, grant 05A17BAA). M.B. was supported by STFC consolidated grant "Astrophysics at Oxford" ST/H002456/1 and ST/K00106X/1. T.A.D. acknowledges support from the UK Science and Technology Facilities Council through grant ST/S00033S/1. This work is based on observations collected at the European Organisation for Astronomical Research in the Southern Hemisphere under ESO programme 60.A-9193(A) and also based on observations made with the NASA/ESA Hubble Space Telescope, obtained from the Hubble Legacy Archive, which is a collaboration between the Space Telescope Science Institute (STScI/NASA), the Space Telescope European Coordinating Facility (STECF/ESA) and the Canadian Astronomy Data Centre (CADC/NRC/CSA). This research has made use of the NASA/IPAC Extragalactic Database(NED) which is operated by the Jet Propulsion Laboratory, California Institute of Technology, under contract with the National Aeronautics and Space Administration.
This research is partially based on data from the MILES project.

\section*{DATA AVAILABILITY}
Raw MUSE data are available in the ESO archive. Kinematics are
available from the author on reasonable request.

\bibliography{papers} 

\begin{thebibliography}{}
\makeatletter
\relax
\def\mn@urlcharsother{\let\do\@makeother \do\$\do\&\do\#\do\^\do\_\do\%\do\~}
\def\mn@doi{\begingroup\mn@urlcharsother \@ifnextchar [ {\mn@doi@}
  {\mn@doi@[]}}
\def\mn@doi@[#1]#2{\def\@tempa{#1}\ifx\@tempa\@empty \href
  {http://dx.doi.org/#2} {doi:#2}\else \href {http://dx.doi.org/#2} {#1}\fi
  \endgroup}
\def\mn@eprint#1#2{\mn@eprint@#1:#2::\@nil}
\def\mn@eprint@arXiv#1{\href {http://arxiv.org/abs/#1} {{\tt arXiv:#1}}}
\def\mn@eprint@dblp#1{\href {http://dblp.uni-trier.de/rec/bibtex/#1.xml}
  {dblp:#1}}
\def\mn@eprint@#1:#2:#3:#4\@nil{\def\@tempa {#1}\def\@tempb {#2}\def\@tempc
  {#3}\ifx \@tempc \@empty \let \@tempc \@tempb \let \@tempb \@tempa \fi \ifx
  \@tempb \@empty \def\@tempb {arXiv}\fi \@ifundefined
  {mn@eprint@\@tempb}{\@tempb:\@tempc}{\expandafter \expandafter \csname
  mn@eprint@\@tempb\endcsname \expandafter{\@tempc}}}

\bibitem[\protect\citeauthoryear{{Ahn} et~al.,}{{Ahn} et~al.}{2018}]{Ahn2018}
{Ahn} C.~P.,  et~al., 2018, \mn@doi [\apj] {10.3847/1538-4357/aabc57}, \href
  {http://adsabs.harvard.edu/abs/2018ApJ...858..102A} {858, 102}

\bibitem[\protect\citeauthoryear{{Annibali}, {Bressan}, {Rampazzo},
  {Zeilinger}, {Vega}  \& {Panuzzo}}{{Annibali} et~al.}{2010}]{Annibali2010}
{Annibali} F.,  {Bressan} A.,  {Rampazzo} R.,  {Zeilinger} W.~W.,  {Vega} O.,
  {Panuzzo} P.,  2010, \mn@doi [\aap] {10.1051/0004-6361/200913774}, \href
  {http://adsabs.harvard.edu/abs/2010A%26A...519A..40A} {519, A40}

\bibitem[\protect\citeauthoryear{{Bacon} et~al.,}{{Bacon}
  et~al.}{2010}]{Bacon2010}
{Bacon} R.,  et~al., 2010, in Ground-based and Airborne Instrumentation for
  Astronomy III. p. 773508, \mn@doi{10.1117/12.856027}

\bibitem[\protect\citeauthoryear{{Bacon} et~al.,}{{Bacon}
  et~al.}{2017}]{Bacon2017}
{Bacon} R.,  et~al., 2017, \mn@doi [\aap] {10.1051/0004-6361/201730833}, \href
  {https://ui.adsabs.harvard.edu/abs/2017A&A...608A...1B} {608, A1}

\bibitem[\protect\citeauthoryear{{Barth}, {Darling}, {Baker}, {Boizelle},
  {Buote}, {Ho}  \& {Walsh}}{{Barth} et~al.}{2016}]{Barth2016}
{Barth} A.~J.,  {Darling} J.,  {Baker} A.~J.,  {Boizelle} B.~D.,  {Buote}
  D.~A.,  {Ho} L.~C.,   {Walsh} J.~L.,  2016, \mn@doi [\apj]
  {10.3847/0004-637X/823/1/51}, \href
  {http://adsabs.harvard.edu/abs/2016ApJ...823...51B} {823, 51}

\bibitem[\protect\citeauthoryear{{Beifiori}, {Courteau}, {Corsini}  \&
  {Zhu}}{{Beifiori} et~al.}{2012}]{Beifiori2012}
{Beifiori} A.,  {Courteau} S.,  {Corsini} E.~M.,   {Zhu} Y.,  2012, \mn@doi
  [\mnras] {10.1111/j.1365-2966.2011.19903.x}, \href
  {https://ui.adsabs.harvard.edu/abs/2012MNRAS.419.2497B} {419, 2497}

\bibitem[\protect\citeauthoryear{{Bellovary}, {Holley-Bockelmann},
  {G{\"u}ltekin}, {Christensen}, {Governato}, {Brooks}, {Loebman}  \&
  {Munshi}}{{Bellovary} et~al.}{2014}]{Bellovary2014}
{Bellovary} J.~M.,  {Holley-Bockelmann} K.,  {G{\"u}ltekin} K.,  {Christensen}
  C.~R.,  {Governato} F.,  {Brooks} A.~M.,  {Loebman} S.,   {Munshi} F.,  2014,
  \mn@doi [\mnras] {10.1093/mnras/stu1958}, \href
  {http://adsabs.harvard.edu/abs/2014MNRAS.445.2667B} {445, 2667}

\bibitem[\protect\citeauthoryear{{Boizelle}, {Barth}, {Walsh}, {Buote},
  {Baker}, {Darling}  \& {Ho}}{{Boizelle} et~al.}{2019}]{Boizelle2019}
{Boizelle} B.~D.,  {Barth} A.~J.,  {Walsh} J.~L.,  {Buote} D.~A.,  {Baker}
  A.~J.,  {Darling} J.,   {Ho} L.~C.,  2019, \mn@doi [\apj]
  {10.3847/1538-4357/ab2a0a}, \href
  {https://ui.adsabs.harvard.edu/abs/2019ApJ...881...10B} {881, 10}

\bibitem[\protect\citeauthoryear{{Boizelle} et~al.,}{{Boizelle}
  et~al.}{2021}]{Boizelle2021}
{Boizelle} B.~D.,  et~al., 2021, \mn@doi [\apj] {10.3847/1538-4357/abd24d},
  \href {https://ui.adsabs.harvard.edu/abs/2021ApJ...908...19B} {908, 19}

\bibitem[\protect\citeauthoryear{{Calzetti}, {Armus}, {Bohlin}, {Kinney},
  {Koornneef}  \& {Storchi-Bergmann}}{{Calzetti} et~al.}{2000}]{Calzetti2000}
{Calzetti} D.,  {Armus} L.,  {Bohlin} R.~C.,  {Kinney} A.~L.,  {Koornneef} J.,
   {Storchi-Bergmann} T.,  2000, \mn@doi [\apj] {10.1086/308692}, \href
  {https://ui.adsabs.harvard.edu/abs/2000ApJ...533..682C} {533, 682}

\bibitem[\protect\citeauthoryear{{Cappellari}}{{Cappellari}}{2002}]{Cappellari2002}
{Cappellari} M.,  2002, \mn@doi [\mnras] {10.1046/j.1365-8711.2002.05412.x},
  \href {https://ui.adsabs.harvard.edu/abs/2002MNRAS.333..400C} {333, 400}

\bibitem[\protect\citeauthoryear{{Cappellari}}{{Cappellari}}{2008}]{Cappellari2008}
{Cappellari} M.,  2008, \mn@doi [\mnras] {10.1111/j.1365-2966.2008.13754.x},
  \href {http://adsabs.harvard.edu/abs/2008MNRAS.390...71C} {390, 71}

\bibitem[\protect\citeauthoryear{{Cappellari}}{{Cappellari}}{2017}]{Cappellari2017}
{Cappellari} M.,  2017, \mn@doi [\mnras] {10.1093/mnras/stw3020}, \href
  {http://adsabs.harvard.edu/abs/2017MNRAS.466..798C} {466, 798}

\bibitem[\protect\citeauthoryear{{Cappellari}}{{Cappellari}}{2020}]{Cappellari2020}
{Cappellari} M.,  2020, \mn@doi [\mnras] {10.1093/mnras/staa959}, \href
  {https://ui.adsabs.harvard.edu/abs/2020MNRAS.494.4819C} {494, 4819}

\bibitem[\protect\citeauthoryear{{Cappellari} \& {Copin}}{{Cappellari} \&
  {Copin}}{2003}]{Cappellari2003}
{Cappellari} M.,  {Copin} Y.,  2003, \mn@doi [\mnras]
  {10.1046/j.1365-8711.2003.06541.x}, \href
  {http://adsabs.harvard.edu/abs/2003MNRAS.342..345C} {342, 345}

\bibitem[\protect\citeauthoryear{{Cappellari} \& {Emsellem}}{{Cappellari} \&
  {Emsellem}}{2004}]{Cappellari2004}
{Cappellari} M.,  {Emsellem} E.,  2004, \mn@doi [\pasp] {10.1086/381875}, \href
  {http://adsabs.harvard.edu/abs/2004PASP..116..138C} {116, 138}

\bibitem[\protect\citeauthoryear{{Cappellari}, {Verolme}, {van der Marel},
  {Verdoes Kleijn}, {Illingworth}, {Franx}, {Carollo}  \& {de
  Zeeuw}}{{Cappellari} et~al.}{2002}]{Cappellari2002b}
{Cappellari} M.,  {Verolme} E.~K.,  {van der Marel} R.~P.,  {Verdoes Kleijn}
  G.~A.,  {Illingworth} G.~D.,  {Franx} M.,  {Carollo} C.~M.,   {de Zeeuw}
  P.~T.,  2002, \mn@doi [\apj] {10.1086/342653}, \href
  {http://adsabs.harvard.edu/abs/2002ApJ...578..787C} {578, 787}

\bibitem[\protect\citeauthoryear{{Cappellari} et~al.,}{{Cappellari}
  et~al.}{2006}]{Cappellari2006}
{Cappellari} M.,  et~al., 2006, \mn@doi [\mnras]
  {10.1111/j.1365-2966.2005.09981.x}, \href
  {http://adsabs.harvard.edu/abs/2006MNRAS.366.1126C} {366, 1126}

\bibitem[\protect\citeauthoryear{{Cappellari} et~al.,}{{Cappellari}
  et~al.}{2007}]{Cappellari2007}
{Cappellari} M.,  et~al., 2007, \mn@doi [\mnras]
  {10.1111/j.1365-2966.2007.11963.x}, \href
  {http://adsabs.harvard.edu/abs/2007MNRAS.379..418C} {379, 418}

\bibitem[\protect\citeauthoryear{{Cappellari} et~al.,}{{Cappellari}
  et~al.}{2009}]{Cappellari2009}
{Cappellari} M.,  et~al., 2009, \mn@doi [\apjl] {10.1088/0004-637X/704/1/L34},
  \href {http://adsabs.harvard.edu/abs/2009ApJ...704L..34C} {704, L34}

\bibitem[\protect\citeauthoryear{{Cappellari} et~al.,}{{Cappellari}
  et~al.}{2012}]{Cappellari2012}
{Cappellari} M.,  et~al., 2012, \mn@doi [\nat] {10.1038/nature10972}, \href
  {http://adsabs.harvard.edu/abs/2012Natur.484..485C} {484, 485}

\bibitem[\protect\citeauthoryear{{Cappellari} et~al.,}{{Cappellari}
  et~al.}{2013}]{Cappellari2013}
{Cappellari} M.,  et~al., 2013, \mn@doi [\mnras] {10.1093/mnras/stt562}, \href
  {http://adsabs.harvard.edu/abs/2013MNRAS.432.1709C} {432, 1709}

\bibitem[\protect\citeauthoryear{{Cohn} et~al.,}{{Cohn}
  et~al.}{2021}]{Cohn2021}
{Cohn} J.~H.,  et~al., 2021, \mn@doi [\apj] {10.3847/1538-4357/ac0f78}, \href
  {https://ui.adsabs.harvard.edu/abs/2021ApJ...919...77C} {919, 77}

\bibitem[\protect\citeauthoryear{{Crocker}, {Bureau}, {Young}  \&
  {Combes}}{{Crocker} et~al.}{2011}]{Crocker2011}
{Crocker} A.~F.,  {Bureau} M.,  {Young} L.~M.,   {Combes} F.,  2011, \mn@doi
  [\mnras] {10.1111/j.1365-2966.2010.17537.x}, \href
  {https://ui.adsabs.harvard.edu/abs/2011MNRAS.410.1197C} {410, 1197}

\bibitem[\protect\citeauthoryear{{Davies}, {M{\"u}ller S{\'a}nchez}, {Genzel},
  {Tacconi}, {Hicks}, {Friedrich}  \& {Sternberg}}{{Davies}
  et~al.}{2007}]{Davies2007a}
{Davies} R.~I.,  {M{\"u}ller S{\'a}nchez} F.,  {Genzel} R.,  {Tacconi} L.~J.,
  {Hicks} E.~K.~S.,  {Friedrich} S.,   {Sternberg} A.,  2007, \mn@doi [\apj]
  {10.1086/523032}, \href
  {https://ui.adsabs.harvard.edu/abs/2007ApJ...671.1388D} {671, 1388}

\bibitem[\protect\citeauthoryear{{Davis} \& {McDermid}}{{Davis} \&
  {McDermid}}{2017}]{Davis2017b}
{Davis} T.~A.,  {McDermid} R.~M.,  2017, \mn@doi [\mnras]
  {10.1093/mnras/stw2366}, \href
  {https://ui.adsabs.harvard.edu/abs/2017MNRAS.464..453D} {464, 453}

\bibitem[\protect\citeauthoryear{{Davis}, {Bureau}, {Cappellari}, {Sarzi}  \&
  {Blitz}}{{Davis} et~al.}{2013}]{Davis2013}
{Davis} T.~A.,  {Bureau} M.,  {Cappellari} M.,  {Sarzi} M.,   {Blitz} L.,
  2013, \mn@doi [\nat] {10.1038/nature11819}, \href
  {http://adsabs.harvard.edu/abs/2013Natur.494..328D} {494, 328}

\bibitem[\protect\citeauthoryear{{Davis}, {Bureau}, {Onishi}, {Cappellari},
  {Iguchi}  \& {Sarzi}}{{Davis} et~al.}{2017}]{Davis2017a}
{Davis} T.~A.,  {Bureau} M.,  {Onishi} K.,  {Cappellari} M.,  {Iguchi} S.,
  {Sarzi} M.,  2017, \mn@doi [\mnras] {10.1093/mnras/stw3217}, \href
  {http://ads.nao.ac.jp/abs/2017MNRAS.468.4675D} {468, 4675}

\bibitem[\protect\citeauthoryear{{Davis} et~al.,}{{Davis}
  et~al.}{2018}]{Davis2018}
{Davis} T.~A.,  et~al., 2018, \mn@doi [\mnras] {10.1093/mnras/stx2600}, \href
  {http://adsabs.harvard.edu/abs/2018MNRAS.473.3818D} {473, 3818}

\bibitem[\protect\citeauthoryear{{Davis} et~al.,}{{Davis}
  et~al.}{2020}]{Davis2020}
{Davis} T.~A.,  et~al., 2020, \mn@doi [\mnras] {10.1093/mnras/staa1567}, \href
  {https://ui.adsabs.harvard.edu/abs/2020MNRAS.496.4061D} {496, 4061}

\bibitem[\protect\citeauthoryear{{Drehmer}, {Storchi-Bergmann}, {Ferrari},
  {Cappellari}  \& {Riffel}}{{Drehmer} et~al.}{2015}]{Drehmer2015}
{Drehmer} D.~A.,  {Storchi-Bergmann} T.,  {Ferrari} F.,  {Cappellari} M.,
  {Riffel} R.~A.,  2015, \mn@doi [\mnras] {10.1093/mnras/stv536}, \href
  {http://adsabs.harvard.edu/abs/2015MNRAS.450..128D} {450, 128}

\bibitem[\protect\citeauthoryear{{Dutton} \& {Macci{\`o}}}{{Dutton} \&
  {Macci{\`o}}}{2014}]{Dutton2014}
{Dutton} A.~A.,  {Macci{\`o}} A.~V.,  2014, \mn@doi [\mnras]
  {10.1093/mnras/stu742}, \href
  {https://ui.adsabs.harvard.edu/abs/2014MNRAS.441.3359D} {441, 3359}

\bibitem[\protect\citeauthoryear{{Falc{\'o}n-Barroso},
  {S{\'a}nchez-Bl{\'a}zquez}, {Vazdekis}, {Ricciardelli}, {Cardiel}, {Cenarro},
  {Gorgas}  \& {Peletier}}{{Falc{\'o}n-Barroso}
  et~al.}{2011}]{Falcon-Barroso2011}
{Falc{\'o}n-Barroso} J.,  {S{\'a}nchez-Bl{\'a}zquez} P.,  {Vazdekis} A.,
  {Ricciardelli} E.,  {Cardiel} N.,  {Cenarro} A.~J.,  {Gorgas} J.,
  {Peletier} R.~F.,  2011, \mn@doi [\aap] {10.1051/0004-6361/201116842}, \href
  {http://adsabs.harvard.edu/abs/2011A%26A...532A..95F} {532, A95}

\bibitem[\protect\citeauthoryear{{Feldmeier-Krause}, {Zhu}, {Neumayer}, {van de
  Ven}, {de Zeeuw}  \& {Sch{\"o}del}}{{Feldmeier-Krause}
  et~al.}{2017}]{FeldmeierKrause2017}
{Feldmeier-Krause} A.,  {Zhu} L.,  {Neumayer} N.,  {van de Ven} G.,  {de Zeeuw}
  P.~T.,   {Sch{\"o}del} R.,  2017, \mn@doi [\mnras] {10.1093/mnras/stw3377},
  \href {https://ui.adsabs.harvard.edu/abs/2017MNRAS.466.4040F} {466, 4040}

\bibitem[\protect\citeauthoryear{{Feldmeier} et~al.,}{{Feldmeier}
  et~al.}{2014}]{Feldmeier2014}
{Feldmeier} A.,  et~al., 2014, \mn@doi [\aap] {10.1051/0004-6361/201423777},
  \href {https://ui.adsabs.harvard.edu/abs/2014A&A...570A...2F} {570, A2}

\bibitem[\protect\citeauthoryear{{Ferrarese}, {Ford}  \& {Jaffe}}{{Ferrarese}
  et~al.}{1996}]{Ferrarese1996}
{Ferrarese} L.,  {Ford} H.~C.,   {Jaffe} W.,  1996, \mn@doi [\apj]
  {10.1086/177876}, \href {http://adsabs.harvard.edu/abs/1996ApJ...470..444F}
  {470, 444}

\bibitem[\protect\citeauthoryear{{Foreman-Mackey}, {Hogg}, {Lang}  \&
  {Goodman}}{{Foreman-Mackey} et~al.}{2013}]{Foreman-Mackey2013}
{Foreman-Mackey} D.,  {Hogg} D.~W.,  {Lang} D.,   {Goodman} J.,  2013, \mn@doi
  [\pasp] {10.1086/670067}, \href
  {http://adsabs.harvard.edu/abs/2013PASP..125..306F} {125, 306}

\bibitem[\protect\citeauthoryear{{Fusco} et~al.,}{{Fusco}
  et~al.}{2020}]{Fusco2020}
{Fusco} T.,  et~al., 2020, \mn@doi [\aap] {10.1051/0004-6361/202037595}, \href
  {https://ui.adsabs.harvard.edu/abs/2020A&A...635A.208F} {635, A208}

\bibitem[\protect\citeauthoryear{{Gao} \& {Ho}}{{Gao} \& {Ho}}{2017}]{Gao2017}
{Gao} H.,  {Ho} L.~C.,  2017, \mn@doi [\apj] {10.3847/1538-4357/aa7da4}, \href
  {http://adsabs.harvard.edu/abs/2017ApJ...845..114G} {845, 114}

\bibitem[\protect\citeauthoryear{{Gebhardt} \& {Thomas}}{{Gebhardt} \&
  {Thomas}}{2009}]{Gebhardt2009}
{Gebhardt} K.,  {Thomas} J.,  2009, \mn@doi [\apj]
  {10.1088/0004-637X/700/2/1690}, \href
  {http://adsabs.harvard.edu/abs/2009ApJ...700.1690G} {700, 1690}

\bibitem[\protect\citeauthoryear{{Gebhardt}, {Adams}, {Richstone}, {Lauer},
  {Faber}, {G{\"u}ltekin}, {Murphy}  \& {Tremaine}}{{Gebhardt}
  et~al.}{2011}]{Gebhardt2011}
{Gebhardt} K.,  {Adams} J.,  {Richstone} D.,  {Lauer} T.~R.,  {Faber} S.~M.,
  {G{\"u}ltekin} K.,  {Murphy} J.,   {Tremaine} S.,  2011, \mn@doi [\apj]
  {10.1088/0004-637X/729/2/119}, \href
  {http://adsabs.harvard.edu/abs/2011ApJ...729..119G} {729, 119}

\bibitem[\protect\citeauthoryear{{Gerhard} \& {Binney}}{{Gerhard} \&
  {Binney}}{1996}]{Gerhard1996}
{Gerhard} O.~E.,  {Binney} J.~J.,  1996, \mn@doi [\mnras]
  {10.1093/mnras/279.3.993}, \href
  {http://adsabs.harvard.edu/abs/1996MNRAS.279..993G} {279, 993}

\bibitem[\protect\citeauthoryear{{Ghez} et~al.,}{{Ghez}
  et~al.}{2008}]{Ghez2008}
{Ghez} A.~M.,  et~al., 2008, \mn@doi [\apj] {10.1086/592738}, \href
  {http://adsabs.harvard.edu/abs/2008ApJ...689.1044G} {689, 1044}

\bibitem[\protect\citeauthoryear{{Gillessen}, {Eisenhauer}, {Trippe},
  {Alexander}, {Genzel}, {Martins}  \& {Ott}}{{Gillessen}
  et~al.}{2009}]{Gillessen2009}
{Gillessen} S.,  {Eisenhauer} F.,  {Trippe} S.,  {Alexander} T.,  {Genzel} R.,
  {Martins} F.,   {Ott} T.,  2009, \mn@doi [\apj]
  {10.1088/0004-637X/692/2/1075}, \href
  {http://adsabs.harvard.edu/abs/2009ApJ...692.1075G} {692, 1075}

\bibitem[\protect\citeauthoryear{{Gillessen} et~al.,}{{Gillessen}
  et~al.}{2017}]{Gillessen2017}
{Gillessen} S.,  et~al., 2017, \mn@doi [\apj] {10.3847/1538-4357/aa5c41}, \href
  {http://adsabs.harvard.edu/abs/2017ApJ...837...30G} {837, 30}

\bibitem[\protect\citeauthoryear{{Girardi}, {Bressan}, {Bertelli}  \&
  {Chiosi}}{{Girardi} et~al.}{2000}]{Girardi2000}
{Girardi} L.,  {Bressan} A.,  {Bertelli} G.,   {Chiosi} C.,  2000, \mn@doi
  [\aaps] {10.1051/aas:2000126}, \href
  {https://ui.adsabs.harvard.edu/abs/2000A&AS..141..371G} {141, 371}

\bibitem[\protect\citeauthoryear{{Greene} et~al.,}{{Greene}
  et~al.}{2016}]{Greene2016}
{Greene} J.~E.,  et~al., 2016, \mn@doi [\apjl] {10.3847/2041-8205/826/2/L32},
  \href {http://adsabs.harvard.edu/abs/2016ApJ...826L..32G} {826, L32}

\bibitem[\protect\citeauthoryear{{Gu{\'e}rou} et~al.,}{{Gu{\'e}rou}
  et~al.}{2017}]{Guerou2017}
{Gu{\'e}rou} A.,  et~al., 2017, \mn@doi [\aap] {10.1051/0004-6361/201730905},
  \href {http://adsabs.harvard.edu/abs/2017A%26A...608A...5G} {608, A5}

\bibitem[\protect\citeauthoryear{Ho, Li, Barth, Seigar  \& Peng}{Ho
  et~al.}{2011}]{Ho2011}
Ho L.~C.,  Li Z.-Y.,  Barth A.~J.,  Seigar M.~S.,   Peng C.~Y.,  2011, \mn@doi
  [The Astrophysical Journal Supplement Series] {10.1088/0067-0049/197/2/21},
  197, 21

\bibitem[\protect\citeauthoryear{{Holtzman}, {Burrows}, {Casertano}, {Hester},
  {Trauger}, {Watson}  \& {Worthey}}{{Holtzman} et~al.}{1995}]{Holtzman1995}
{Holtzman} J.~A.,  {Burrows} C.~J.,  {Casertano} S.,  {Hester} J.~J.,
  {Trauger} J.~T.,  {Watson} A.~M.,   {Worthey} G.,  1995, \mn@doi [\pasp]
  {10.1086/133664}, \href {http://adsabs.harvard.edu/abs/1995PASP..107.1065H}
  {107, 1065}

\bibitem[\protect\citeauthoryear{Huang, Ho, Peng, Li  \& Barth}{Huang
  et~al.}{2013}]{Huang2013}
Huang S.,  Ho L.~C.,  Peng C.~Y.,  Li Z.-Y.,   Barth A.~J.,  2013, \mn@doi [The
  Astrophysical Journal] {10.1088/0004-637x/766/1/47}, 766, 47

\bibitem[\protect\citeauthoryear{{Husser}, {Wende-von Berg}, {Dreizler},
  {Homeier}, {Reiners}, {Barman}  \& {Hauschildt}}{{Husser}
  et~al.}{2013}]{Husser2013}
{Husser} T.~O.,  {Wende-von Berg} S.,  {Dreizler} S.,  {Homeier} D.,  {Reiners}
  A.,  {Barman} T.,   {Hauschildt} P.~H.,  2013, \mn@doi [\aap]
  {10.1051/0004-6361/201219058}, \href
  {https://ui.adsabs.harvard.edu/abs/2013A&A...553A...6H} {553, A6}

\bibitem[\protect\citeauthoryear{{Knapen}, {Comer{\'o}n}  \& {Seidel}}{{Knapen}
  et~al.}{2019}]{Knapen2019}
{Knapen} J.~H.,  {Comer{\'o}n} S.,   {Seidel} M.~K.,  2019, \mn@doi [\aap]
  {10.1051/0004-6361/201834669}, \href
  {https://ui.adsabs.harvard.edu/abs/2019A&A...621L...5K} {621, L5}

\bibitem[\protect\citeauthoryear{Kormendy \& Ho}{Kormendy \&
  Ho}{2013}]{Kormendy2013}
Kormendy J.,  Ho L.~C.,  2013, \mn@doi [Annu. Rev. Astro. Astrophys.]
  {10.1146/annurev-astro-082708-101811}, 51, 511–653

\bibitem[\protect\citeauthoryear{{Krajnovi{\'c}}, {Cappellari}, {Emsellem},
  {McDermid}  \& {de Zeeuw}}{{Krajnovi{\'c}} et~al.}{2005}]{Krajnovic2005}
{Krajnovi{\'c}} D.,  {Cappellari} M.,  {Emsellem} E.,  {McDermid} R.~M.,   {de
  Zeeuw} P.~T.,  2005, \mn@doi [\mnras] {10.1111/j.1365-2966.2005.08715.x},
  \href {http://adsabs.harvard.edu/abs/2005MNRAS.357.1113K} {357, 1113}

\bibitem[\protect\citeauthoryear{{Krajnovi{\'c}}, {Cappellari}, {de Zeeuw}  \&
  {Copin}}{{Krajnovi{\'c}} et~al.}{2006}]{Krajnovic2006}
{Krajnovi{\'c}} D.,  {Cappellari} M.,  {de Zeeuw} P.~T.,   {Copin} Y.,  2006,
  \mn@doi [\mnras] {10.1111/j.1365-2966.2005.09902.x}, \href
  {http://adsabs.harvard.edu/abs/2006MNRAS.366..787K} {366, 787}

\bibitem[\protect\citeauthoryear{{Krajnovi{\'c}}, {McDermid}, {Cappellari}  \&
  {Davies}}{{Krajnovi{\'c}} et~al.}{2009}]{Krajnovic2009}
{Krajnovi{\'c}} D.,  {McDermid} R.~M.,  {Cappellari} M.,   {Davies} R.~L.,
  2009, \mn@doi [\mnras] {10.1111/j.1365-2966.2009.15415.x}, \href
  {http://adsabs.harvard.edu/abs/2009MNRAS.399.1839K} {399, 1839}

\bibitem[\protect\citeauthoryear{{Krajnovi{\'c}} et~al.,}{{Krajnovi{\'c}}
  et~al.}{2018}]{Krajnovic2018}
{Krajnovi{\'c}} D.,  et~al., 2018, \mn@doi [\mnras] {10.1093/mnras/sty778},
  \href {http://adsabs.harvard.edu/abs/2018MNRAS.477.3030K} {477, 3030}

\bibitem[\protect\citeauthoryear{{Krist} \& {Hook}}{{Krist} \&
  {Hook}}{2001}]{Krist2001}
{Krist} J.,  {Hook} R.,  2001, {The Tiny Tim User's Manual, version 6.3}

\bibitem[\protect\citeauthoryear{{Kroupa}}{{Kroupa}}{2001}]{Kroupa2001}
{Kroupa} P.,  2001, \mn@doi [\mnras] {10.1046/j.1365-8711.2001.04022.x}, \href
  {http://adsabs.harvard.edu/abs/2001MNRAS.322..231K} {322, 231}

\bibitem[\protect\citeauthoryear{{Kuo} et~al.,}{{Kuo} et~al.}{2011}]{Kuo2011}
{Kuo} C.~Y.,  et~al., 2011, \mn@doi [\apj] {10.1088/0004-637X/727/1/20}, \href
  {http://adsabs.harvard.edu/abs/2011ApJ...727...20K} {727, 20}

\bibitem[\protect\citeauthoryear{{Lablanche} et~al.,}{{Lablanche}
  et~al.}{2012}]{Lablanche2012}
{Lablanche} P.-Y.,  et~al., 2012, \mn@doi [\mnras]
  {10.1111/j.1365-2966.2012.21343.x}, \href
  {http://adsabs.harvard.edu/abs/2012MNRAS.424.1495L} {424, 1495}

\bibitem[\protect\citeauthoryear{{Laurikainen}, {Salo}, {Buta}, {Knapen}  \&
  {Comer{\'o}n}}{{Laurikainen} et~al.}{2010}]{Laurikainen2010}
{Laurikainen} E.,  {Salo} H.,  {Buta} R.,  {Knapen} J.~H.,   {Comer{\'o}n} S.,
  2010, \mn@doi [\mnras] {10.1111/j.1365-2966.2010.16521.x}, \href
  {http://adsabs.harvard.edu/abs/2010MNRAS.405.1089L} {405, 1089}

\bibitem[\protect\citeauthoryear{{Lelli}, {McGaugh}, {Schombert}  \&
  {Pawlowski}}{{Lelli} et~al.}{2017}]{Lelli2017}
{Lelli} F.,  {McGaugh} S.~S.,  {Schombert} J.~M.,   {Pawlowski} M.~S.,  2017,
  \mn@doi [\apj] {10.3847/1538-4357/836/2/152}, \href
  {http://adsabs.harvard.edu/abs/2017ApJ...836..152L} {836, 152}

\bibitem[\protect\citeauthoryear{Li, Ho, Barth  \& Peng}{Li
  et~al.}{2011}]{Li2011}
Li Z.-Y.,  Ho L.~C.,  Barth A.~J.,   Peng C.~Y.,  2011, \mn@doi [The
  Astrophysical Journal Supplement Series] {10.1088/0067-0049/197/2/22}, 197,
  22

\bibitem[\protect\citeauthoryear{{Li}, {Sellwood}  \& {Shen}}{{Li}
  et~al.}{2017}]{Li2017}
{Li} Z.,  {Sellwood} J.~A.,   {Shen} J.,  2017, \mn@doi [\apj]
  {10.3847/1538-4357/aa9377}, \href
  {http://adsabs.harvard.edu/abs/2017ApJ...850...67L} {850, 67}

\bibitem[\protect\citeauthoryear{{Li} et~al.,}{{Li} et~al.}{2018}]{Li2018b}
{Li} H.,  et~al., 2018, \mn@doi [\mnras] {10.1093/mnras/sty334}, \href
  {https://ui.adsabs.harvard.edu/abs/2018MNRAS.476.1765L} {476, 1765}

\bibitem[\protect\citeauthoryear{{Lipka} \& {Thomas}}{{Lipka} \&
  {Thomas}}{2021}]{Lipka2021}
{Lipka} M.,  {Thomas} J.,  2021, \mn@doi [\mnras] {10.1093/mnras/stab1092},
  \href {https://ui.adsabs.harvard.edu/abs/2021MNRAS.504.4599L} {504, 4599}

\bibitem[\protect\citeauthoryear{{Lyubenova} et~al.,}{{Lyubenova}
  et~al.}{2016}]{Lyubenova2016}
{Lyubenova} M.,  et~al., 2016, \mn@doi [\mnras] {10.1093/mnras/stw2434}, \href
  {https://ui.adsabs.harvard.edu/abs/2016MNRAS.463.3220L} {463, 3220}

\bibitem[\protect\citeauthoryear{{Madore}, {Freedman}  \& {Bothun}}{{Madore}
  et~al.}{2004}]{Madore2004}
{Madore} B.~F.,  {Freedman} W.~L.,   {Bothun} G.~D.,  2004, \mn@doi [\apj]
  {10.1086/383486}, \href
  {https://ui.adsabs.harvard.edu/abs/2004ApJ...607..810M} {607, 810}

\bibitem[\protect\citeauthoryear{{Malin} \& {Carter}}{{Malin} \&
  {Carter}}{1983}]{Malin1983}
{Malin} D.~F.,  {Carter} D.,  1983, \mn@doi [\apj] {10.1086/161467}, \href
  {https://ui.adsabs.harvard.edu/abs/1983ApJ...274..534M} {274, 534}

\bibitem[\protect\citeauthoryear{{McConnell}, {Chen}, {Ma}, {Greene}, {Lauer}
  \& {Gebhardt}}{{McConnell} et~al.}{2013}]{McConnell2013b}
{McConnell} N.~J.,  {Chen} S.-F.~S.,  {Ma} C.-P.,  {Greene} J.~E.,  {Lauer}
  T.~R.,   {Gebhardt} K.,  2013, \mn@doi [\apjl] {10.1088/2041-8205/768/1/L21},
  \href {http://adsabs.harvard.edu/abs/2013ApJ...768L..21M} {768, L21}

\bibitem[\protect\citeauthoryear{{McDermid} et~al.,}{{McDermid}
  et~al.}{2006}]{McDermid2006}
{McDermid} R.~M.,  et~al., 2006, \mn@doi [\mnras]
  {10.1111/j.1365-2966.2006.11065.x}, \href
  {http://adsabs.harvard.edu/abs/2006MNRAS.373..906M} {373, 906}

\bibitem[\protect\citeauthoryear{{McGaugh}, {Lelli}  \& {Schombert}}{{McGaugh}
  et~al.}{2016}]{McGaugh2016}
{McGaugh} S.~S.,  {Lelli} F.,   {Schombert} J.~M.,  2016, \mn@doi [Physical
  Review Letters] {10.1103/PhysRevLett.117.201101}, \href
  {http://adsabs.harvard.edu/abs/2016PhRvL.117t1101M} {117, 201101}

\bibitem[\protect\citeauthoryear{{Mehrgan}, {Thomas}, {Saglia}, {Mazzalay},
  {Erwin}, {Bender}, {Kluge}  \& {Fabricius}}{{Mehrgan}
  et~al.}{2019}]{Mehrgan2019}
{Mehrgan} K.,  {Thomas} J.,  {Saglia} R.,  {Mazzalay} X.,  {Erwin} P.,
  {Bender} R.,  {Kluge} M.,   {Fabricius} M.,  2019, \mn@doi [\apj]
  {10.3847/1538-4357/ab5856}, \href
  {https://ui.adsabs.harvard.edu/abs/2019ApJ...887..195M} {887, 195}

\bibitem[\protect\citeauthoryear{{Mitzkus}, {Cappellari}  \&
  {Walcher}}{{Mitzkus} et~al.}{2017}]{Mitzkus2017}
{Mitzkus} M.,  {Cappellari} M.,   {Walcher} C.~J.,  2017, \mn@doi [\mnras]
  {10.1093/mnras/stw2677}, \href
  {https://ui.adsabs.harvard.edu/abs/2017MNRAS.464.4789M} {464, 4789}

\bibitem[\protect\citeauthoryear{Miyoshi, Moran, Herrnstein, Greenhill, Nakai,
  Diamond  \& Inoue}{Miyoshi et~al.}{1995}]{Miyoshi1995}
Miyoshi M.,  Moran J.,  Herrnstein J.,  Greenhill L.,  Nakai N.,  Diamond P.,
  Inoue M.,  1995, \mn@doi [Nature] {10.1038/373127a0}, 373, 127–129

\bibitem[\protect\citeauthoryear{{Moffat}}{{Moffat}}{1969}]{Moffat1969}
{Moffat} A.~F.~J.,  1969, \aap, \href
  {https://ui.adsabs.harvard.edu/abs/1969A&A.....3..455M} {3, 455}

\bibitem[\protect\citeauthoryear{Moster, Naab  \& White}{Moster
  et~al.}{2013}]{Moster2013}
Moster B.~P.,  Naab T.,   White S. D.~M.,  2013, \mn@doi [\mnras]
  {10.1093/mnras/sts261}, 428, 3121

\bibitem[\protect\citeauthoryear{{Navarro}, {Frenk}  \& {White}}{{Navarro}
  et~al.}{1996}]{Navarro1996}
{Navarro} J.~F.,  {Frenk} C.~S.,   {White} S.~D.~M.,  1996, \mn@doi [\apj]
  {10.1086/177173}, \href {http://adsabs.harvard.edu/abs/1996ApJ...462..563N}
  {462, 563}

\bibitem[\protect\citeauthoryear{Neumayer, Cappellari, Reunanen, Rix, van~der
  Werf, de Zeeuw  \& Davies}{Neumayer et~al.}{2007}]{Neumayer2007}
Neumayer N.,  Cappellari M.,  Reunanen J.,  Rix H.,  van~der Werf P.~P.,  de
  Zeeuw P.~T.,   Davies R.~I.,  2007, \mn@doi [The Astrophysical Journal]
  {10.1086/523039}, 671, 1329–1344

\bibitem[\protect\citeauthoryear{{Nguyen} et~al.,}{{Nguyen}
  et~al.}{2017}]{Nguyen2017a}
{Nguyen} D.~D.,  et~al., 2017, \mn@doi [\apj] {10.3847/1538-4357/aa5cb4}, \href
  {http://adsabs.harvard.edu/abs/2017ApJ...836..237N} {836, 237}

\bibitem[\protect\citeauthoryear{{Nguyen} et~al.,}{{Nguyen}
  et~al.}{2018}]{Nguyen2018}
{Nguyen} D.~D.,  et~al., 2018, \mn@doi [\apj] {10.3847/1538-4357/aabe28}, \href
  {http://adsabs.harvard.edu/abs/2018ApJ...858..118N} {858, 118}

\bibitem[\protect\citeauthoryear{{Nguyen} et~al.,}{{Nguyen}
  et~al.}{2019}]{Nguyen2019a}
{Nguyen} D.~D.,  et~al., 2019, \mn@doi [\apj] {10.3847/1538-4357/aafe7a}, \href
  {http://adsabs.harvard.edu/abs/2019ApJ...872..104N} {872, 104}

\bibitem[\protect\citeauthoryear{{Nguyen} et~al.,}{{Nguyen}
  et~al.}{2020}]{Nguyen2020}
{Nguyen} D.~D.,  et~al., 2020, \mn@doi [\apj] {10.3847/1538-4357/ab77aa}, \href
  {https://ui.adsabs.harvard.edu/abs/2020ApJ...892...68N} {892, 68}

\bibitem[\protect\citeauthoryear{{Nguyen} et~al.,}{{Nguyen}
  et~al.}{2021a}]{Nguyen2021b}
{Nguyen} D.~D.,  et~al., 2021a, \mn@doi [\mnras] {10.1093/mnras/stab3016},
  \href {https://ui.adsabs.harvard.edu/abs/2021MNRAS.tmp.2741N} {}

\bibitem[\protect\citeauthoryear{{Nguyen} et~al.,}{{Nguyen}
  et~al.}{2021b}]{Nguyen2021}
{Nguyen} D.~D.,  et~al., 2021b, \mn@doi [\mnras] {10.1093/mnras/stab1002},
  \href {https://ui.adsabs.harvard.edu/abs/2021MNRAS.504.4123N} {504, 4123}

\bibitem[\protect\citeauthoryear{{Onishi}, {Iguchi}, {Sheth}  \&
  {Kohno}}{{Onishi} et~al.}{2015}]{Onishi2015}
{Onishi} K.,  {Iguchi} S.,  {Sheth} K.,   {Kohno} K.,  2015, \mn@doi [\apj]
  {10.1088/0004-637X/806/1/39}, \href
  {https://ui.adsabs.harvard.edu/abs/2015ApJ...806...39O} {806, 39}

\bibitem[\protect\citeauthoryear{{Onishi}, {Iguchi}, {Davis}, {Bureau},
  {Cappellari}, {Sarzi}  \& {Blitz}}{{Onishi} et~al.}{2017}]{Onishi2017}
{Onishi} K.,  {Iguchi} S.,  {Davis} T.~A.,  {Bureau} M.,  {Cappellari} M.,
  {Sarzi} M.,   {Blitz} L.,  2017, \mn@doi [\mnras] {10.1093/mnras/stx631},
  \href {http://ads.nao.ac.jp/abs/2017MNRAS.468.4663O} {468, 4663}

\bibitem[\protect\citeauthoryear{{Onken} et~al.,}{{Onken}
  et~al.}{2014}]{Onken2014}
{Onken} C.~A.,  et~al., 2014, \mn@doi [\apj] {10.1088/0004-637X/791/1/37},
  \href {http://adsabs.harvard.edu/abs/2014ApJ...791...37O} {791, 37}

\bibitem[\protect\citeauthoryear{{Pastorini} et~al.,}{{Pastorini}
  et~al.}{2007}]{Pastorini2007}
{Pastorini} G.,  et~al., 2007, \mn@doi [\aap] {10.1051/0004-6361:20066784},
  \href {http://adsabs.harvard.edu/abs/2007A%26A...469..405P} {469, 405}

\bibitem[\protect\citeauthoryear{{Rusli} et~al.,}{{Rusli}
  et~al.}{2013}]{Rusli2013}
{Rusli} S.~P.,  et~al., 2013, \mn@doi [\aj] {10.1088/0004-6256/146/3/45}, \href
  {http://adsabs.harvard.edu/abs/2013AJ....146...45R} {146, 45}

\bibitem[\protect\citeauthoryear{{Rybicki}}{{Rybicki}}{1987}]{Rybicki1987}
{Rybicki} G.~B.,  1987, in {de Zeeuw} P.~T.,  ed.,  IAU Symposium Vol. 127,
  Structure and Dynamics of Elliptical Galaxies. p.~397

\bibitem[\protect\citeauthoryear{{Saglia} et~al.,}{{Saglia}
  et~al.}{2016}]{Saglia2016}
{Saglia} R.~P.,  et~al., 2016, \mn@doi [\apj] {10.3847/0004-637X/818/1/47},
  \href {http://adsabs.harvard.edu/abs/2016ApJ...818...47S} {818, 47}

\bibitem[\protect\citeauthoryear{{Sahu}, {Graham}  \& {Davis}}{{Sahu}
  et~al.}{2019}]{Sahu2019b}
{Sahu} N.,  {Graham} A.~W.,   {Davis} B.~L.,  2019, \mn@doi [\apj]
  {10.3847/1538-4357/ab50b7}, \href
  {https://ui.adsabs.harvard.edu/abs/2019ApJ...887...10S} {887, 10}

\bibitem[\protect\citeauthoryear{{Salpeter}}{{Salpeter}}{1955}]{Salpeter1955}
{Salpeter} E.~E.,  1955, \mn@doi [\apj] {10.1086/145971}, \href
  {http://adsabs.harvard.edu/abs/1955ApJ...121..161S} {121, 161}

\bibitem[\protect\citeauthoryear{{S{\'a}nchez-Bl{\'a}zquez}
  et~al.,}{{S{\'a}nchez-Bl{\'a}zquez} et~al.}{2006}]{Sanchez-Blazquez2006}
{S{\'a}nchez-Bl{\'a}zquez} P.,  et~al., 2006, \mn@doi [\mnras]
  {10.1111/j.1365-2966.2006.10699.x}, \href
  {http://adsabs.harvard.edu/abs/2006MNRAS.371..703S} {371, 703}

\bibitem[\protect\citeauthoryear{{Sandage} \& {Bedke}}{{Sandage} \&
  {Bedke}}{1994}]{Sandage1994}
{Sandage} A.,  {Bedke} J.,  1994, {The Carnegie Atlas of Galaxies. Volumes I,
  II.}

\bibitem[\protect\citeauthoryear{{Saraiva}, {Ferrari}  \&
  {Pastoriza}}{{Saraiva} et~al.}{1999}]{Saraiva1999}
{Saraiva} M.~F.,  {Ferrari} F.,   {Pastoriza} M.~G.,  1999, \aap, \href
  {http://adsabs.harvard.edu/abs/1999A%26A...350..399S} {350, 399}

\bibitem[\protect\citeauthoryear{{Saraiva}, {Bica}, {Pastoriza}  \&
  {Bonatto}}{{Saraiva} et~al.}{2001}]{Saraiva2001}
{Saraiva} M.~F.,  {Bica} E.,  {Pastoriza} M.~G.,   {Bonatto} C.,  2001, \mn@doi
  [\aap] {10.1051/0004-6361:20010905}, \href
  {http://adsabs.harvard.edu/abs/2001A%26A...376...43S} {376, 43}

\bibitem[\protect\citeauthoryear{{Sarzi} et~al.,}{{Sarzi}
  et~al.}{2006}]{Sarzi2006}
{Sarzi} M.,  et~al., 2006, \mn@doi [\mnras] {10.1111/j.1365-2966.2005.09839.x},
  \href {http://adsabs.harvard.edu/abs/2006MNRAS.366.1151S} {366, 1151}

\bibitem[\protect\citeauthoryear{{Schlafly} \& {Finkbeiner}}{{Schlafly} \&
  {Finkbeiner}}{2011}]{Schlafly2011}
{Schlafly} E.~F.,  {Finkbeiner} D.~P.,  2011, \mn@doi [\apj]
  {10.1088/0004-637X/737/2/103}, \href
  {http://ads.nao.ac.jp/abs/2011ApJ...737..103S} {737, 103}

\bibitem[\protect\citeauthoryear{{Schulze} \& {Wisotzki}}{{Schulze} \&
  {Wisotzki}}{2011}]{Schulze2011}
{Schulze} A.,  {Wisotzki} L.,  2011, \mn@doi [\aap]
  {10.1051/0004-6361/201117564}, \href
  {http://adsabs.harvard.edu/abs/2011A%26A...535A..87S} {535, A87}

\bibitem[\protect\citeauthoryear{{Schwarzschild}}{{Schwarzschild}}{1979}]{Schwarzschild1979}
{Schwarzschild} M.,  1979, \mn@doi [\apj] {10.1086/157282}, \href
  {http://adsabs.harvard.edu/abs/1979ApJ...232..236S} {232, 236}

\bibitem[\protect\citeauthoryear{{Shapiro}, {Cappellari}, {de Zeeuw},
  {McDermid}, {Gebhardt}, {van den Bosch}  \& {Statler}}{{Shapiro}
  et~al.}{2006}]{Shapiro2006}
{Shapiro} K.~L.,  {Cappellari} M.,  {de Zeeuw} T.,  {McDermid} R.~M.,
  {Gebhardt} K.,  {van den Bosch} R.~C.~E.,   {Statler} T.~S.,  2006, \mn@doi
  [\mnras] {10.1111/j.1365-2966.2006.10537.x}, \href
  {http://adsabs.harvard.edu/abs/2006MNRAS.370..559S} {370, 559}

\bibitem[\protect\citeauthoryear{Smith, Evans  \& An}{Smith
  et~al.}{2009}]{Smith2009}
Smith M.~C.,  Evans N.~W.,   An J.,  2009, \mn@doi
  [Astrophys.J.698:1110-1116,2009] {10.1088/0004-637X/698/2/1110}

\bibitem[\protect\citeauthoryear{Ströbele et~al.,}{Ströbele
  et~al.}{2012}]{Stroebele2012}
Ströbele S.,  et~al., 2012, GALACSI system design and analysis,
  \mn@doi{10.1117/12.926110}

\bibitem[\protect\citeauthoryear{{Tal}, {van Dokkum}, {Nelan}  \&
  {Bezanson}}{{Tal} et~al.}{2009}]{Tal2009}
{Tal} T.,  {van Dokkum} P.~G.,  {Nelan} J.,   {Bezanson} R.,  2009, \mn@doi
  [\aj] {10.1088/0004-6256/138/5/1417}, \href
  {http://adsabs.harvard.edu/abs/2009AJ....138.1417T} {138, 1417}

\bibitem[\protect\citeauthoryear{{Thater} et~al.,}{{Thater}
  et~al.}{2017}]{Thater2017}
{Thater} S.,  et~al., 2017, \mn@doi [\aap] {10.1051/0004-6361/201629480}, \href
  {http://adsabs.harvard.edu/abs/2017A%26A...597A..18T} {597, A18}

\bibitem[\protect\citeauthoryear{{Thater}, {Krajnovi{\'c}}, {Cappellari},
  {Davis}, {de Zeeuw}, {McDermid}  \& {Sarzi}}{{Thater}
  et~al.}{2019}]{Thater2019}
{Thater} S.,  {Krajnovi{\'c}} D.,  {Cappellari} M.,  {Davis} T.~A.,  {de Zeeuw}
  P.~T.,  {McDermid} R.~M.,   {Sarzi} M.,  2019, \mn@doi [\aap]
  {10.1051/0004-6361/201834808}, \href
  {https://ui.adsabs.harvard.edu/abs/2019A&A...625A..62T} {625, A62}

\bibitem[\protect\citeauthoryear{{Thater}, {Krajnovi{\'c}}, {Nguyen}, {Iguchi}
  \& {Weilbacher}}{{Thater} et~al.}{2020}]{Thater2020}
{Thater} S.,  {Krajnovi{\'c}} D.,  {Nguyen} D.~D.,  {Iguchi} S.,   {Weilbacher}
  P.~M.,  2020, in {Valluri} M.,  {Sellwood} J.~A.,  eds, , Vol.~353, Galactic
  Dynamics in the Era of Large Surveys.
pp 199--202, \mn@doi{10.1017/S1743921319008445}

\bibitem[\protect\citeauthoryear{{Thatte}}{{Thatte}}{2009}]{Thatte2009}
{Thatte} D.,  2009, {NICMOS Data Handbook v. 8.0}

\bibitem[\protect\citeauthoryear{{Tortora}, {Napolitano}, {Romanowsky},
  {Jetzer}, {Cardone}  \& {Capaccioli}}{{Tortora} et~al.}{2011}]{Tortora2011}
{Tortora} C.,  {Napolitano} N.~R.,  {Romanowsky} A.~J.,  {Jetzer} P.,
  {Cardone} V.~F.,   {Capaccioli} M.,  2011, \mn@doi [\mnras]
  {10.1111/j.1365-2966.2011.19438.x}, \href
  {http://adsabs.harvard.edu/abs/2011MNRAS.418.1557T} {418, 1557}

\bibitem[\protect\citeauthoryear{{Vazdekis}, {Koleva}, {Ricciardelli},
  {R{\"o}ck}  \& {Falc{\'o}n-Barroso}}{{Vazdekis} et~al.}{2016}]{Vazdekis2016}
{Vazdekis} A.,  {Koleva} M.,  {Ricciardelli} E.,  {R{\"o}ck} B.,
  {Falc{\'o}n-Barroso} J.,  2016, \mn@doi [\mnras] {10.1093/mnras/stw2231},
  \href {https://ui.adsabs.harvard.edu/abs/2016MNRAS.463.3409V} {463, 3409}

\bibitem[\protect\citeauthoryear{{Verdoes Kleijn}, {van der Marel}, {de Zeeuw},
  {Noel-Storr}  \& {Baum}}{{Verdoes Kleijn} et~al.}{2002}]{VerdoesKleijn2002}
{Verdoes Kleijn} G.~A.,  {van der Marel} R.~P.,  {de Zeeuw} P.~T.,
  {Noel-Storr} J.,   {Baum} S.~A.,  2002, \mn@doi [\aj] {10.1086/344073}, \href
  {http://adsabs.harvard.edu/abs/2002AJ....124.2524V} {124, 2524}

\bibitem[\protect\citeauthoryear{{Walsh}, {van den Bosch}, {Barth}  \&
  {Sarzi}}{{Walsh} et~al.}{2012}]{Walsh2012}
{Walsh} J.~L.,  {van den Bosch} R.~C.~E.,  {Barth} A.~J.,   {Sarzi} M.,  2012,
  \mn@doi [\apj] {10.1088/0004-637X/753/1/79}, \href
  {https://ui.adsabs.harvard.edu/abs/2012ApJ...753...79W} {753, 79}

\bibitem[\protect\citeauthoryear{{Walsh}, {Barth}, {Ho}  \& {Sarzi}}{{Walsh}
  et~al.}{2013}]{Walsh2013}
{Walsh} J.~L.,  {Barth} A.~J.,  {Ho} L.~C.,   {Sarzi} M.,  2013, \mn@doi [\apj]
  {10.1088/0004-637X/770/2/86}, \href
  {http://adsabs.harvard.edu/abs/2013ApJ...770...86W} {770, 86}

\bibitem[\protect\citeauthoryear{{Weilbacher} et~al.,}{{Weilbacher}
  et~al.}{2020}]{Weilbacher2020}
{Weilbacher} P.~M.,  et~al., 2020, \mn@doi [\aap]
  {10.1051/0004-6361/202037855}, \href
  {https://ui.adsabs.harvard.edu/abs/2020A&A...641A..28W} {641, A28}

\bibitem[\protect\citeauthoryear{{Willmer}}{{Willmer}}{2018}]{Willmer2018}
{Willmer} C.~N.~A.,  2018, \mn@doi [\apjs] {10.3847/1538-4365/aabfdf}, \href
  {http://adsabs.harvard.edu/abs/2018ApJS..236...47W} {236, 47}

\bibitem[\protect\citeauthoryear{{Yoon}}{{Yoon}}{2017}]{Yoon2017}
{Yoon} I.,  2017, \mn@doi [\mnras] {10.1093/mnras/stw3171}, \href
  {http://adsabs.harvard.edu/abs/2017MNRAS.466.1987Y} {466, 1987}

\bibitem[\protect\citeauthoryear{{de Francesco}, {Capetti}  \& {Marconi}}{{de
  Francesco} et~al.}{2006}]{deFrancesco2006}
{de Francesco} G.,  {Capetti} A.,   {Marconi} A.,  2006, \mn@doi [\aap]
  {10.1051/0004-6361:20065826}, \href
  {http://adsabs.harvard.edu/abs/2006A%26A...460..439D} {460, 439}

\bibitem[\protect\citeauthoryear{{de Vaucouleurs}, {de Vaucouleurs}, {Corwin},
  {Buta}, {Paturel}  \& {Fouqu{\'e}}}{{de Vaucouleurs}
  et~al.}{1991}]{deVaucouleurs1991rc3}
{de Vaucouleurs} G.,  {de Vaucouleurs} A.,  {Corwin} Jr. H.~G.,  {Buta} R.~J.,
  {Paturel} G.,   {Fouqu{\'e}} P.,  1991, {Third Reference Catalogue of Bright
  Galaxies. Volume I: Explanations and references. Volume II: Data for galaxies
  between 0$^{h}$ and 12$^{h}$. Volume III: Data for galaxies between 12$^{h}$
  and 24$^{h}$.}

\bibitem[\protect\citeauthoryear{{den Brok}, {Krajnovi{\'c}}, {Emsellem},
  {Brinchmann}  \& {Maseda}}{{den Brok} et~al.}{2021}]{Brok2021}
{den Brok} M.,  {Krajnovi{\'c}} D.,  {Emsellem} E.,  {Brinchmann} J.,
  {Maseda} M.,  2021, arXiv e-prints, \href
  {https://ui.adsabs.harvard.edu/abs/2021arXiv210914640D} {p. arXiv:2109.14640}

\bibitem[\protect\citeauthoryear{van~den Bosch}{van~den
  Bosch}{2016}]{Bosch2016}
van~den Bosch R. C.~E.,  2016, \mn@doi [The Astrophysical Journal]
  {10.3847/0004-637x/831/2/134}, 831, 134

\bibitem[\protect\citeauthoryear{{van den Bosch} \& {de Zeeuw}}{{van den Bosch}
  \& {de Zeeuw}}{2010}]{vandenBosch2010}
{van den Bosch} R.~C.~E.,  {de Zeeuw} P.~T.,  2010, \mn@doi [\mnras]
  {10.1111/j.1365-2966.2009.15832.x}, \href
  {http://adsabs.harvard.edu/abs/2010MNRAS.401.1770V} {401, 1770}

\bibitem[\protect\citeauthoryear{{van den Bosch} \& {van de Ven}}{{van den
  Bosch} \& {van de Ven}}{2009}]{vandenBosch2009}
{van den Bosch} R. C.~E.,  {van de Ven} G.,  2009, \mn@doi [\mnras]
  {10.1111/j.1365-2966.2009.15177.x}, \href
  {https://ui.adsabs.harvard.edu/abs/2009MNRAS.398.1117V} {398, 1117}

\bibitem[\protect\citeauthoryear{{van der Marel}, {Cretton}, {de Zeeuw}  \&
  {Rix}}{{van der Marel} et~al.}{1998}]{vanderMarel1998}
{van der Marel} R.~P.,  {Cretton} N.,  {de Zeeuw} P.~T.,   {Rix} H.-W.,  1998,
  \mn@doi [\apj] {10.1086/305147}, \href
  {http://adsabs.harvard.edu/abs/1998ApJ...493..613V} {493, 613}

\makeatother
\end{thebibliography}
\bibliographystyle{mnras}

\section{SUPPORTING INFORMATION}
Supplementary figures 1-5 are available at MNRAS online.

\appendix
\section{MGE of the F814W WFPC2 + {\it i}-band CGS images}\label{ss:mge_f814}

\begin{figure*}
  \centering
      \includegraphics[width=0.36\textwidth]{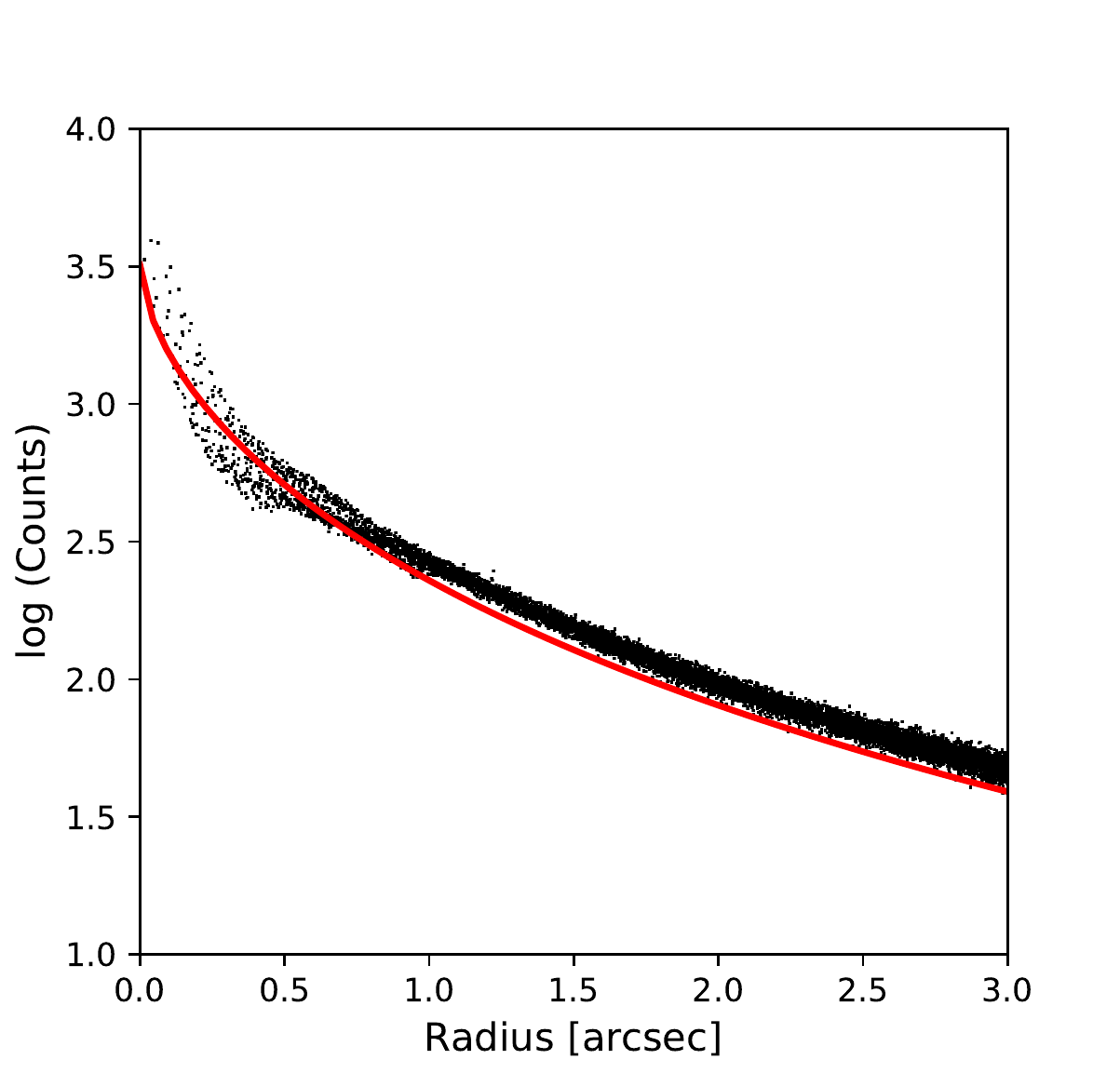}
      \includegraphics[width=0.63\textwidth]{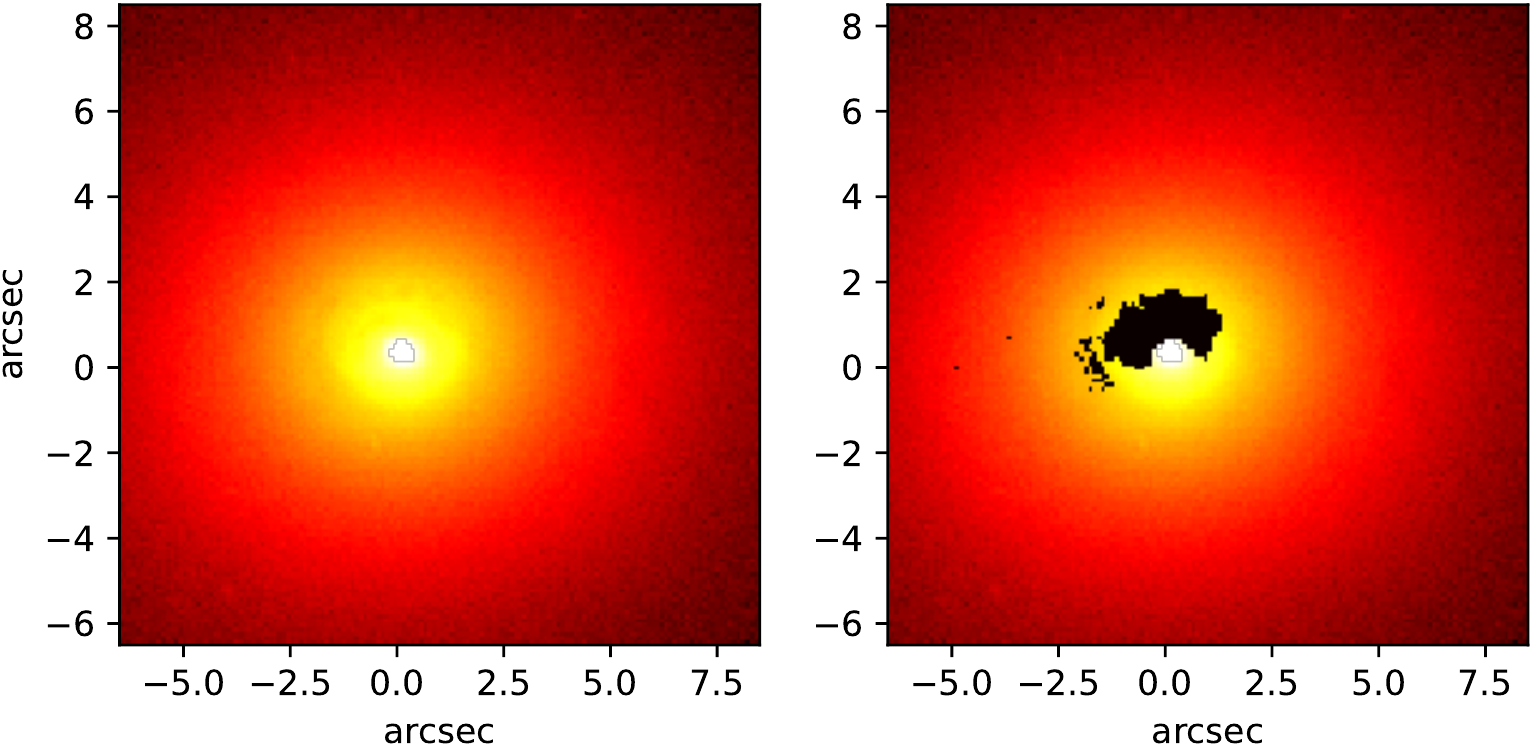}
      \caption{Dust-masking of the F814W WFPC2 image. Left panel: Every point is a pixel of the WFPC2 image, the lower envelope fit of the surface brightness is given by
the red solid line. Each pixel below the red line is masked when creating the MGE. Middle and right panel: Central part of the HST/F814W image with and without dust mask.}
      \label{ff:dust_mask}
\end{figure*}

We followed a similar approach as in Section~\ref{ss:mge_nicmos} to obtain a light model for NGC 6958 in the \textit{i-}band. As the \textit{i}-band is more affected by dust extinction than the \textit{H}-band, this light model required a careful treatment of the dust-affected galaxy centre. We therefore created a dust mask following the procedure given in \cite{Thater2017} and \cite{Thater2019}. We generated the surface brightness profile and iteratively  fitted the lower envelope of the not-dust-affected regions with a  4-parameter logistic function (see Fig. \ref{ff:dust_mask}). Masked were all pixels which had a surface brightness below this envelope fit. We then applied the 
Multi Gaussian Expansion (MGE) routine \citep{Cappellari2002} simultaneously on the dust-masked WFPC2 and the CGS image as described in Section~\ref{ss:mge_nicmos}. In the central 10 arcsec, the MGE was constrained by the WFPC2 image, while the CGS image constrains the photometry at larger radii (Fig. 1 of the supplementary material). The WFPC2 image was used for the photometric calibration. For the conversion, we used 4.53 mag for the absolute AB-magnitude of the sun \citep{Willmer2018} in the F814W band and A$_{\mathrm{F814}}=0.07$ mag for the Galactic extinction (NED).
This MGE model was used in Section~\ref{ss:muse_psf} to derive the PSF of the MUSE data and in Section 4.4.3 where we tested the effect of different mass models on the robustness of our black hole mass measurement.

\begin{table}
\caption{{\it HST}/WFPC2 F814W + {\it i}-band CGS MGE model.}
\centering
\begin{tabular}{lcccr}
\hline\hline
j &   $\log\,(I_j$) & $\sigma_j$ & q$_j$ & $\log\,(M_{j}^{\rm const}$)  \\
  & (L$_{\sun , H}$ pc$^{-2}$) & (arcsec) & & (M$_{\sun}$)  \\
(1)  & (2) &  (3) & (4) & (5) \\
\hline
1  &  5.136 & 0.041 & 0.91 & 8.186 \\
2  &  4.699 & 0.146 & 0.90 & 8.853 \\
3  &  4.169 & 0.416 & 0.91 & 9.239 \\
4  &  3.912 & 0.924 & 0.89 & 9.664 \\
5  &  3.609 & 1.807 & 0.87 & 9.934 \\
6  &  2.833 & 3.806 & 0.86 & 9.803 \\
7  &  2.654 & 7.209 & 0.86 & 10.179 \\
8  &  2.056 & 15.081 & 0.88 & 10.227 \\
9  &  1.39 & 31.787 & 0.91 & 10.226 \\
10  &  -1.263 & 83.174 & 0.91 & 8.408 \\

\hline
\end{tabular}
\\
{Note - Column 1: Index of the Gaussian component. Column 2: Surface brightness. Column 3: Projected gaussian width along the major axis. Column 4: Projected axial ratio for each Gaussian component. Column 5 and 6: Total mass of Gaussian component. In column (5) the constant dynamical $M/L$ = 4.1\,M$_{\odot}$/L$_{\odot}$ from the Schwarzschild modelling (Section 4.4.3) was used to determine the mass of each Gaussian component.} 
\label{tt:mge}
\end{table}

\section{Double Gaussian parametrisations of the MUSE PSF star}

\begin{figure*}
  \centering
    \includegraphics[width=0.255\textwidth]{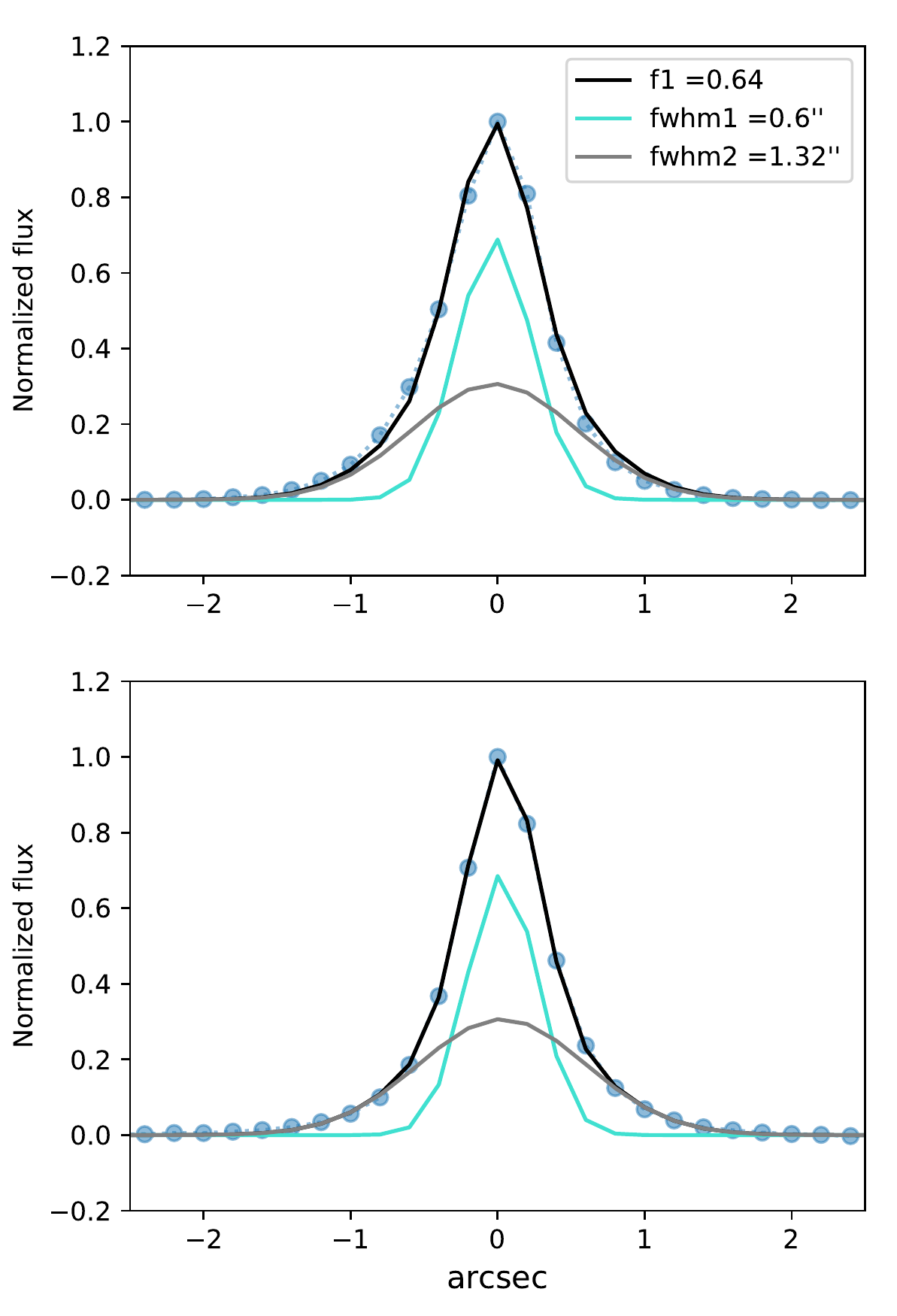}
    \includegraphics[width=0.24\textwidth]{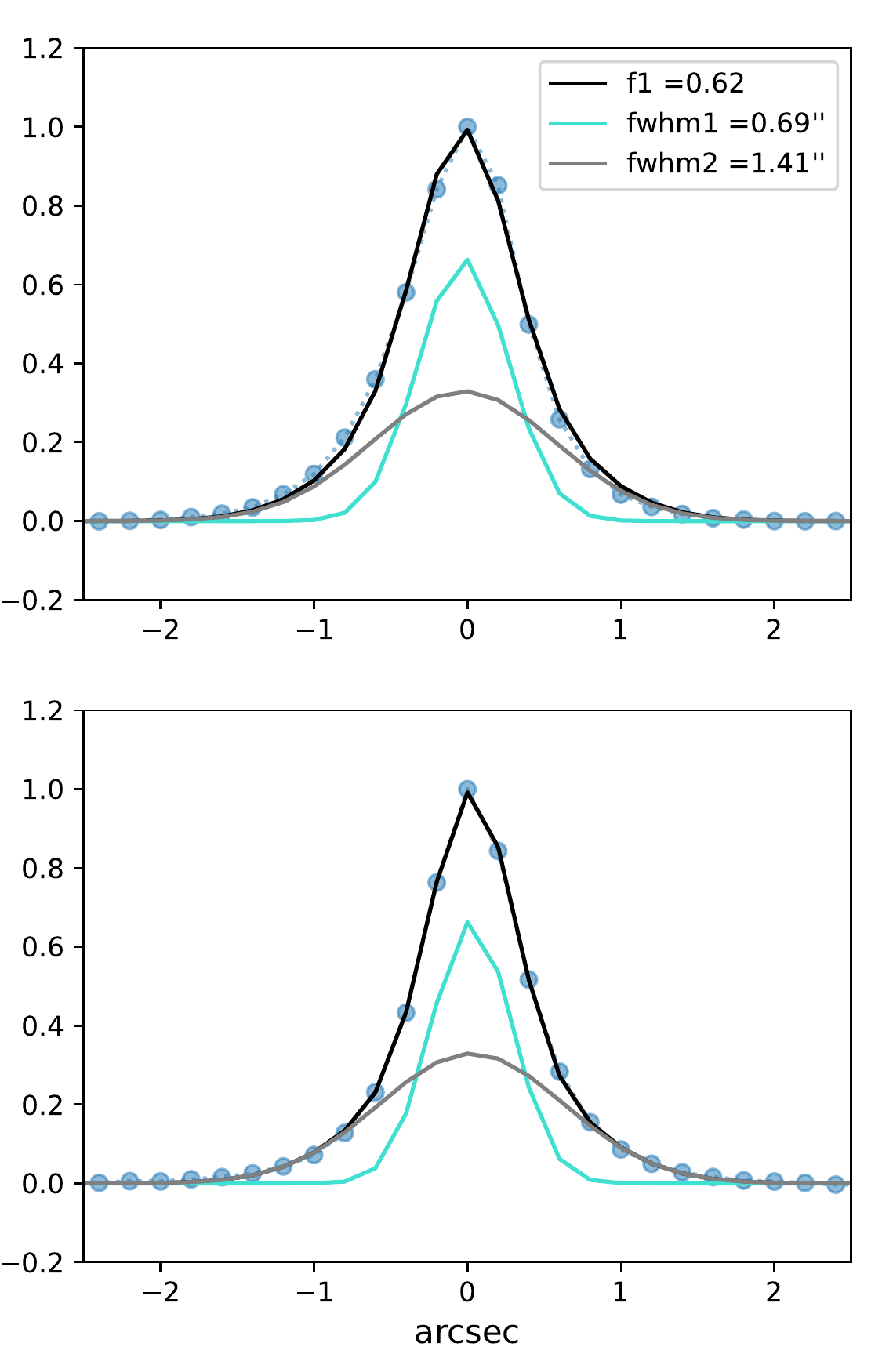}
    \includegraphics[width=0.24\textwidth]{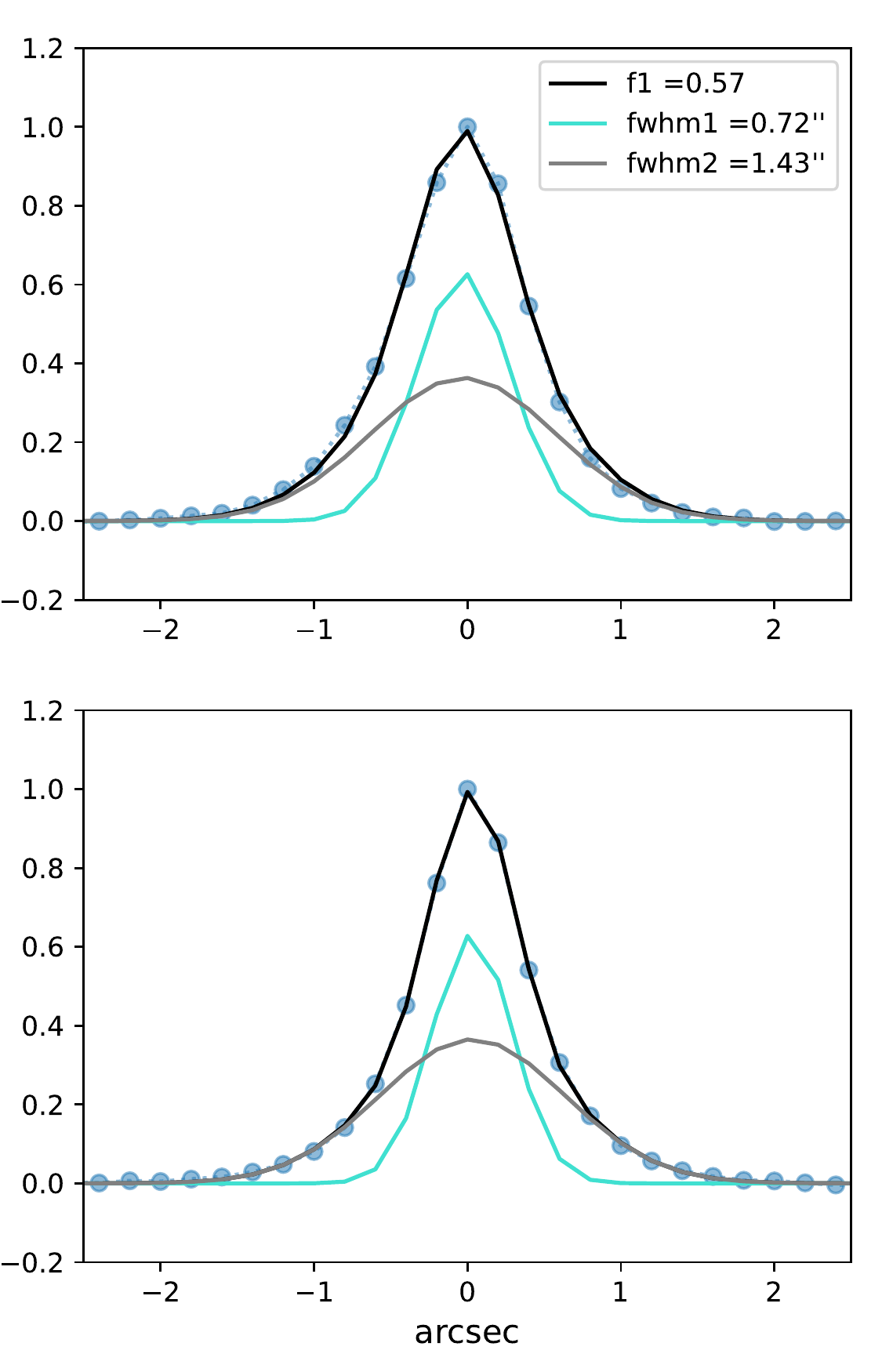}
    \includegraphics[width=0.24\textwidth]{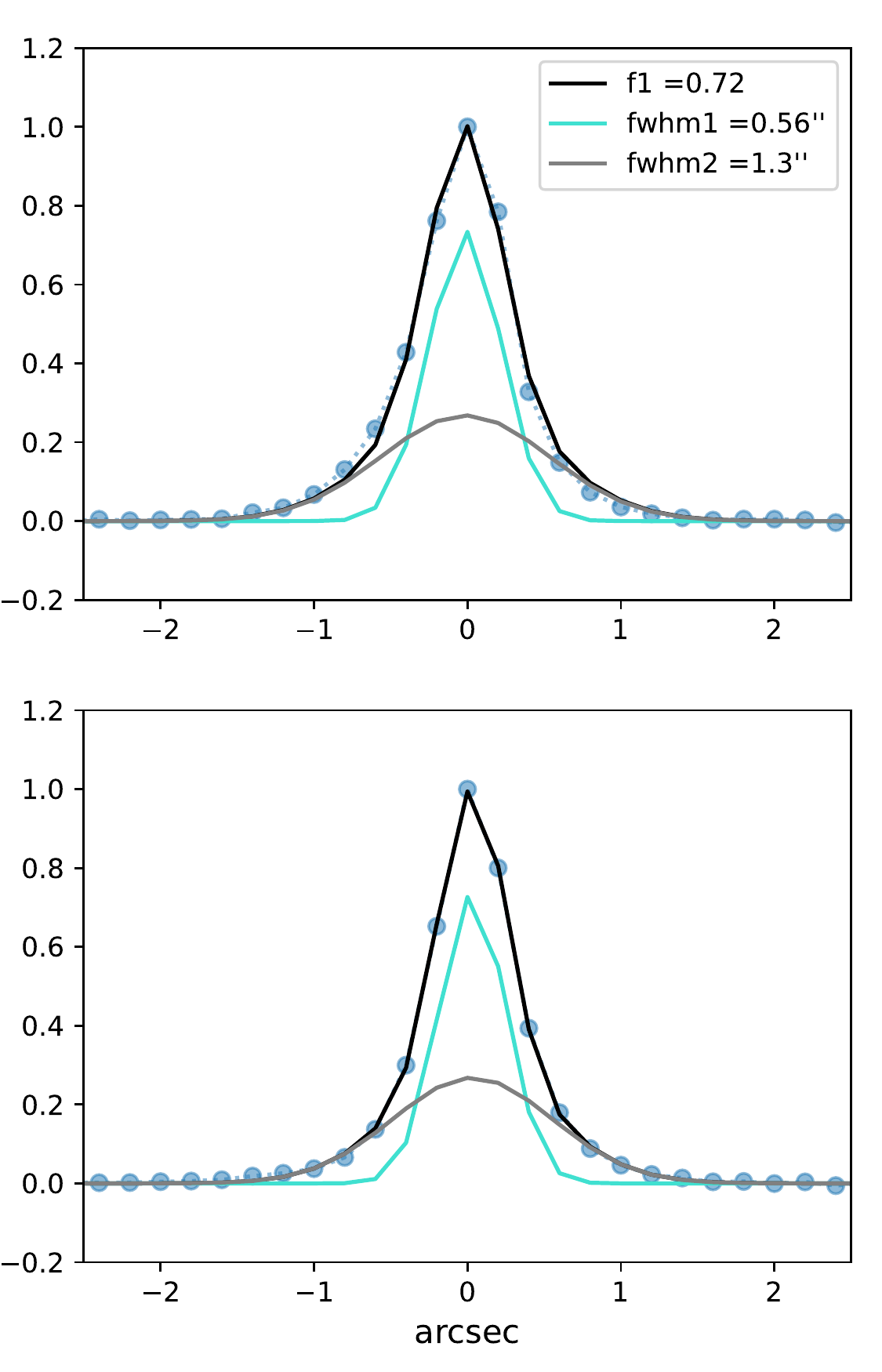}
            
      \caption{PSF parametrisation of the PSF star close to NGC 6958 using a double Gaussian function. From left to right: White-light, optical, blue and CaT wavelength image. Top is the cut along the major axis, bottom along the minor axis. The blue circles show the data points of the PSF star.
      The fits in different colors show the double Gaussian (black), the narrow (cyan) and the broad (grey) Gaussian component of the PSF. The Gaussians are characterised by their full width at half maximum (fwhm) and the relative flux of the narrow Gaussian (f1).}
      \label{ff:psf_parametrization}
\end{figure*}

\section{JAM result validation}

  \begin{table*}
\caption{Results of the JAM measurement testing different systematics}
\centering
\begin{tabular}{lccc|cccc}
\hline\hline
 Velocity ellipsoid & Spectral range &    $i$ & Additional change & $M_{\rm BH}$  &  $M/L$ & $\beta$ &  $\chi^2$/d.o.f.\\
    & & & & ($\times 10^8\,$M$_{\odot}$) & (M$_{\odot}/$L$_{\rm \odot ,H}$) & & \\
\hline

cyl. alignment  & optical &  45 & fiducial model & 8.6$^{+ 0.8}_{-0.8}$   & 0.83$^{\pm 0.02}$  & -0.02$^{\pm 0.06}$ &   0.24 \\
cyl. alignment  & optical   & 45 & masked emission lines   & 8.7$^{+0.8}_{-0.8}$   & 0.84$^{\pm 0.02}$  & -0.01$^{\pm 0.06}$ &   0.25 \\
cyl. alignment  & blue &  45 & --   & 6.9$^{+0.7}_{-0.7}$   & 0.83$^{\pm 0.02}$  & -0.02$^{\pm 0.06}$ &  0.35 \\
cyl. alignment  & CaT &  45 &  --   & 8.1$^{+ 1.1}_{-1.0}$   & 0.85$^{\pm 0.02}$  & -0.01$^{\pm 0.12}$ &   0.13 \\
cyl. alignment  & optical &  89 &  --   & 9.0$^{+0.8}_{-0.8}$   & 0.82$^{\pm 0.03}$  & -0.04$^{\pm 0.04}$ &   0.28\\
cyl. alignment  & optical &  45 &  F814 mass model  & 8.6$^{+ 1.2}_{-1.3}$   & $^*$3.33$^{\pm 0.12}$  & -0.02$^{\pm 0.09}$ &   0.33 \\
cyl. alignment  & optical &  45 &  radially-varying $M/L$  & 7.3$^{+ 1.2}_{-1.1}$   & 0.80$^{\pm 0.02}$  & -0.02$^{\pm 0.07}$ &   0.20 \\
sph. alignment & optical & 45  & fiducial model & 4.6$^{+2.5}_{-2.7}$   & 0.86$^{\pm 0.02}$  & 0.38$^{\pm 0.17}$ &    0.21\\
sph. alignment  & optical   & 45 & masked emission lines   & 4.7$^{+2.1}_{-2.3}$   & 0.87$^{\pm 0.02}$  & 0.42$^{\pm 0.18}$ &   0.21 \\
sph. alignment  & blue &  45 & --   & 2.9$^{+ 1.8}_{-2.1}$   & 0.86$^{\pm 0.02}$  & 0.40$^{\pm 0.18}$ &   0.27 \\
sph. alignment  & CaT &  45 &  --   & 4.1$^{+2.3}_{-2.4}$   & 0.87$^{\pm 0.03}$  & 0.40$^{\pm 0.16}$ &   0.10 \\
sph. alignment  & optical &  89 &  --   & 4.8$^{+1.6}_{-1.7}$  & 0.84$^{\pm 0.03}$  & 0.35 $^{\pm 0.14}$ &   0.25 \\
sph. alignment  & optical &  45 &  F814 mass model  & 4.3$^{+ 1.8}_{-1.8}$   & $^*$3.45$^{\pm 0.12}$  & 0.36$^{\pm 0.15}$ &  0.28 \\
sph. alignment  & optical &  45 & radially-varying $M/L$  & 4.1$^{+2.2}_{-1.8}$   & 0.82$^{\pm 0.02}$  & 0.30$^{\pm 0.16}$ &   0.18 \\
\hline
\end{tabular}
\\
{\textbf{Notes.} Column 1-4 show the inputs of the dynamical models, column 5-8 the JAM results. The tests are described in detail in Section 4.4.3. For each JAM run, we used the setup explained in Section 4.3 with the kinematic uncertainties ($\delta_{\rm kin}$) enhanced for R> 0.5 arcsec. The reduced $\chi^2$ values were calculated using the enhanced kinematic errors. 
$^*$ This $M/L$ is in the {\it i}-band.} 
\label{tt:mge_results}
\end{table*}

\begin{figure*}
  \centering
    \includegraphics[width=0.48\textwidth]{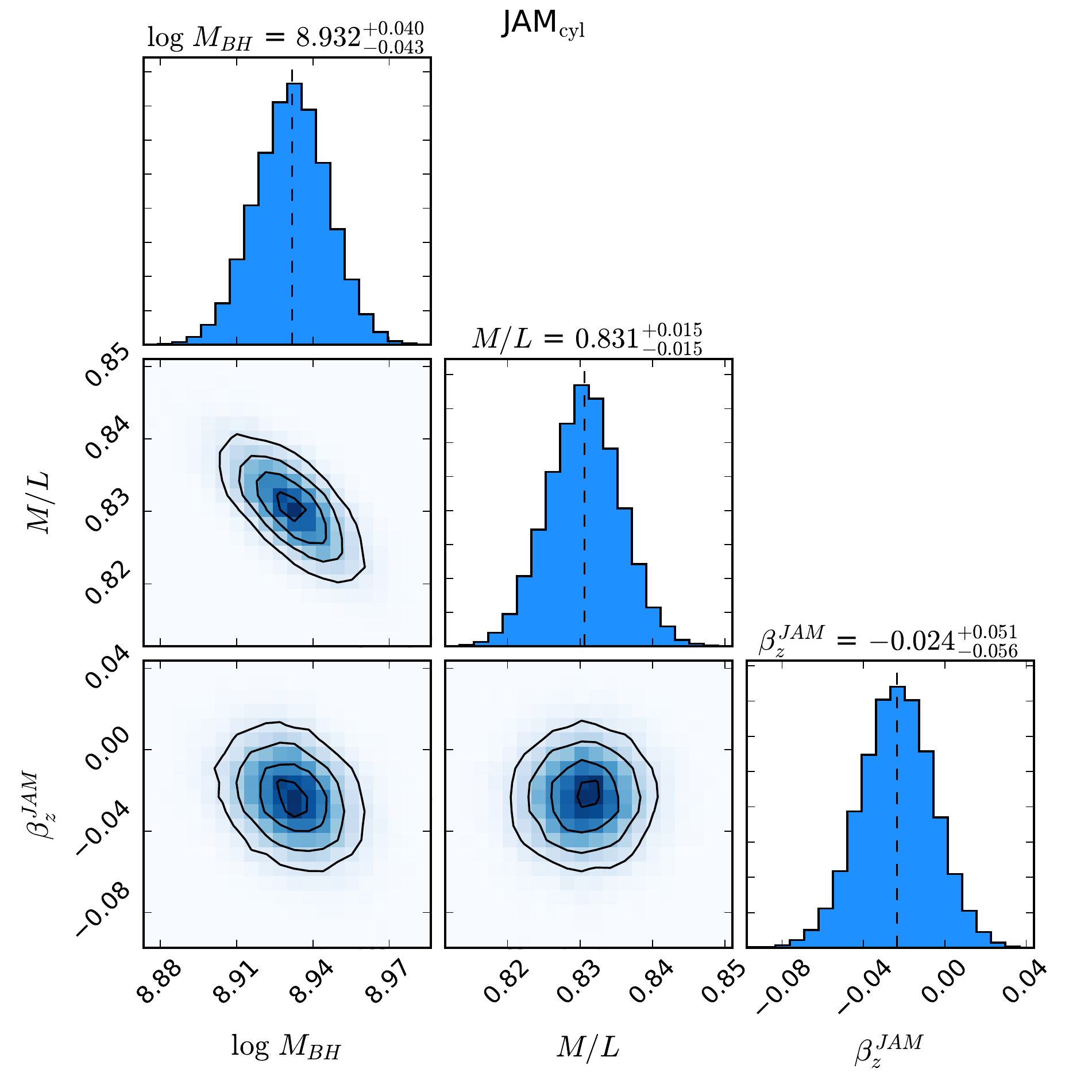}
    \includegraphics[width=0.48\textwidth]{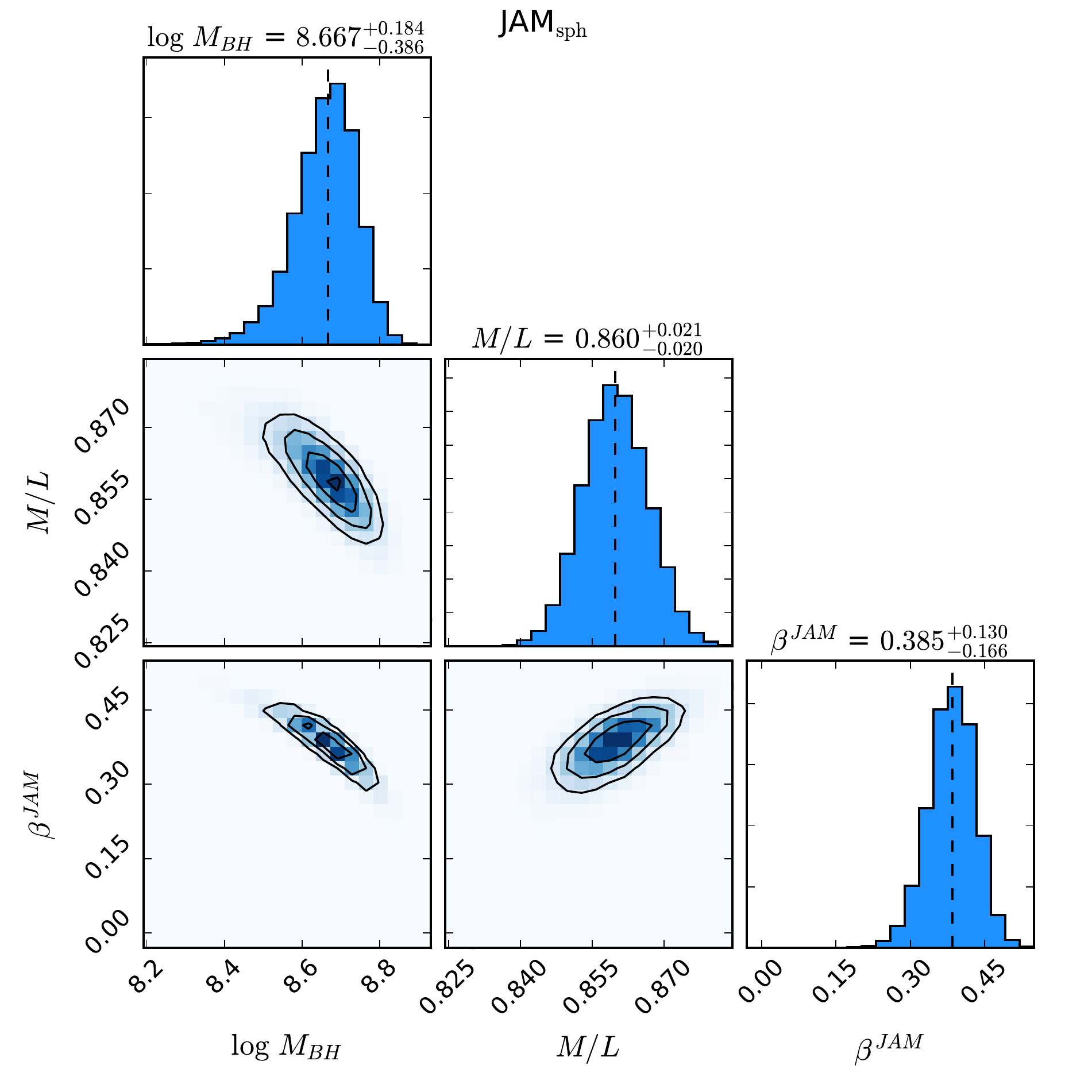}
      \caption{MCMC posterior probability distributions of the fiducial JAM models assuming a cylindrical (left) and spherical-aligned (right) velocity ellipsoid. The contour plots show the two-dimensional distributions for each parameter ($M_{\rm BH}$, $M/L$, $\beta$) combination, the histograms show the projected one-dimensional distributions. The errors above the histogramms are at 99.73 per cent confidence intervals which corresponds to $3\sigma$ significance.}
      \label{ff:jam_corner_plots}
\end{figure*}

\section{Schwarzschild + dark matter grids }
\begin{figure*}

    \includegraphics[width=0.33\textwidth]{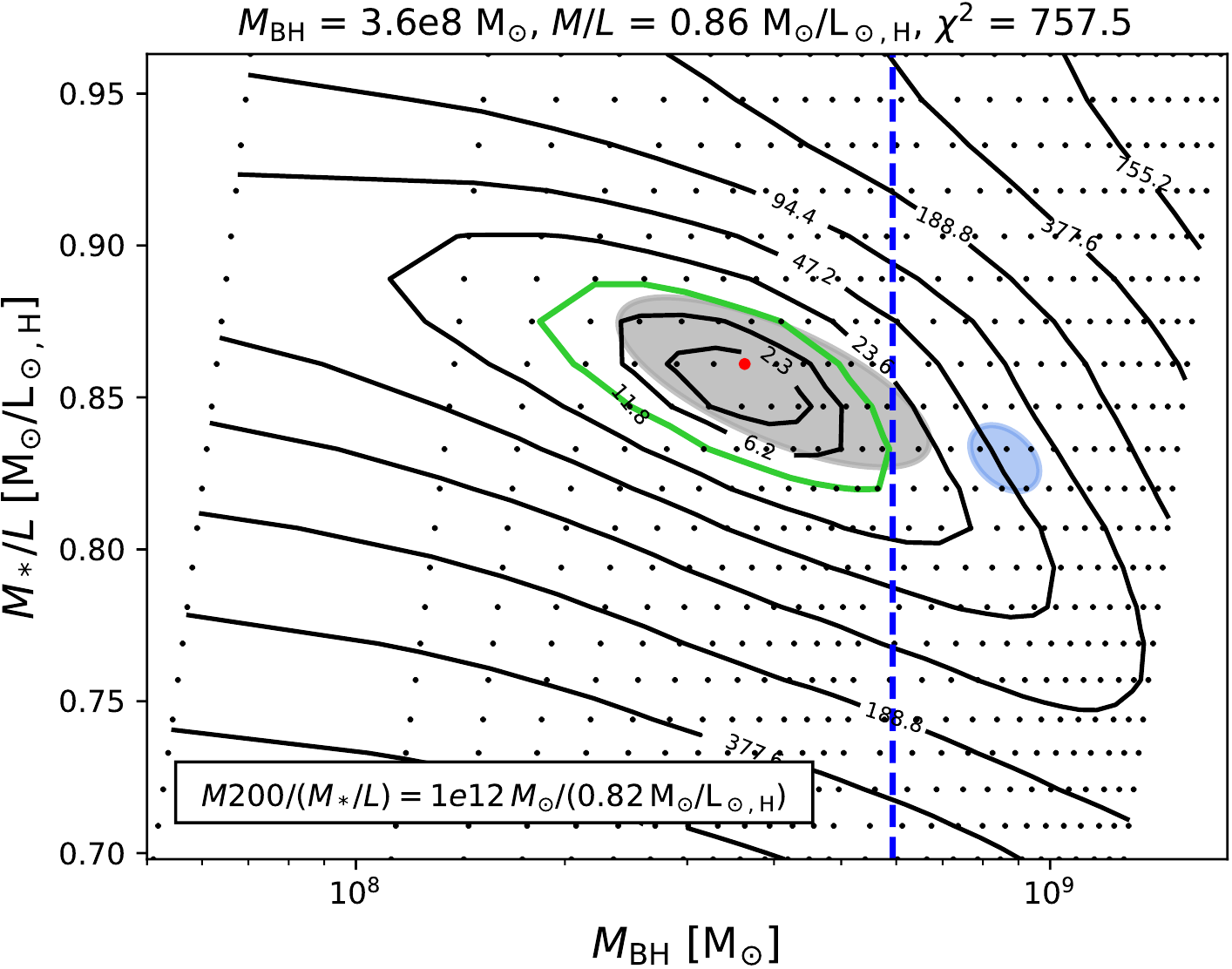}
    \includegraphics[width=0.33\textwidth]{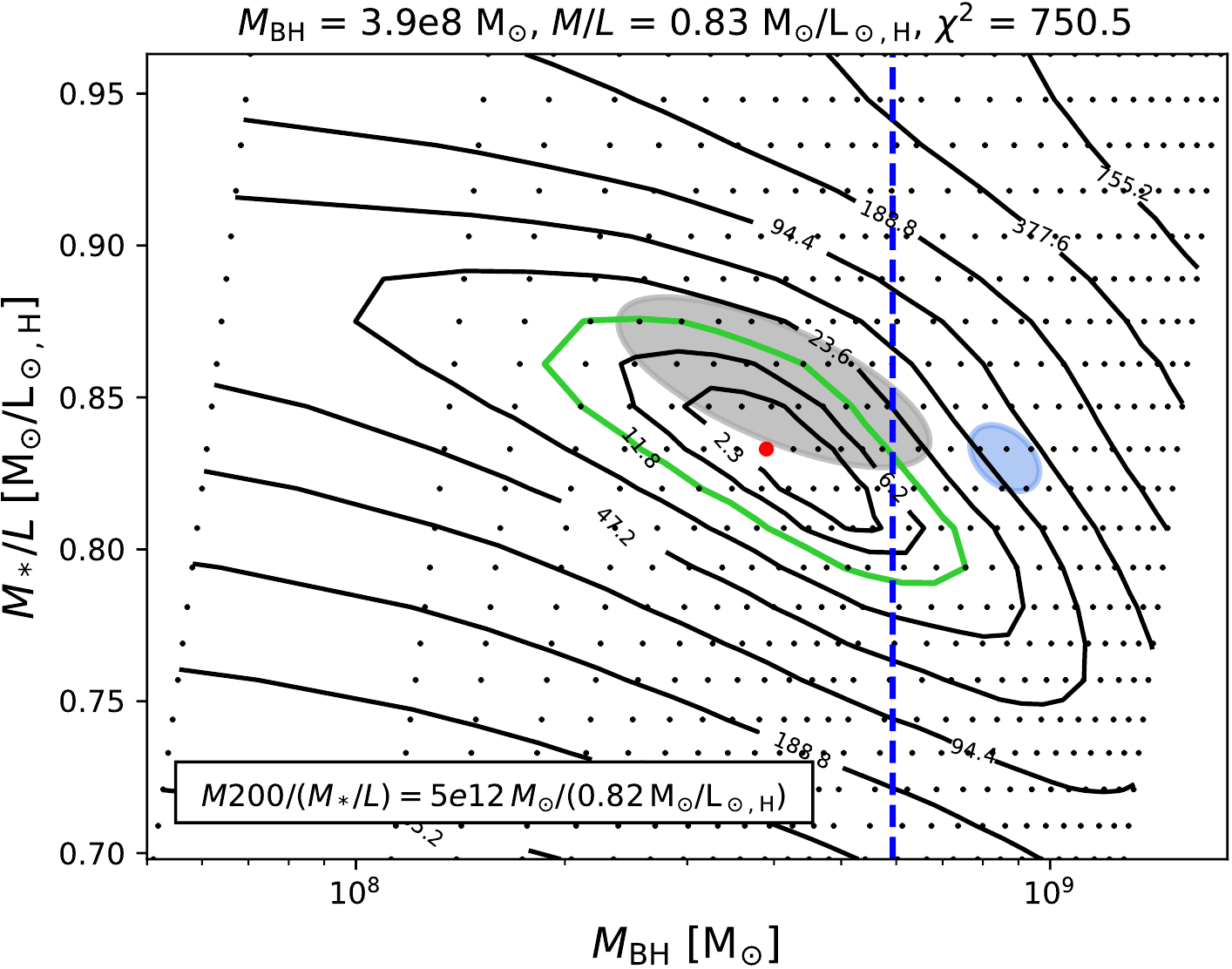}
    \includegraphics[width=0.33\textwidth]{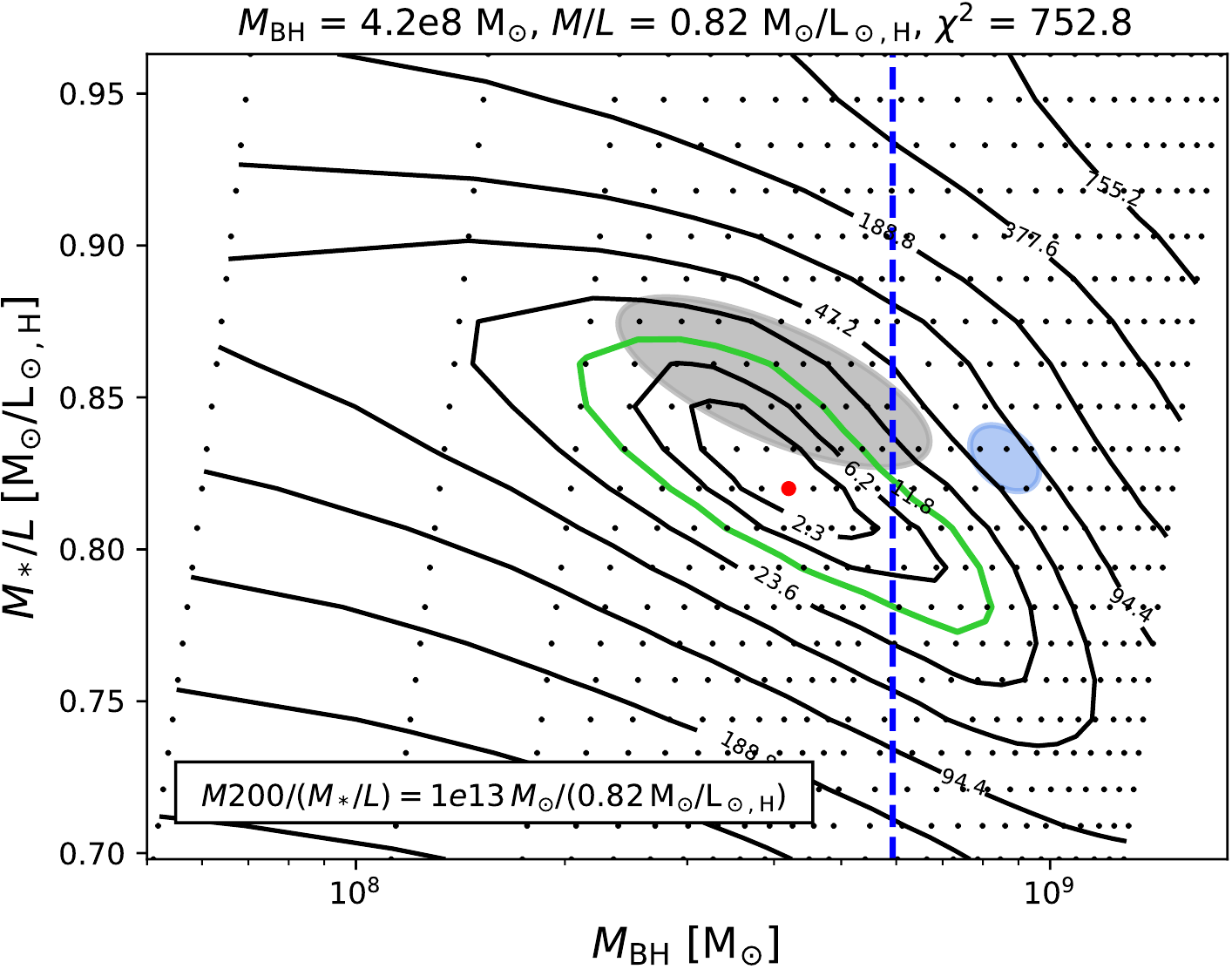}
    
    \includegraphics[width=0.33\textwidth]{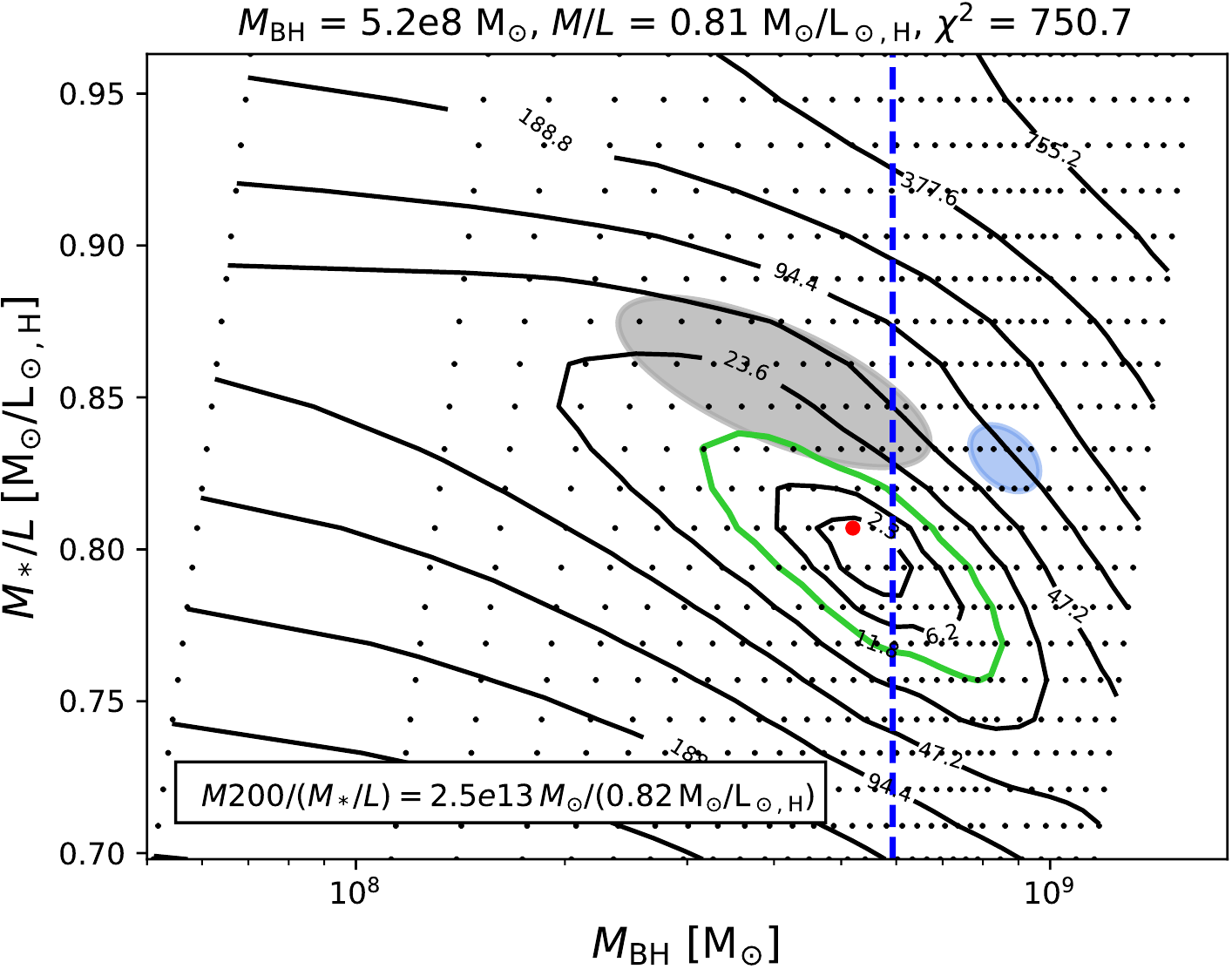}
    \includegraphics[width=0.33\textwidth]{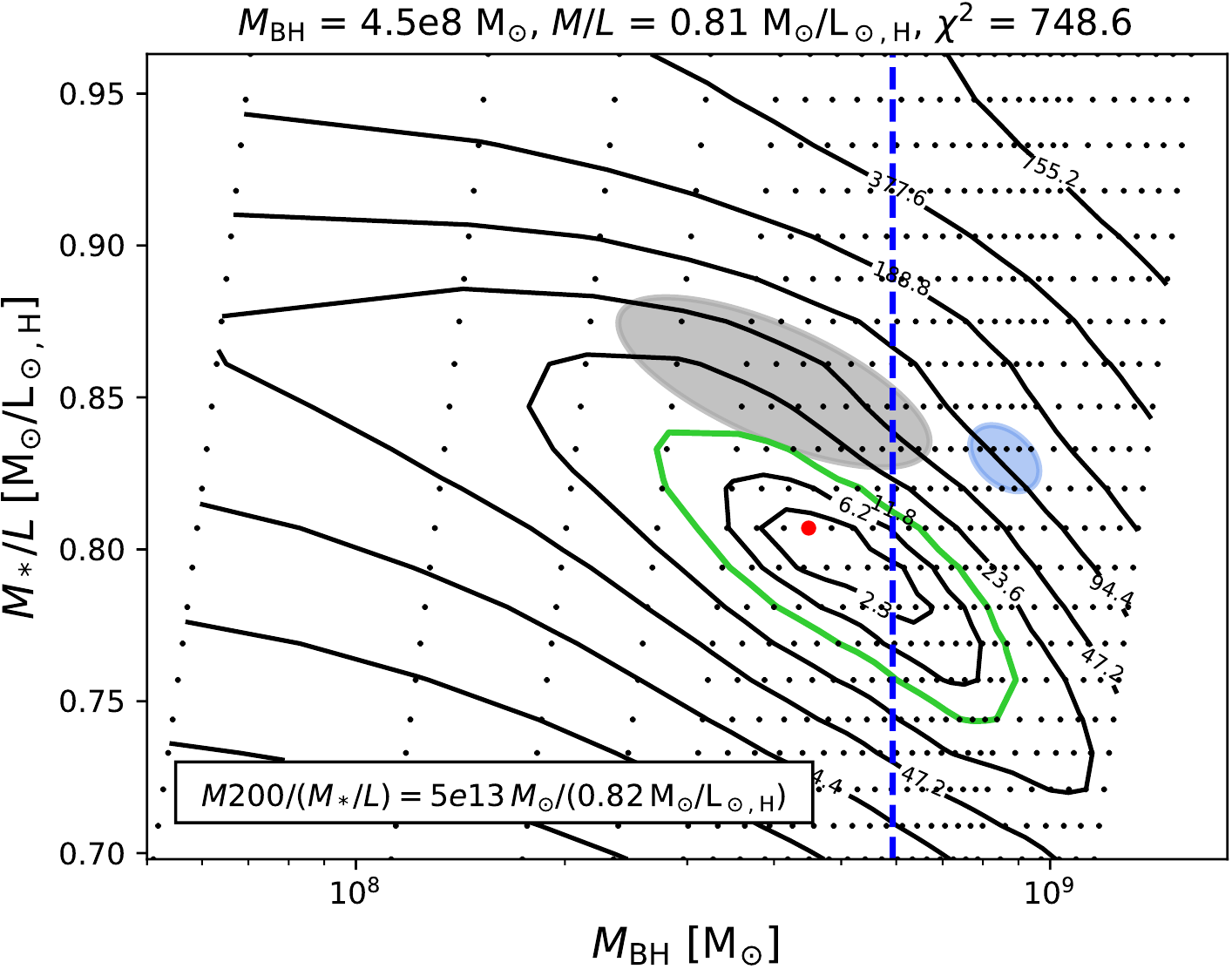}
    \includegraphics[width=0.33\textwidth]{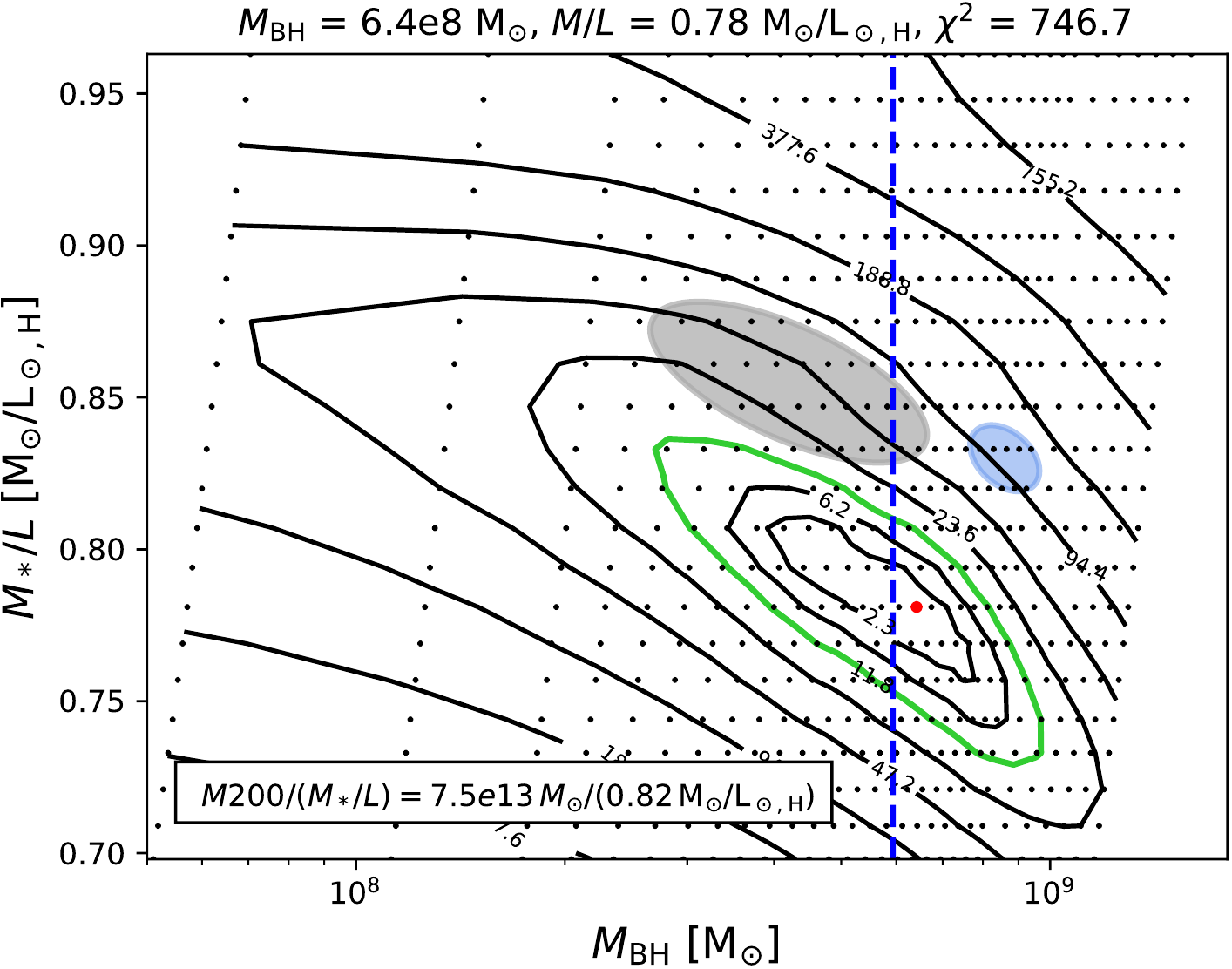}
    
    \includegraphics[width=0.33\textwidth]{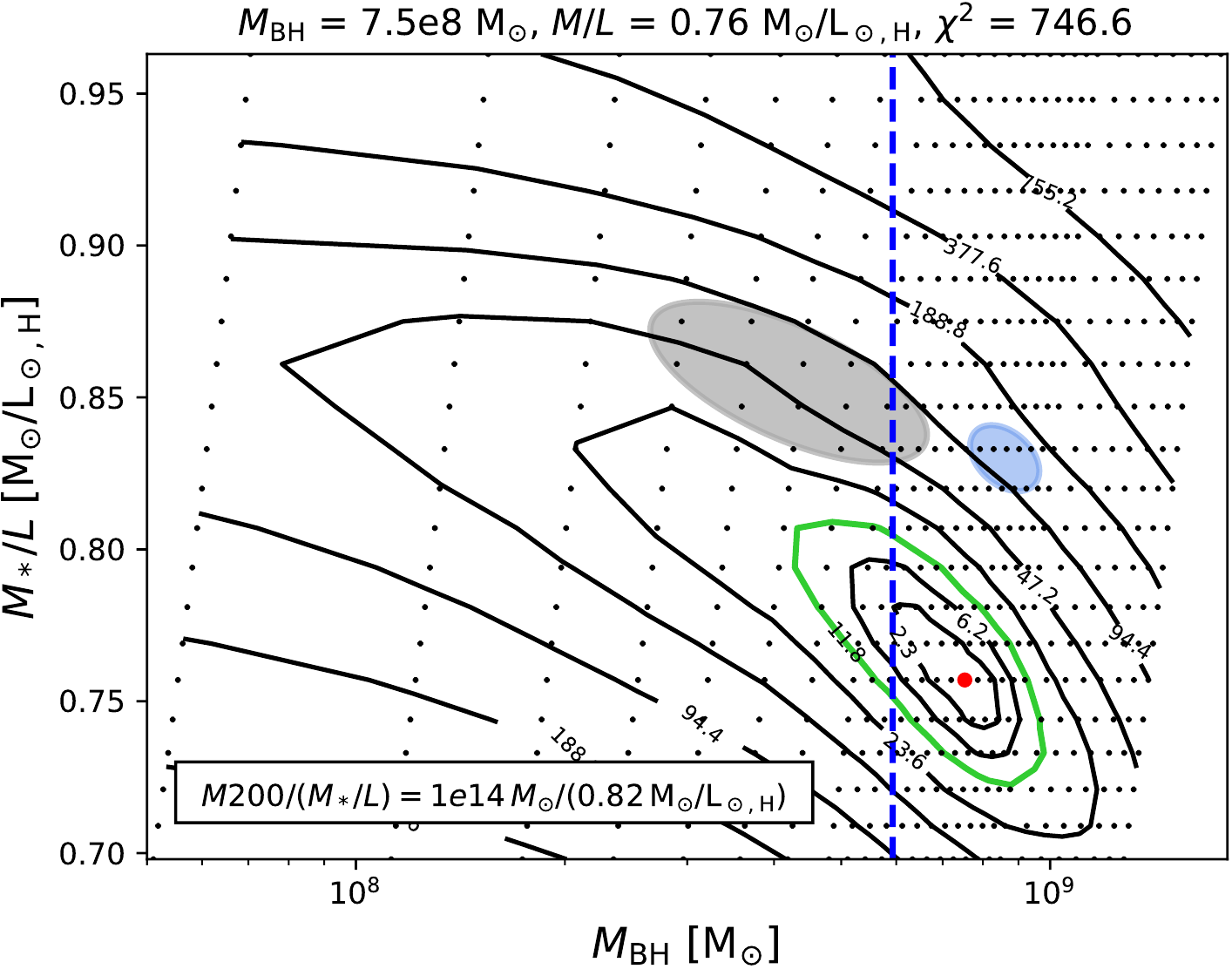}
    \includegraphics[width=0.33\textwidth]{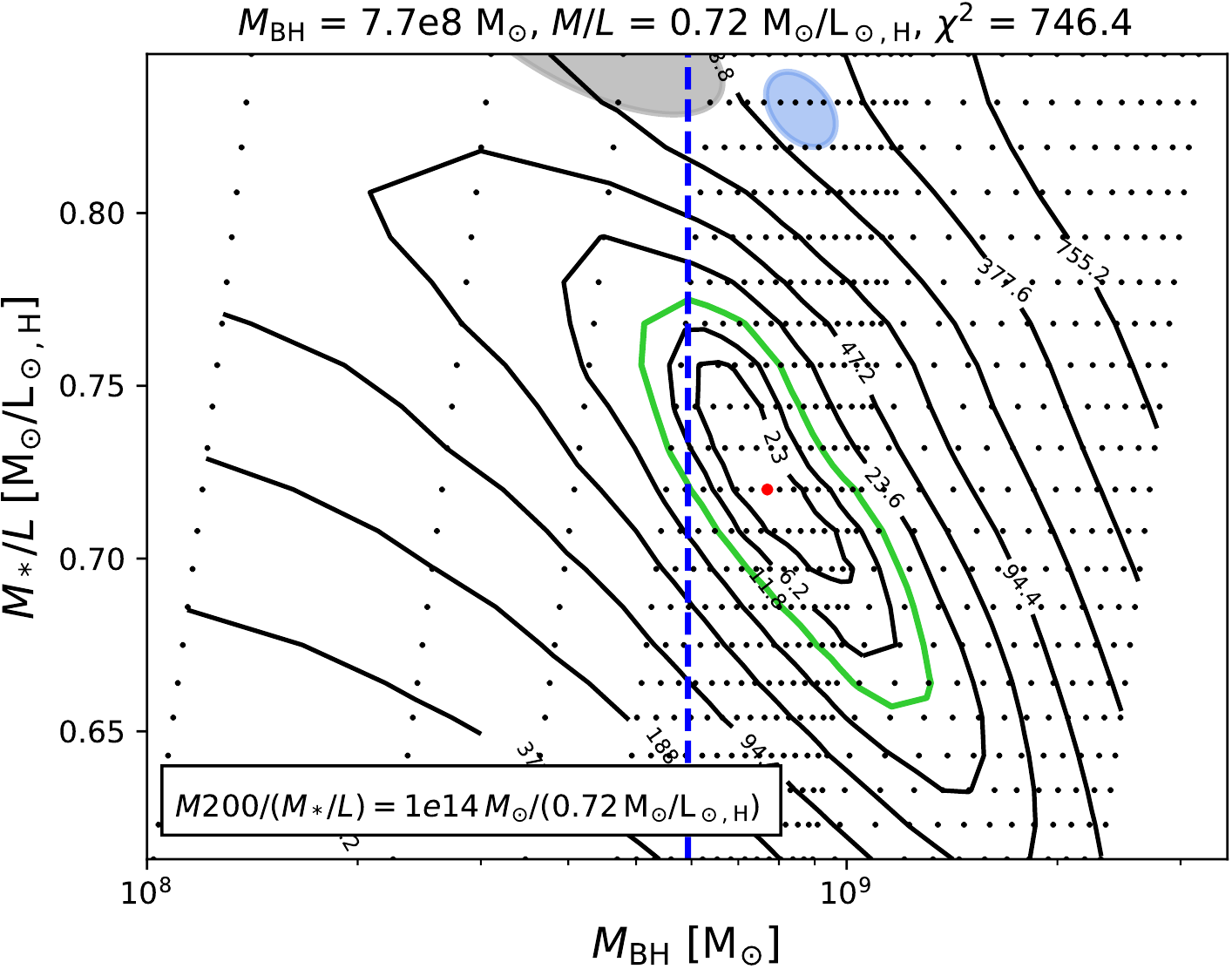}
    \includegraphics[width=0.33\textwidth]{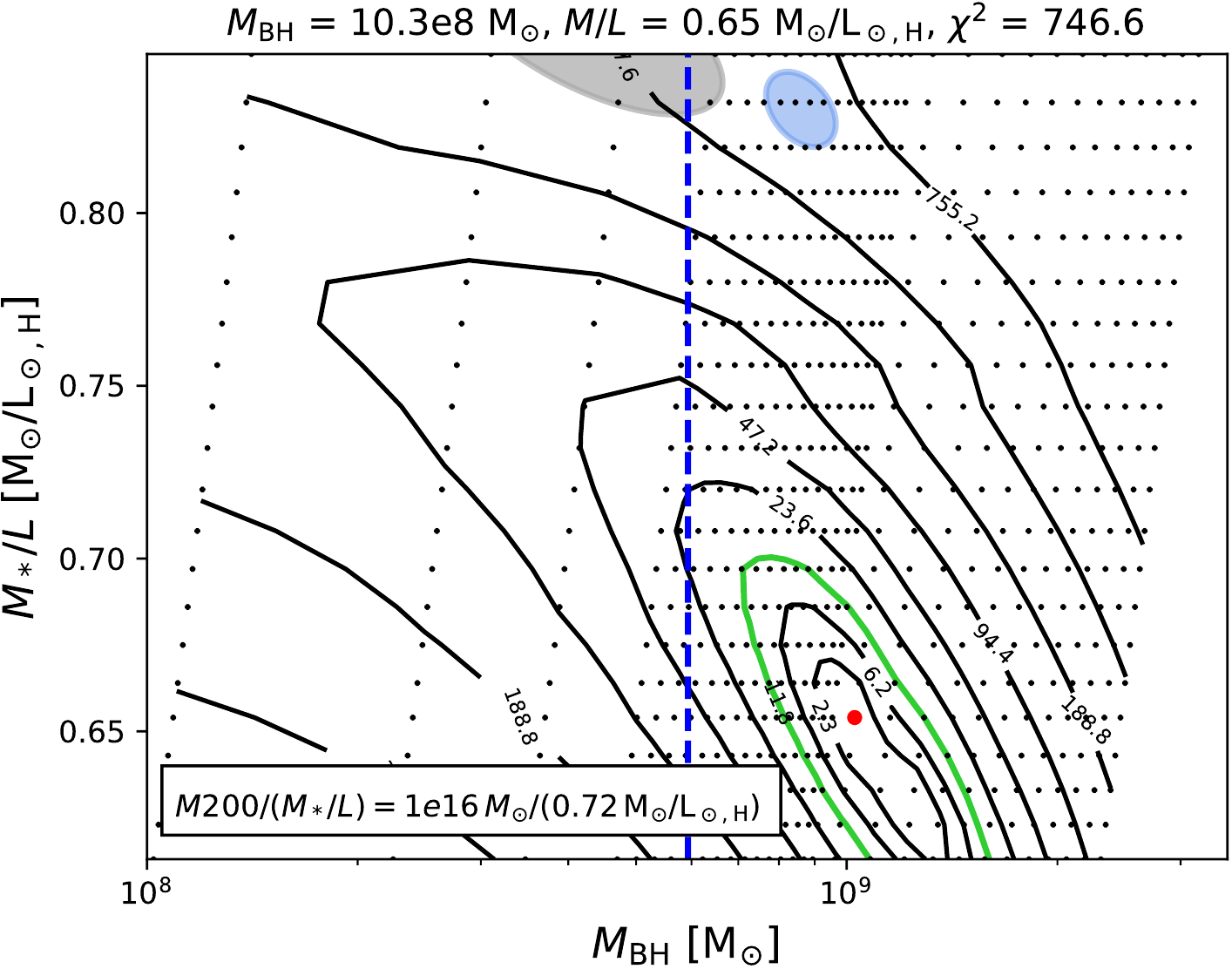}
      \caption{Results of our Schwarzschild run including a dark matter halo. In the runs, we have varied $M_{\rm BH}, M_{*}/L$ and $M_{200}$. Each of the panels shows the $M_{\rm BH} - M_{*}/L$ grid (like Figure 8) for an increasing dark matter fraction. While running Schwarzschild models, both $M_{\rm BH}$ and $M_{200}$ need to be re-scaled for each model. The scale factor is the ($M_{*}/L$) of the grid divided by the starting ($M_{*}/L$) of the models. The confidence level were derived for each $M_{\rm BH} - M_{*}/L$ grid independently. As expected, with increasing dark matter fraction, the black hole mass increases and the $M_{*}/L$ decreases. As expected, with increasing dark matter fraction, the black hole mass increases and the M/L decreases. Contrary to Fig. 8, here the M/L from JAM is larger or equal to that from Schwarzschild's models. This is because the former represent the total M/L, while the latter is that of the stellar component alone. When dark matter increases the stellar M/L must decrease with respect to the total one, as observed.}
      \label{ff:dm_grid}
\end{figure*}

\section{Additional online material for the publication}

\begin{figure*}
  \centering
    \includegraphics[width=0.48\textwidth]{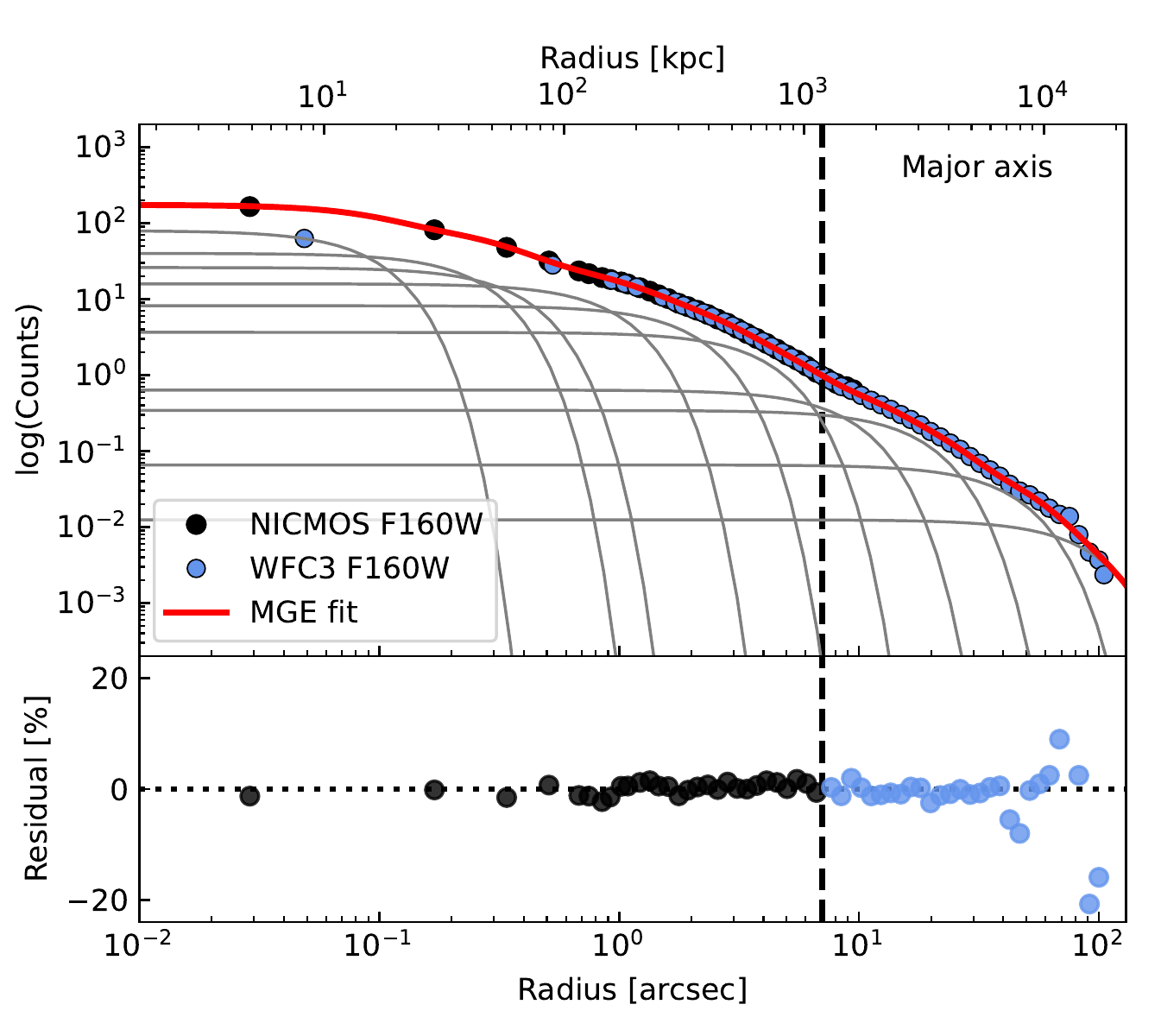}
    \includegraphics[width=0.48\textwidth]{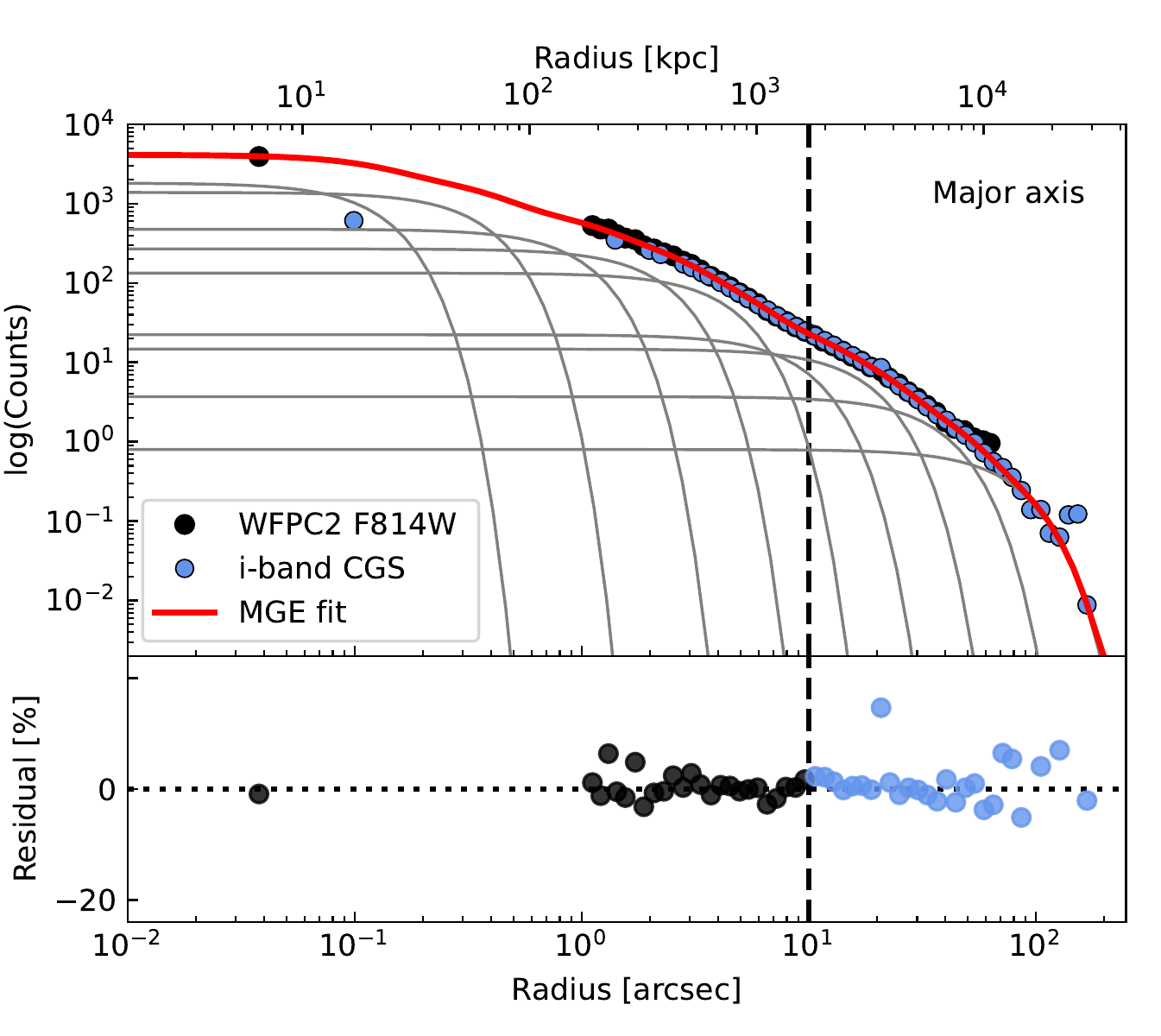}
      \caption{Count profiles of the photometric data along the major axis overplotted with the best-fitting MGE models. The left MGE is described in the main text, the right MGE is described in Appendix A. Top: Count profiles of the photometric data along the major axis. Overplotted is the best-fitting MGE model (red line) and each Gaussian component (grey lines). The MGE model was built from the combined photometric information of NICMOS ($R \leq$ 7\arcsec) and wide-field WFC3 ($R >$  7\arcsec) data.  The vertical dashed line marks the 7 arcsec boundary. Bottom: Residuals of the MGE model: (Data-Model)/Data.}
      \label{ff:mge_alt}
\end{figure*}

\begin{figure*}
  \centering
    \includegraphics[width=0.98\textwidth]{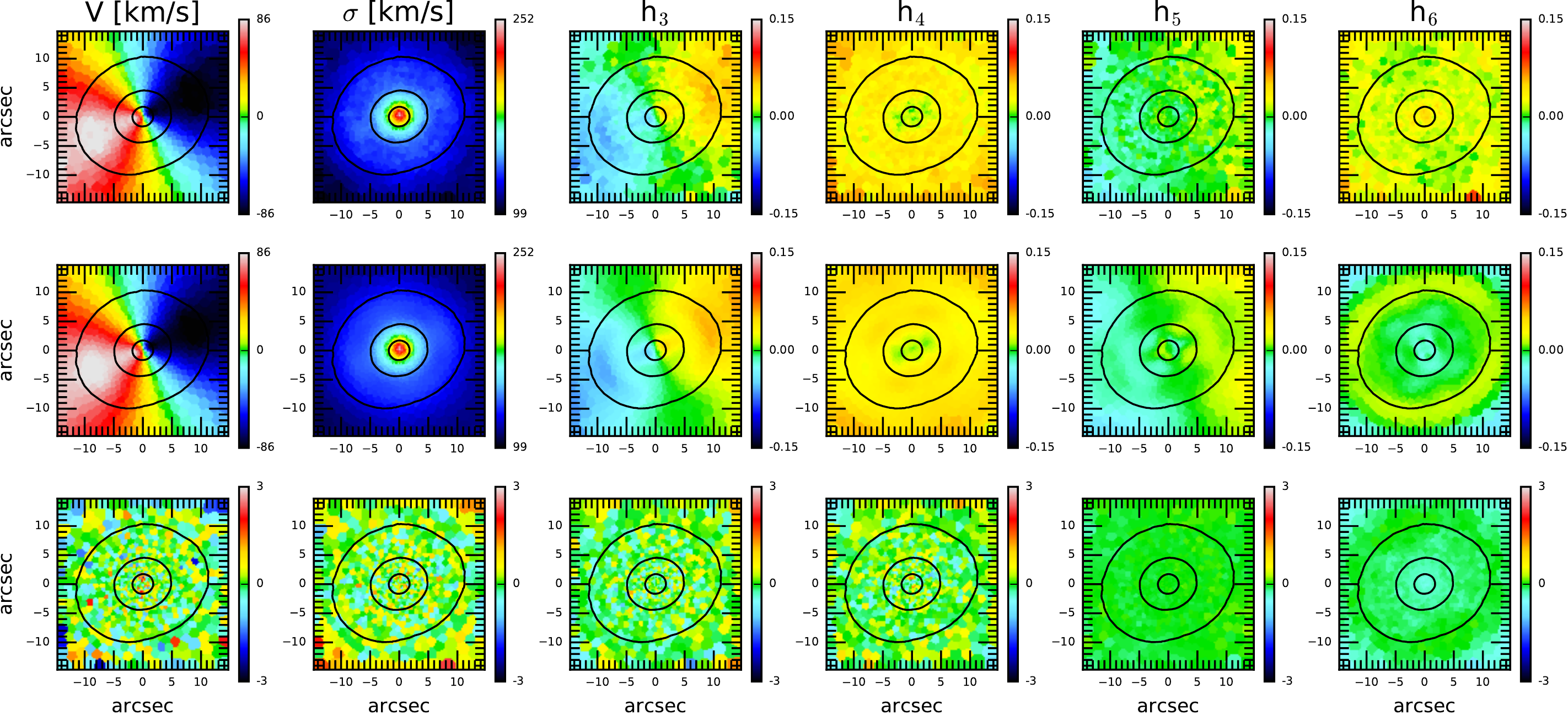}
      \caption{Comparison between symmetrized MUSE kinematics and best-fitting Schwarzschild model for $M_{\rm BH} = 3.6 \times 10^8$ M$_{\odot}$ and $M/L$ = 0.91 M$_{\odot}$/L$_{\rm \odot, H}$ of the fiducial run.. From left to right: Mean velocity, velocity dispersion, $h_{3}$, $h_4$, $h_{5}$ and $h_6$ Gauss-Hermite moments. From top to bottom: Symmetrized data, best-fitting Schwarzschild model and residual maps. Residuals are defined as difference between the Schwarzschild model and observed kinematics divided by the observational errors. North is up and east to the left.}
      \label{ff:schwarschild_model}
\end{figure*}

\begin{figure*}
  \centering
    \includegraphics[width=0.98\textwidth]{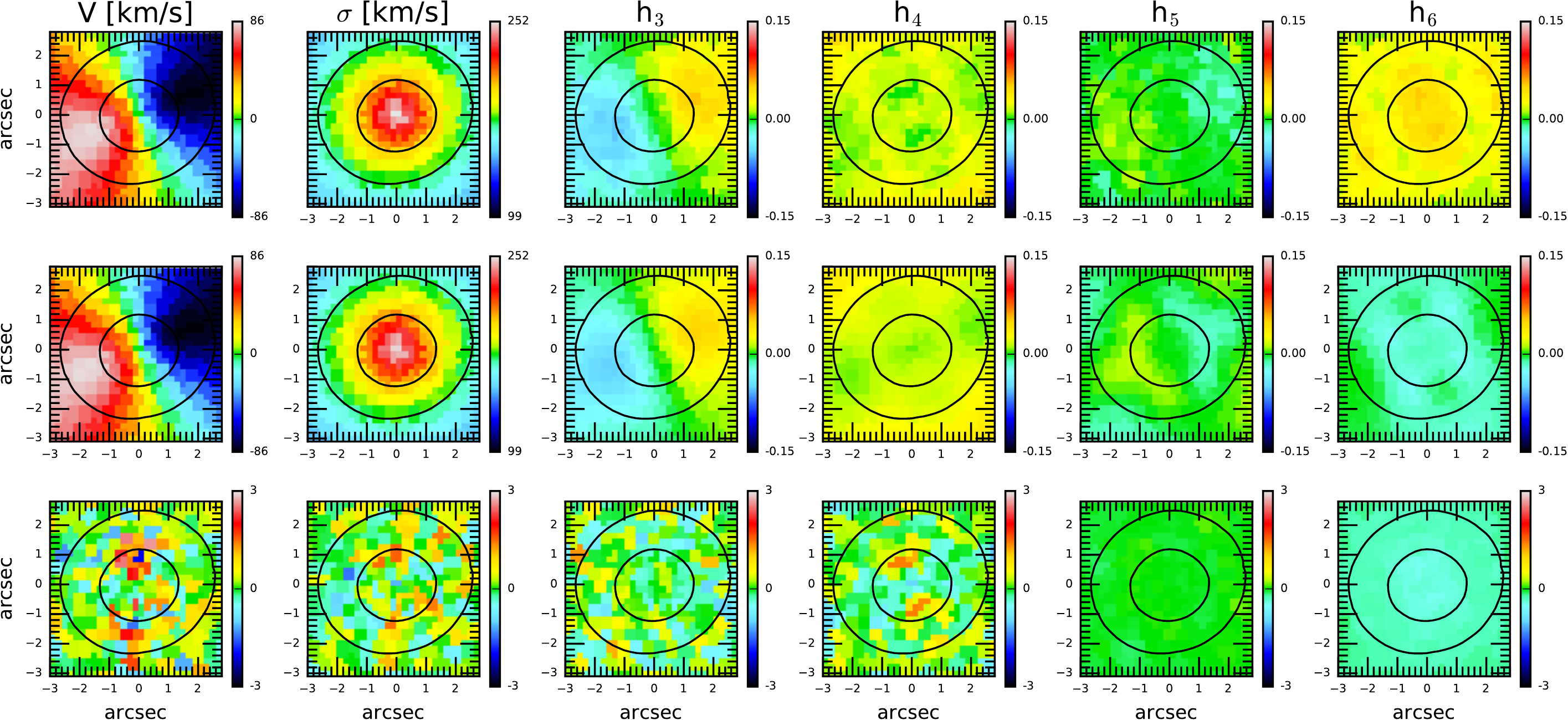}
      \caption{Same as Fig. C1, but zoom-in to the centre.}
      \label{ff:schwarschild_model2}
\end{figure*}

\begin{figure*}

    \includegraphics[width=0.49\textwidth]{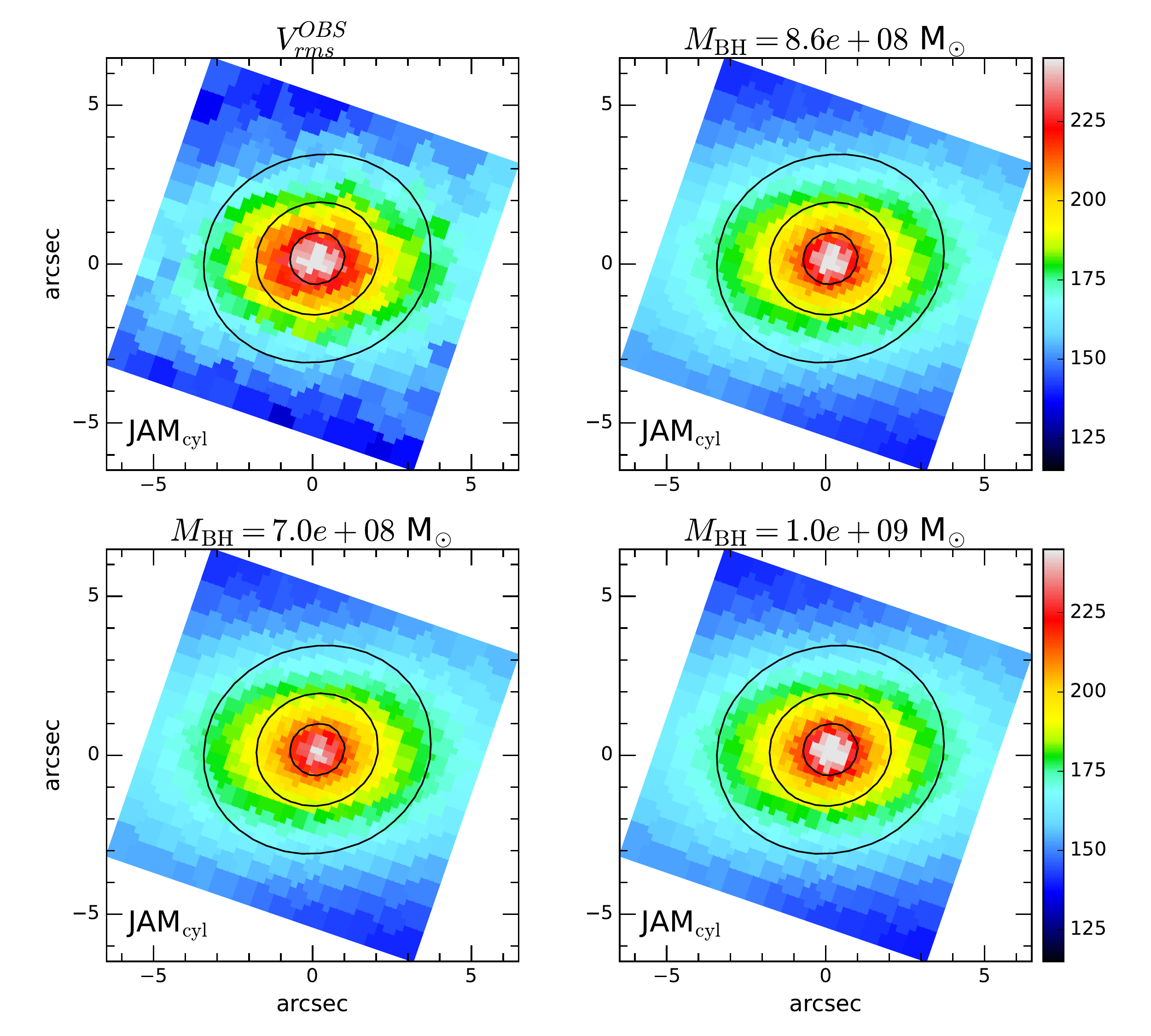}
     \includegraphics[width=0.49\textwidth]{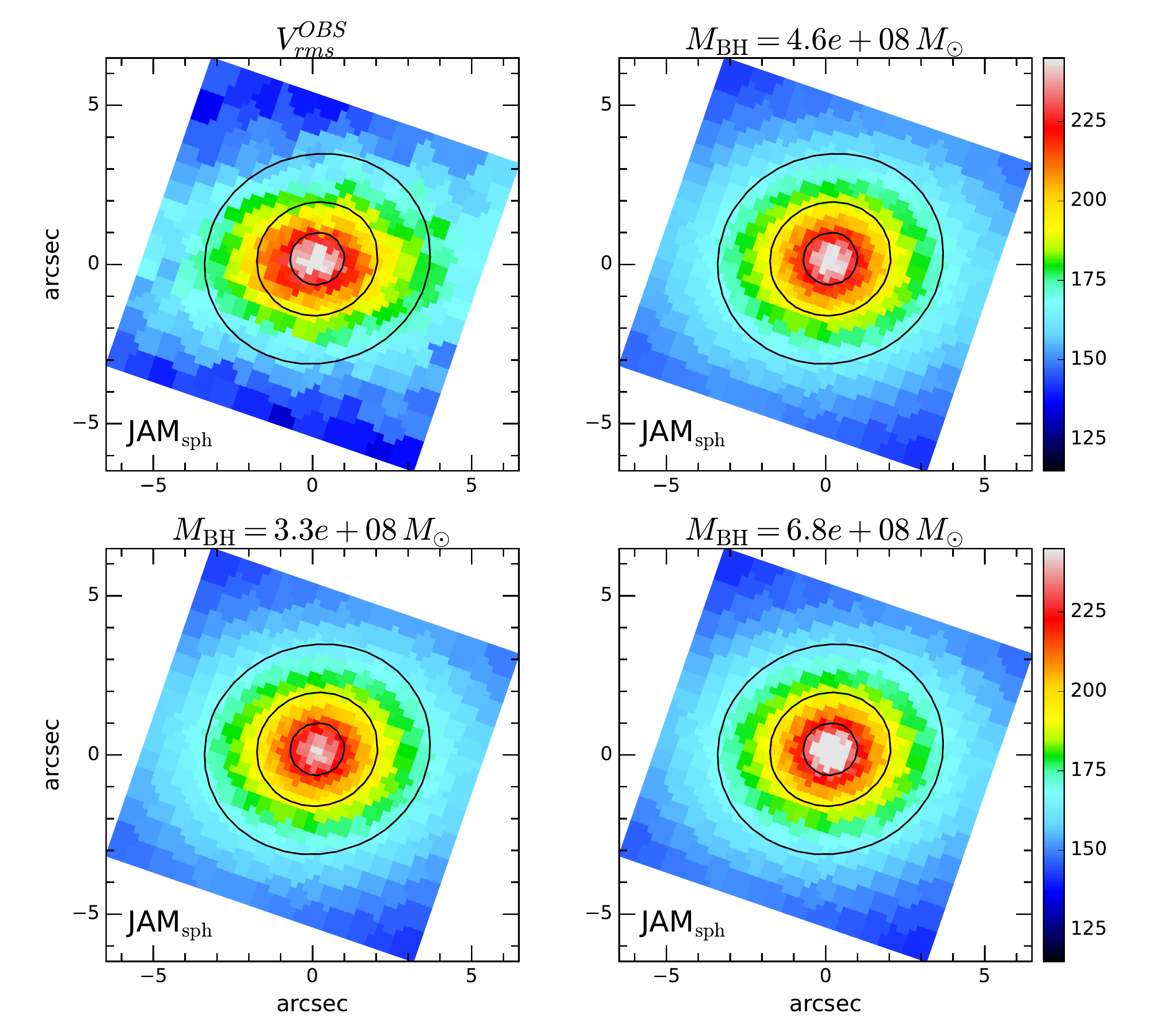}
      \caption{Results of the JAM modelling with cylindrical (left) and spherical (right) velocity ellipsoid alignment ordered into $2\times 2$ panels. The top left panel shows the observed $V_{\rm rms}$ 
       of NGC 6958, the top right panel shows the best-fit JAM model obtained from MCMC (fiducial model). Model and data are mostly in good agreement with each other, but the central $V_{\rm rms}$ is slightly too high. We also show models for a formally too low (bottom left) and too high $M_{\rm BH}$ (bottom right). Note that for JAM$_{\rm cyl}$, the lower-mass black hole recovers the central $V_{\rm rms}$ quite well. Furthermore, both M$_{\rm cyl}$ and M$_{\rm sph}$ models can recover the observed $V_{\rm rms}$ equally well.}
      \label{ff:jeans_maps}
\end{figure*}

\begin{figure*}

    \includegraphics[width=\textwidth]{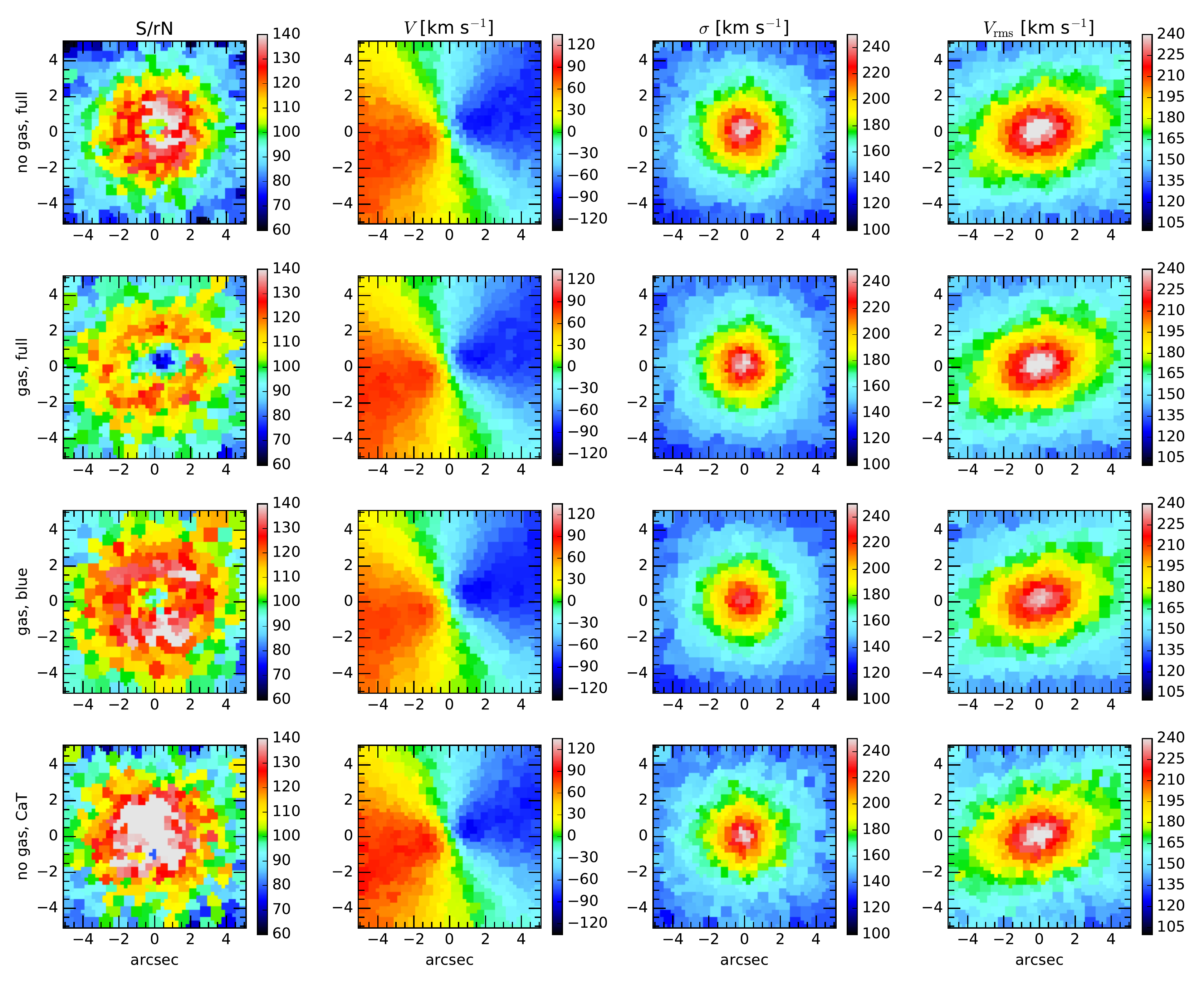}

      \caption{Comparison of the kinematic maps for the different kinematics extractions discussed in Section 4.4.1. From left to right, the panels show the signal-to-residual noise, the rotational velocity after subtracting the systemic velocity, velocity dispersion and $V_{\rm rms}$. From top to bottom, we show the kinematics extractions masking the gas emission lines, simultaneously fitting gas and stellar kinematics, "blue" spectral region and CaT spectral region. The second row is the extraction that we used for the main result of this paper. However, it is clear that the different approaches did not change the extracted kinematics significantly (except for "blue"). North is up and east to the left.}
      \label{ff:kin_maps}
\end{figure*}

\end{document}